\documentclass[printer]{aa}
\usepackage{graphicx}
\usepackage{txfonts}
\usepackage{color}
\usepackage[dvipsnames]{xcolor}
\usepackage{float}
\usepackage{lscape}
\usepackage[colorlinks=true,linkcolor=blue,filecolor=blue,citecolor=blue,urlcolor=blue, draft=False]{hyperref}
\usepackage{natbib}
\usepackage{multirow}

\usepackage{siunitx}
\usepackage{caption}
\DeclareCaptionFormat{cont}{#1#2#3\par}

\newcommand{\citexp}[1]{(\citeauthor{#1}\,\citeyear{#1})}
\newcommand{\citex}[1]{\citeauthor{#1}\,(\citeyear{#1})}
\newcommand{\citexl}[1]{\citeauthor{#1}\,\citeyear{#1}}
\newcommand{\Msun}[1]{\,M$_{\odot}$}
\newcommand{\kmpersec}[1]{\,km.s$^{-1}$}

\usepackage{amstext}

\begin{document} 

   \titlerunning{A catalogue of high-mass X-ray binaries in the Galaxy}
   \authorrunning{F.\,Fortin et al.}

   \title{A catalogue of high-mass X-ray binaries in the Galaxy: from the \emph{INTEGRAL} to the \emph{Gaia} era.\thanks{An online version of the catalogue is publicly available at \url{https://binary-revolution.github.io/HMXBwebcat} and the database in the associated GitHub repository will be continuously updated based on community inputs.}}

   \author{
           Francis Fortin \inst{1} \and
           Federico Garc\'ia \inst{2} \and
           Adolfo Simaz~Bunzel \inst{2} \and
           Sylvain Chaty \inst{1}
          }

   \institute{Universit\'e Paris Cit\'e, CNRS, Astroparticule et Cosmologie, F-75013 Paris, France
     \and
     Instituto Argentino de Radioastronom\'ia (CCT La Plata, CONICET; CICPBA; UNLP), C.C.5, (1894) Villa Elisa, Buenos Aires, Argentina
   }

   \date{received 17/10/2022 ; accepted 28/01/2023}

 
  \abstract
    {
    High-mass X-ray binaries (HMXBs) are a particular class of high-energy sources that require multi-wavelength observational efforts to be properly characterised. New identifications and the refinement of previous measurements are regularly published in the literature by independent teams of researchers and might, when they are collected in a catalogue, offer a tool for facilitating further studies of HMXBs.
    }
    {
    We update previous instances of HMXB catalogues in the Galaxy and provide the community easy access to the most complete set of observables on Galactic HMXBs. In addition to the fixed version that is available in Vizier, we also aim to host and maintain a dynamic version that can be updated upon request from users. Any modification will be logged in this version.
    }
    {
    Using previous HMXB catalogues supplemented by listings of hard X-ray sources detected in the past 20 years, we produced a base set of HMXBs and candidates by means of identifier and sky coordinate cross matches. We queried in Simbad for unreferenced HMXBs. We searched for as many hard X-ray, soft X-ray, optical, and infrared counterparts to the HMXBs as we could in well-known catalogues and compiled their coordinates. Each HMXB was subjected to a meticulous search in the literature to find relevant measurements and the original reference.
    }
    {
    We provide a catalogue of 152 HMXBs in the Galaxy with their best known coordinates, the spectral type of the companion star, systemic radial velocities, component masses, orbital period, eccentricity, and spin period when available. We also provide the coordinates and identifiers for each counterpart we found from hard X-rays to the near-infrared, including 111 counterparts from the recent \textit{Gaia} DR3 catalogue.
    }
    {}


   \maketitle
%

\section{Introduction}\label{sect:introduction}

Since the birth of X-ray astronomy after the discovery of the first extrasolar X-ray source in the early 1960s, thousands of high-energy astrophysical objects were observed and revealed to be of various nature (see the broad review of accreting binaries in \citexl{2022abn..book.....C}). Of these, high-mass X-ray binaries (HMXBs) are powered by the accretion of material from a massive donor star (M$\geq$8\,\Msun\,) onto a compact object, usually a neutron star (NS) and rarely a black hole (BH). HMXBs are usually divided into subclasses, of which BeHXMBs (see review by \citexl{2013AARv..21...69R}) host a fast-rotating Be star, and sgHMXBs (see the review by \citexl{2013AdSpR..52.2132C}) host a supergiant companion. Before the launch of INTErnational Gamma-Ray Astrophysics Laboratory (\textit{INTEGRAL}), sgHMXBs used to be outnumbered about 1 to 10 compared to BeHMXBs. BeHMXBs transfer matter through the interaction of a compact object with a decretion disk, while sgHMXBs generally transfer mass via an intense stellar wind. In some rare instances, accretion in sgHMXBs may take place via Roche-lobe overflow, which produces higher X-ray luminosities than wind-accreting systems. This is the case for Cen X-3, and was more recently suggested for IGR J08408-4503  at periastron \citexp{2019A&A...631A.135D}. Accretion through a Be disk is much more efficient at transporting angular momentum than via wind. The spin of the compact object spin is therefore correlated to the orbital period in BeHMXBs, but not in sgHMXBs (see e.g. \citexl{1984AA...141...91C}).

The \textit{INTEGRAL} satellite (see numbers given in \citexl{2016ApJS..223...15B}) has a higher sensitivity at high energies than previous generations of hard X-ray observatories. sgHMXBs are therefore no longer a minority. Notably, \textit{INTEGRAL} allowed the discovery of highly obscured sgHMXBs \citexp{2004ApJ...616..469F} and supergiant fast X-ray transients (SFXTs; \citexl{2006ESASP.604..165N}).

The discovery and subsequent unambiguous identification of an HMXB requires several observations at various wavelengths. This is usually performed by independent teams of astronomers, and it can take several years before an HMXB is securely associated with a hard X-ray source. This is mainly due to the difficulty of associating soft X-ray, optical, infrared, and radio counterparts with high-energy detections as the astrometrical precision of hard X-ray observatories, which are physically unable to focus the radiation, is systematically outperformed by at least an order of magnitude compared to focusing observatories.

This leads to a lag in the information available on HMXBs and candidate HMXBs, which is spread within the literature; the more time passes, the more tedious it is to recover valuable parameters characterising the binaries, such as the various counterparts, the spectral type of the companion star, the orbital solution, or the detection of a pulse period. Collecting this information in a single place is necessary for a proper overview of the current observational knowledge on HMXBs, and catalogues dedicated to these peculiar sources have therefore been assembled in the past.

The first edition of such a catalogue was compiled by \citex{1983ARAA..21...13B}. Following this, \citex{1995xrbi.nasa..536V} proposed a second edition, which was then further improved by a third \citexp{2000AAS..147...25L}. Eventually, \citex{2006AA...455.1165L} compiled the fourth and latest edition to date of the catalogue (although we note that \citexl{2005AAT...24..151R} proposed a similar work immediately before). We hereby present a catalogue of HMXBs in the Galaxy that covers new information brought during the era from \textit{INTEGRAL} to \textit{Gaia} (2006--2022).

We can identify various arguments towards the necessity of building an updated catalogue of HMXBs. Firstly, the aforementioned catalogues are still being used today, even though they have not been updated for more than 15 years. However, the absence of any recent update pushed us to begin compiling recent information on HMXBs in \citex{2022AA...665A..31F} to infer natal kick properties not only on individual systems, but on the population of BeHMXBs and sgHMXBs. We are still missing crucial parameters on many binaries, however, which narrows the number of systems available for population studies. This shows the need for such catalogues to identify good HMXB candidates to follow up and which information to look for in order to complete our knowledge on these sources. Secondly, with the arrival of the new generation of observing facilities dedicated to high-energy and/or transient astronomy such as the Space Variable Objects Monitor (\textit{SVOM}), the Large Synoptic Survey Telescope (LSST), or eROSITA (for the latter, see e.g. \citexl{2023A&A...669A..30M} for a study of a new BeHMXB in the Large Magellanic Cloud, LMC) and the nascent gravitational astronomy with LIGO\footnote{Laser Interferometer Gravitational-Wave Observatory}, Virgo, KAGRA\footnote{Kamioka Gravitational Wave Detector}, and \textit{LISA}\footnote{Laser Interferometer Space Antenna}, having a contemporaneous view of the current HMXB landscape would be interesting in the scope of population studies. Catalogues of HMXBs have already been used to constrain their properties as a population (see e.g. \citexl{2013AA...560A.108C} and \citexl{2022AA...665A..69F}). HMXBs are also representatives of a source category that is directly related to supernova explosions as well as to compact binaries that finally merge as gravitational wave sources (see a recent review by \citexl{2019IAUS..346....1V}). Comparing the current population of HMXBs with the population of gravitational mergers that is going to build up in the years to come may yield insightful results on stellar evolution in general.

We therefore suggest that for an evolutionary snapshot of the current population of HMXBs, it is necessary to compile measurements on intrinsic binary parameters (orbital period, eccentricity, and systemic radial velocity) as well as measurements of the individual components such as the mass of the compact object (Mx) and its spin period, and of the mass of the optical companion (Mo), its spectral type, and its luminosity class. The latest data release of the \textit{Gaia} satellite \citexp{2022yCat.1355....0G} has made the distances to Galactic binaries now widely available, giving access to their 3D spatial distribution and therefore their place in the Galactic ecology.

We note that many HMXBs are known in the Magellanic Clouds (MCs) and that previous catalogues \citexp{2005AA...442.1135L} may also benefit from an update. As stated in \citex{2006AA...455.1165L}, the sheer number of new data justifies splitting these works, especially in our case, where \textit{Gaia} plays an essential role in the Galaxy (distance determination) that is not applicable to the MCs. The data-mining strategy to recover information about MC HMXBs should also be adapted. Lastly, the population of MC HMXBs is known to be quite different from the Galactic population and therefore deserves a dedicated discussion and paper.

In this paper, we build an updated catalogue of HMXBs and candidate HMXBs in the Galaxy.
We also include systems identified as high-mass gamma-ray binaries (HMGB), which are thought to be powered by the spin-down of a pulsar and not by direct accretion onto it (see e.g. the review in \citexl{2013AARv..21...64D}).
Since the publication of the last HMXB catalogues, high-energy observations (e.g. \textit{INTEGRAL}, \textit{Chandra}, \textit{XMM-Newton}, \textit{Swift}, the Monitor of All-sky X-ray Image \textit{MAXI}, the Nuclear Spectroscopic Telescope Array \textit{NuSTAR}, \textit{Suzaku,} or \textit{Fermi}) and optical/near-infrared (nIR) follow-ups allowed astronomers to discover new HMXBs. Many of the parameters mentioned above, such as spectral type, period, or eccentricity, were accurately determined. While the catalogue contents proposed here will remain fixed (last updated in September 2022), we also host a dynamic version of the catalogue online that is regularly updated when new observations are performed on HMXBs to add new systems or complete the list of known parameters. We strive to find the original references for each measurement we present, and not just reference previous catalogues. In Section\,\ref{sect:catalogue} we describe how the catalogue is built and how we attempted to automatise the search for the multi-wavelength counterparts to HMXBs. We briefly discuss the resulting catalogue and its uses in Section\,\ref{sect:byproducts} before we conclude in Section\,\ref{sect:conclusion}.

\section{Building the catalogue}\label{sect:catalogue}
We describe in this section the steps we took in order to build the catalogue. We first used existing catalogues dedicated to HMXBs and cross-matched them with more recent catalogues of hard X-ray sources. We used the services of the Centre de Données Astronomiques de Strasbourg (CDS), namely Simbad \citexp{2000AAS..143....9W} and Vizier \citexp{2000AAS..143...23O}, to search for updated content on the sources and searched for missing HMXBs. We semi-automatically searched for known counterparts from hard X-rays to the near-infrared. To complete this, we manually compiled all the known parameters available on the HMXBs that we were able to find in the literature and list the proper reference to the original papers.

The following Section\,\ref{subsect:base} is quite similar to what is described in a previous work \citexp{2022AA...665A..31F}, in which we built what can be seen as a precursor to this catalogue. We provide a summary of what has been done and focus on the additions brought in the present work.

\subsection{Reference catalogues}\label{subsect:base}

\citex{2006AA...455.1165L} is the most commonly referred catalogue of HMXBs, listing 114 systems in the Galaxy (including candidates). To build a working base, we added the sources seen by \textit{INTEGRAL} as of 2016 to this catalogue \citexp{2016ApJS..223...15B}. Many of the 939 hard X-ray sources presented in this catalogue are already identified, and nearly 40\% are active galactic nuclei. We thus only added the sources labelled HMXBs, low-mass X-ray binaries (LMXBs), cataclysmic variables (CVs), or still unidentified. Misidentification in the exact type of X-ray binary is not unheard of, therefore we kept all X-ray binaries in this step, and discarded non-HMXB sources only after reviewing the new results published in the literature since then. We performed a positional cross-match using Topcat \citexp{2005ASPC..347...29T} to find the bulk of sources common to both catalogues, and we manually confirmed any duplicates or sources that were left out because of poor astrometrical constraints. Identifiers of the sources were especially useful in this task because they are often similar from one catalogue to the next. This produced a working base of 128 HMXBs.

In parallel, we queried the Simbad database for sources of the type (or subtype) labelled HXB, the identification associated with HMXBs in Simbad. We retrieved 1288 sources in this way. Most of them were extragalactic; they are usually bundled in very tight regions of the sky associated with close-by galaxies, forming dense patches of extragalactic HMXBs. A simple way to automatically detect and remove them is to discard sources with neighbours closer than 6\arcmin\,. We verified that even in the Galactic plane, the sources we retrieved from \citex{2006AA...455.1165L} and \citex{2016ApJS..223...15B} are typically twice as separated (around 15\arcmin\,). This left us with 175 sources, several of which are isolated extragalactic HMXBs, which we discarded later. We note that only 109 of the base HMXBs were found in this way in Simbad; the remaining 19 are simply not labelled HXB. We individually investigated the 66 additional Simbad HXBs in order to supplement our catalogue.

In effect, a majority of these 66 Simbad HXBs are actually LMXBs. Their primary type in Simbad is still set to HMXBs, however, even though a spectral type is available many times and clearly corresponds to a cool main-sequence star. We discarded them, but kept the remaining entries even when no precise information on spectral type was available in Simbad because we performed a thorough manual search for this information later. At this point, we had a set of 145 HMXBs and candidate HMXBs.

\subsection{Finding an unambiguous chain of counterparts}\label{subsect:counterparts}

We considered that a secure identification of an HMXB partly comes from having an unambiguous list of its detections from hard X-rays down to the near-infrared. This ensures that none of them are blended with close-by high-energy sources, and it efficiently removes sources listed as HXMBs in the literature that were detected only once in hard X-rays 40 years ago and have had no new detection since then.

Hence, we verified each of the HMXBs in the present catalogue for their counterparts at various wavelengths. In increasing typical astrometric precision, we cross-matched the available position of HMXBs with the catalogues listed in Table\,\ref{tab:counterparts}. Independently of the origin of the positional data that were retrieved, we first queried each catalogue in a cone whose angular size varied depending on 1) the typical astrometrical accuracy of the queried catalogue and 2) the accuracy of the initial positional data. If the positional data were more accurate than the queried catalogue, the cone size was set to the radii given in Table\,\ref{tab:counterparts}, which are about twice of the worst astrometric performance in the corresponding catalogue. If the astrometric precision of the queried catalogue was more accurate than the positional data, the cone size was set to the error available in the positional data.

\begin{table}
    \centering
    \caption{List of queried catalogues for the counterpart search.}
    \label{tab:counterparts}
    \begin{tabular}{lll}
        \hline\hline\\[-2ex]
        Catalogue & Reference & Radius \\
        \hline\\[-2ex]
        \textit{HEAO 1} & \citex{1984ApJS...56..507W} & 20\arcmin \\
        \textit{Uhuru 4}  & \citex{1978ApJS...38..357F} & 20\arcmin\\
        \textit{Ariel V 3} &  \citex{1981MNRAS.197..865W} & 20\arcmin \\
        \textit{INTEGRAL}  & \citex{2016ApJS..223...15B} & 20\arcmin \\
        \textit{Fermi}  & \citex{2022ApJS..260...53A} & 20\arcmin \\
        \textit{BeppoSAX}  & \citex{2011ApJS..195....9C} & 6\arcmin  \\
        \textit{Einstein 2E} &  \citex{1990EObsC...2.....H} & 4\arcmin \\
        \textit{ROSAT}  & \citex{2000yCat.9031....0W} & 35\arcsec \\
        \textit{Swift} 2SXPS  & \citex{2020ApJS..247...54E} & 8\arcsec \\
        4XMM DR11 &  \citex{2020AA...641A.136W} & 4\arcsec \\
        \textit{Chandra} CSC 2 &  \citex{2019yCat.9057....0E} & 3\arcsec \\
        2MASS &  \citex{2003yCat.2246....0C} & 120\,mas \\
        \textit{Gaia} DR3 &  \citex{2022yCat.1355....0G} & 20\,mas \\
        \hline
    \end{tabular}
\end{table}

Then, after reviewing the counterparts found at high energies, we performed a recursive search, from poorly accurate counterparts to the most accurate catalogues (2MASS and \textit{Gaia}). This allowed us to recover the chain of detection from high energies down to the optical/nIR wavelengths, as well as the soft X-ray detections whose astrometrical accuracies (particularly from \textit{Chandra} and \textit{XMM-Newton}) can rival optical telescopes.

There is a limit in this process because this automatic query can generate false counterparts because the typical astrometrical accuracies that we used are based on the worst performance of each facility, so that any systematic errors in the astrometric calibration between catalogues could be taken into account. Systematic errors in astrometry appear to be especially large in older catalogues (\textit{Uhuru}, the High Energy Astronomy Observatory \textit{HEAO}, or \textit{Ariel V}) because we often find that the historical detections of high-energy sources are not exactly compatible with more recent detections (e.g. \textit{INTEGRAL} or \textit{Swift}) when considering their 90\% positional uncertainty. We also note that for observing facilities with astrometrical accuracies of about 1\arcsec\, or lower (\textit{Swift}, \textit{XMM-Newton}, \textit{Chandra}, 2MASS, and \textit{Gaia}), we added 0.5\arcsec\, to the positional uncertainty when validating the chain of counterparts. For instance, some \textit{XMM-Newton}, \textit{Chandra}, 2MASS, and \textit{Gaia} detections of the same source can be so precise that they are not technically compatible with one another; for Galactic sources, even when we look towards the Galactic plane in crowded regions, it is unlikely that two separate sources lie closer than 0.5\arcsec. Using this value of systematic error was already successful in \citex{2022AA...665A..31F}, who searched for unambiguous \textit{Gaia} counterparts to 2MASS sources.

We verified each individual result of this automatic counterpart search. We manually removed false detections of counterparts, and searched for actual counterparts in the literature when necessary. When we manually input coordinates from specific publications, we added a reference towards it in the online catalogue; they usually come from Astronomer's Telegrams\footnote{\url{https://www.astronomerstelegram.org/}} and are therefore not necessarily present in the queried catalogues.

\subsection{Retrieving binary parameters and new HMXBs}\label{subsect:parameters}

We made extensive use of NASA's Astrophysics Data System\footnote{\url{https://ui.adsabs.harvard.edu/\#}} (ADS) to recover the parameters and their corresponding references. Some papers greatly facilitated the process as they already listed information on some HMXBs in our catalogue. Orbital periods, spin periods, and spectral types are found in \citex{2009ApJ...707..870B}, spin periods of pulsars  are reported in \citex{2010AA...520A..76A}, spectroscopic information on Ae/Be stars is given in \citex{2015MNRAS.453..976F}, tabled data on BeHMXBs is presented in \citex{2017MNRAS.470..126T} and \citex{2017AA...598A..16R}, HMXBs detected by \textit{INTEGRAL} are reported in \citex{2018MNRAS.481.2779S}, an overview of SFXT candidates is given in \citex{2020MNRAS.491.4543S}, much information on radio pulsars is collected in \citex{2021MNRAS.507.3899V}, \textit{XMM-Newton} and \textit{Swift} observations of sgHMXBs are reported in \citex{2022AA...664A..99F}, and HMXBs seen by \textit{Fermi} are presented in \citex{2022MNRAS.512.1141H}.

For each information we compiled (spectral type, systemic radial velocity, masses, orbital period, spin period, and eccentricity), we provide the reference to the paper that first reported the measurements. While the articles listed above greatly sped up the process, we still manually checked each and every listed source in ADS and Simbad to search for any missing measurement and/or reference. This step is crucial not only to gather the most complete set of data on HMXBs in one place, but also to ensure that we do not cite papers in which no actual measurement was made. This facilitates determining the original source.

Furthermore, we also searched for papers announcing the detection of new HMXBs between 2016 and 2022, and added any new entry to the catalogue after performing the same precautionary steps described in this section. We mention for instance HD\,96670, which was recently identified as new BH HMXB candidate in \citex{2021ApJ...913...48G}.

\subsection{Contents of the catalogue}\label{subsect:contents}

In Table\,\ref{tab:cat:general} we provide a single identifier that is either the historical name of the HMXBs, the most commonly used (e.g. for \textit{INTEGRAL} sources), or the main identifier as queried in Simbad. This service can be used to retrieve other identifiers available for the HMXBs. The "Spectype" column refers to the spectral type of the donor star in the binary. We also provide an indication of the subclass of the HMXBs: Be, supergiant (sg), supergiant fast X-ray transient (SFXT), and a few peculiar subclasses such as sgB[e] or Wolf-Rayet (WR). Most of the subclass information comes from the spectral type of the companion; if no spectral type is provided, a reference may be available to motivate the choice of subclass. The sky coordinates of the most accurate counterpart we found are listed alongside their 90\% positional error. We also include distance inferences from \citex{2021AJ....161..147B} when a \textit{Gaia} DR3 counterpart is available. These distances are based on \textit{Gaia} EDR3, and as a result, they cannot be directly retrieved using the \textit{Gaia} DR3 identifiers we provide in the full catalogue; instead, we retrieved the \textit{Gaia} EDR3 identifiers first using a cone sky match, and then queried the distances in \citex{2021AJ....161..147B}. Finally, Table\,\ref{tab:cat:general} provides a variability flag ("\textit{Var}") that summarises whether the HMXBs were flagged as variable sources in the \textit{INTEGRAL}, 4XMM DR11, or \textit{Chandra} catalogues, or if the ratio of the peak to mean flux in the \textit{Swift} 2SXPS catalogue is greater than 5. The detailed information about individual variability flags is given in the on-line version of the catalogue.

In Table\,\ref{tab:cat:orbital} we introduce the orbital characteristics of the catalogued HMXBs. We have separated this information from Table\,\ref{tab:cat:general} for readability, but the full on-line catalogue contains information from both tables together\footnote{Tables \ref{tab:cat:general} and \ref{tab:cat:orbital} are available as a single electronic table
at the CDS via anonymous ftp to cdsarc.cds.unistra.fr (130.79.128.5)
or via https://cdsarc.cds.unistra.fr/cgi-bin/qcat?J/A+A/}. First are given indications on the mass of the compact object (Mx) and the companion star (Mo). Companion masses that were broadly inferred from the spectral type by us are labelled with a $\text{dagger}$; we used the atlas of Be stars from \citex{1996MNRAS.280L..31P} and the stellar parameters for O stars available in \citex{2005AA...436.1049M}. The orbital period, eccentricity, spin period, and radial systemic velocity are given as available in the literature.

In addition to all the information in Tables\,\ref{tab:cat:general} and\,\ref{tab:cat:orbital}, the on-line version of the catalogue provides a list of the multi-wavelength counterparts to each HMXB. For each counterpart, we provide the right ascension and declination in J2000, the 90\% positional error, and the identifier as listed in the queried catalogues. This can facilitate any further cross match because sky matches can produce false associations, and identifiers help to identify any mistake in this matter.

The full catalogue content is available on Vizier in a fixed version. We also host it in a dynamic version that can be browsed online\footnote{Available at \url{https://binary-revolution.github.io/HMXBwebcat5/}}, and which will be updated upon the request of users. New versions will be regularly published on the website and will be available for download in various formats.

\section{Results, discussions, and byproducts}\label{sect:byproducts}

\begin{figure}[h]
    \centering
    \includegraphics[width=\columnwidth]{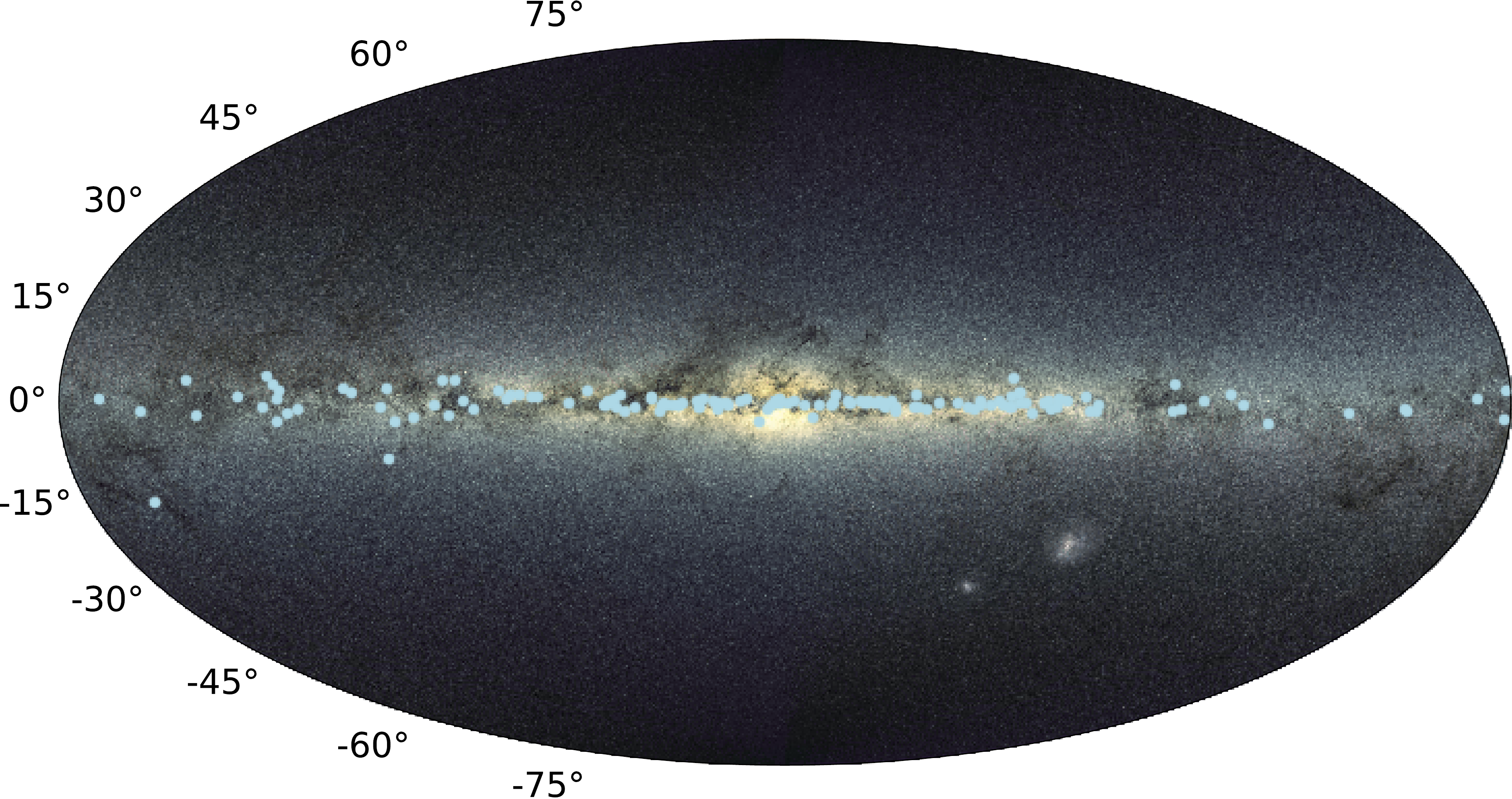}
    \caption{Edge-on view of the 152 HMXBs in the Galaxy. Galactic latitudes are indicated in degrees. \textit{Background image credits: ESA/Gaia/DPAC.}}
    \label{fig:edgeon}
\end{figure}

\begin{figure}
    \centering
    \includegraphics[width=\columnwidth]{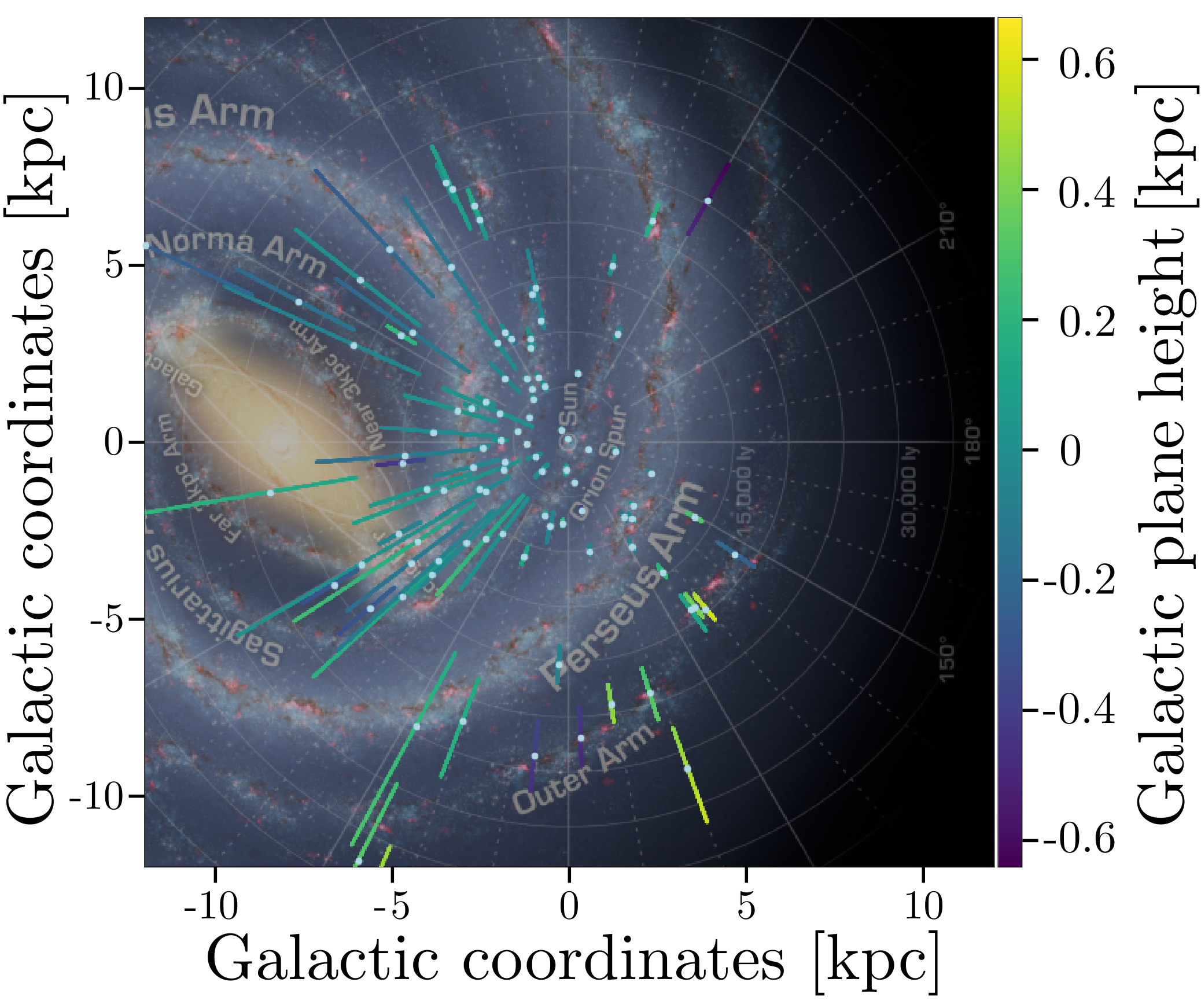}
    \caption{Face-on view of the 107 Galactic HMXBs with \textit{Gaia} parallaxes. Bars indicate the 68\%\,confidence interval in distance. \textit{Background image credits: NASA/JPL-Caltech/R. Hurt (SSC/Caltech)}}
    \label{fig:faceon}
\end{figure}

In this catalogue, we present 152 HMXBs and candidates in the Galaxy. This is a 33\% increase from \citex{2006AA...455.1165L} for the whole sample. We can also compare the increase in securely identified HMXBs because the 2006 catalogue mentions that only 63 were confirmed systems, the remaining 51 were candidates at that time. In the current catalogue, if we consider HMXBs for which we have a spectral type indicative of a massive star as confirmed, then we count 126 confirmed HMXBs. If we add to this those with a detected orbital period and spin period, this pushes the number of confirmed HMXBs to 134, more than twice the number of \citex{2006AA...455.1165L}. The Galactic sky map of HMXBs is shown in Figure\,\ref{fig:edgeon}. We note that 111 of the HMXBs have a \textit{Gaia} DR3 counterpart, of which 4 do not have a parallax estimation. We show the face-on Galactic distribution of HMXBs seen by \textit{Gaia} in Figure\,\ref{fig:faceon}, which indicates a positional correlation between Galactic spiral arms and HMXBs that was recently explored in \citex{2022AA...665A..69F} along with Galactic stellar clusters to retrieve the possible birthplaces and age of the binaries.

We find that the current number of BeHMXBs in the Galaxy is 74; there are 52 sgHMXBs, of which 21 are SFXT candidates, and 5 are sgB[e] systems. Two HMXBs have a Wolf-Rayet companion. We also note that the spectral type of the companion is poorly constrained in 28 HMXBs, which indicates that optical/nIR identification campaigns are still very necessary. \citex{2006AA...455.1165L} listed 50 BeHMXBs and 16 sgHMXBs according to the listed spectral types. This means that we improve the census of these subclasses by 50\% and more than 200\%, respectively. The dramatic increase in known sgHMXBs over the past 15 years is associated with the performances of the \textit{INTEGRAL} satellite at high energies, which has at least in part lifted the observational bias we had towards BeHMXBs.

\subsection{Examples of use}\label{subsect:examples}

This catalogue was compiled to facilitate the retrieval of information on HMXBs and to allow for considerations to be made not only on individual systems, but on Galactic HMXBs as a population. We provide two examples of how this catalogue can be used for this purpose.

\begin{figure}
    \centering
    \includegraphics[width=\columnwidth]{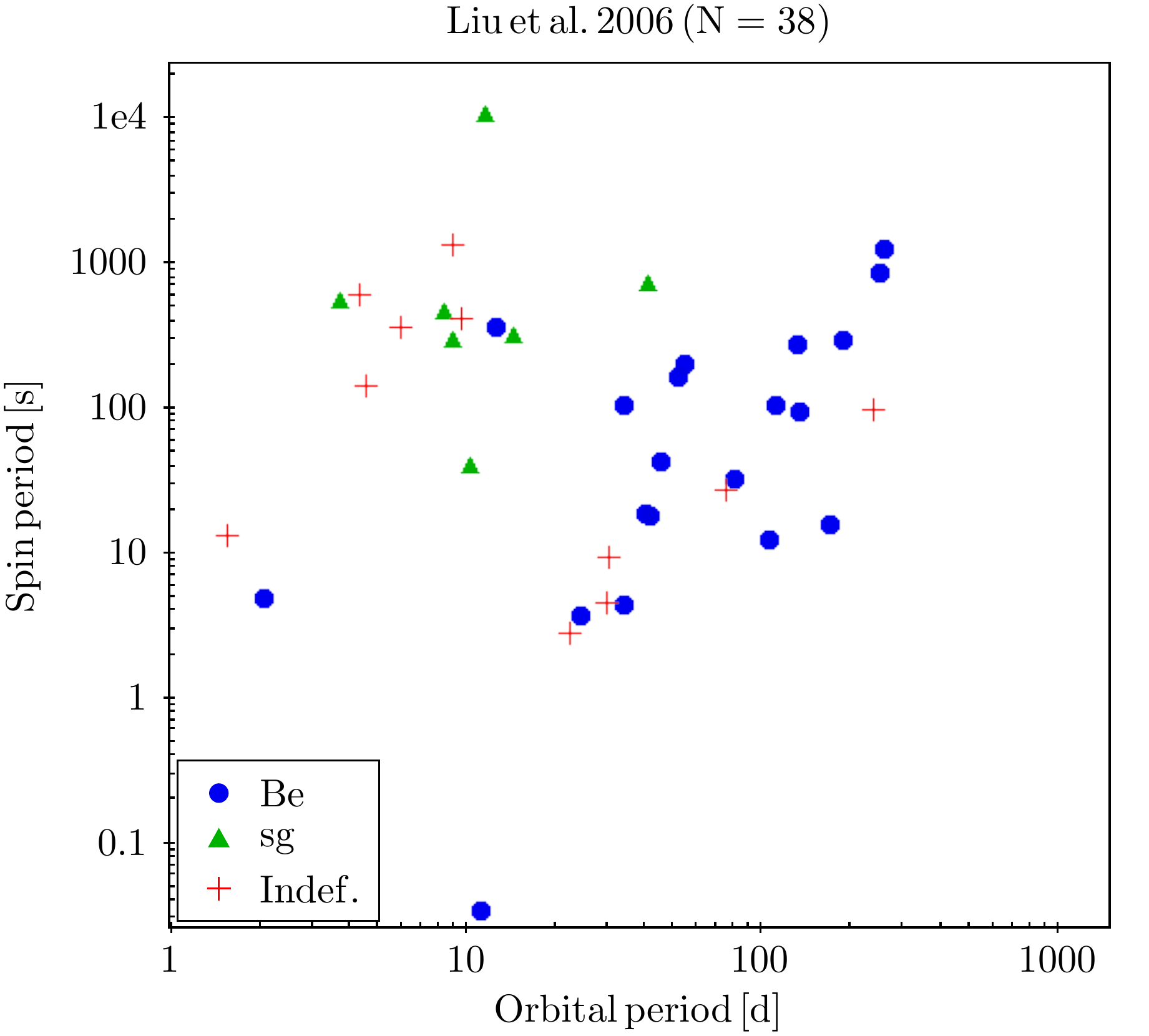}
    \bigbreak
    \includegraphics[width=\columnwidth]{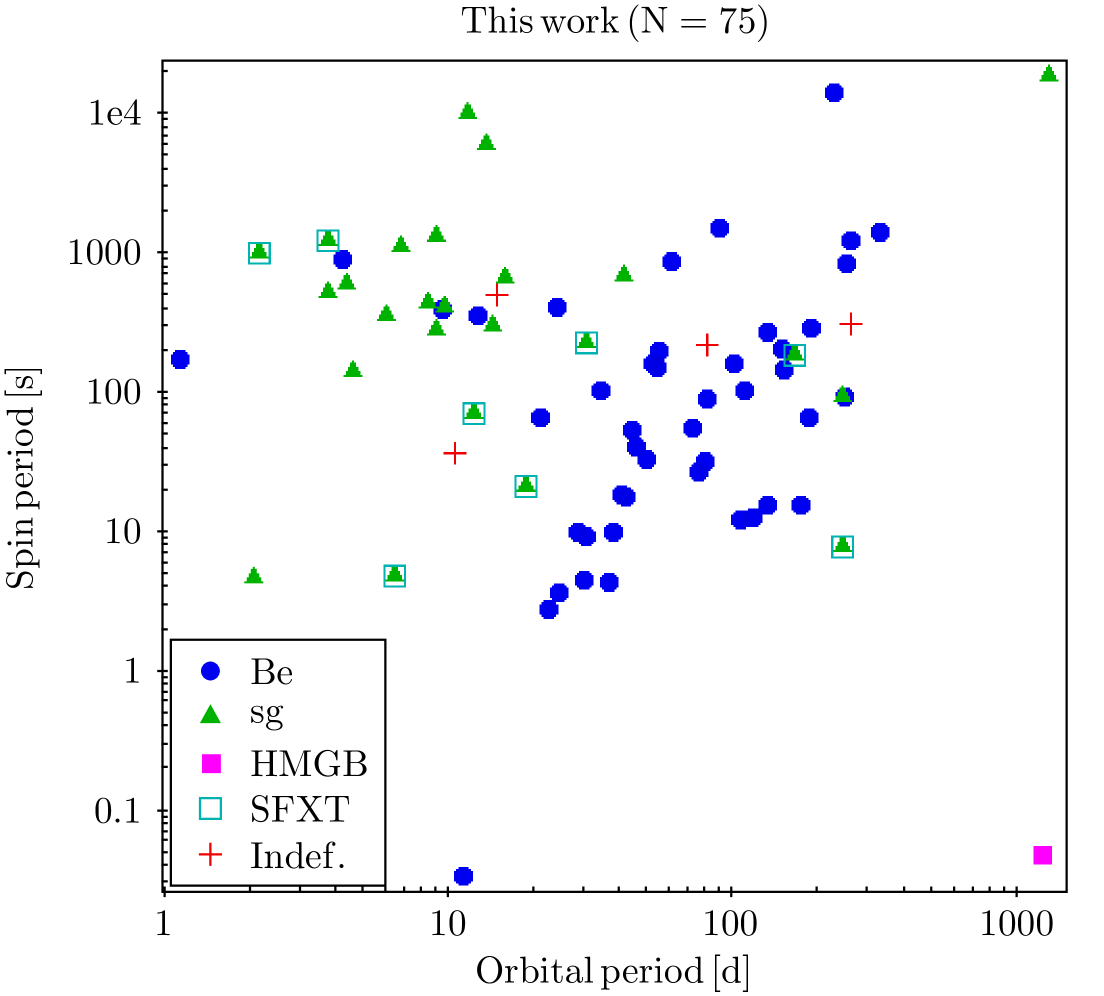}
    \caption{Corbet diagram of the 38 HMXBs in the \citex{2006AA...455.1165L} catalogue (top panel) and the 75 HMXBs in the current catalogue (bottom panel). BeHMXBs are shown as blue dots, sgHMXBs (SFXTs) are shown as green triangles (squares), HMGB are shown as pink squares, and the remaining HMXBs with a peculiar and/or unclear spectral type are shown as red crosses ("Indef").}
    \label{fig:corbet}
\end{figure}

As a first example of how the data may be used is to build a Corbet diagram, shown in Figure\,\ref{fig:corbet} with the 38 HMXBs in \citex{2006AA...455.1165L} (top panel) and the 75 HMXBs from the current catalogue (bottom panel), for which orbital period and spin period are determined.
It is a great tool for visualising the effect of mass transfer in wind accretion versus decretion disk. Because the angular momentum is transferred very efficiently when an NS interacts with a decretion disk, BeHMXBs present a strong correlation (P$_{spin}$\,$\propto$\,P$_{orb}^2$, \citexl{1984AA...141...91C}); on the other hand, sgHMXBs do not show a significant correlation as wind accretion is inefficient at angular momentum transfer.

As expected, the updated Corbet diagram shows a dichotomy between sgHMXBs, which tend to have shorter orbital periods and host more slowly spinning NSs, and BeHMXBs with longer orbital periods but slightly faster-spinning NSs. Even with the greatly improved census on sgHMXBs, they do not show any particular correlation in the Corbet diagram, as opposed to BeHMXBs (see e.g. \citexl{2014ApJ...786..128C}), whose orbital period generally increases with spin period. A few remarkable systems can be quickly identified: the two millisecond pulsars SAX\,J0635.2+0533 and PSR\,B1259-63 (the latter orbiting its companion in more than 1000\,d), and at the opposite end, the very slowly rotating 1A\,0114+650, IGR\,J19140+0951, 1H\,1249-637, and 4U\,1954+31. The last system is also peculiar because it is the only HMXB in the Galaxy with a confirmed M\,I massive supergiant donor star \citexp{2020ApJ...904..143H}.

\begin{figure}
    \centering
    \includegraphics[width=\columnwidth]{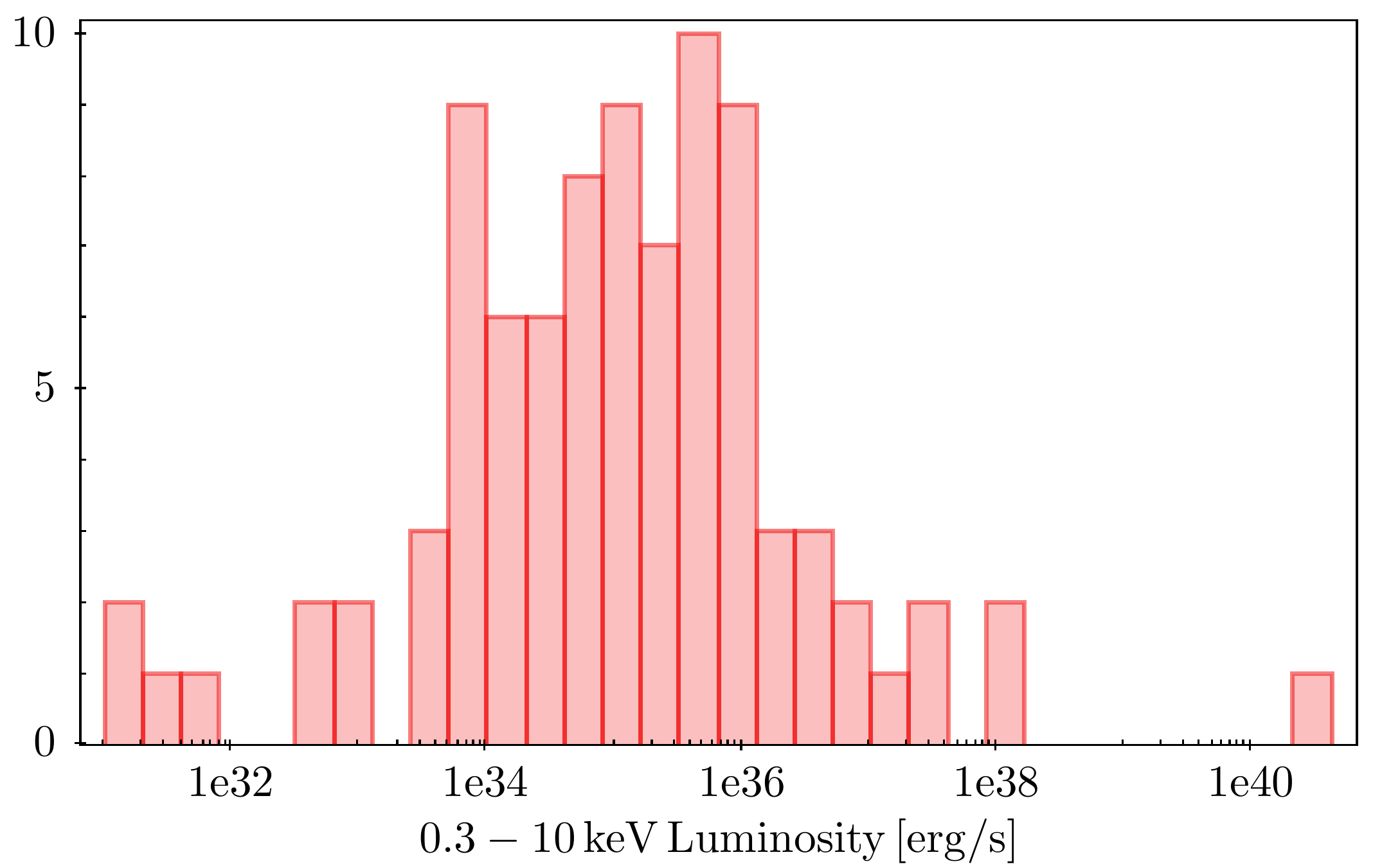}
    \caption{Distribution of soft X-ray luminosities of the Galactic HMXBs seen by \textit{Swift} and \textit{Gaia} (N=89).}
    \label{fig:Xray_luminosity}
\end{figure}

As a second example, we built a distribution of soft X-ray luminosities of HMXBs in the Galaxy. The current catalogue does not list the common high-energy information such as X-ray fluxes, hardness ratio, or hydrogen column density. HMXBs can be variable sources or be obscured, and the modelling of their high-energy emission requires a case-by-case approach, which is why we do not provide such high-level information. However, with the provided list of their counterparts in soft and hard X-ray catalogues, users can easily find this information. First, we use the distances in Table\,\ref{tab:cat:general}, which were queried in \citex{2021AJ....161..147B} using the \textit{Gaia} DR3 positions. Then, we query the \textit{Swift} 2SXPS catalogue using their available \textit{Swift} identifiers, and fetch the value of the unabsorbed flux in the 0.3--10\,keV band (column \textit{apec\_flux\_b}). In Figure\,\ref{fig:Xray_luminosity} we present the distribution of X-ray luminosities that can be derived from \textit{Swift} and \textit{Gaia} data. We note that the source showing extreme X-ray luminosity at $>$10$^{40}$\,erg/s is IGR J16318-4848, the prime example of an absorbed sgB[e]HMXBs (see \citexl{2020ApJ...894...86F} for recent broadband observations of this binary). This luminosity should clearly be considered with caution because of the uncertainty on the very high absorption in the line of sight and on the distance to the source. This is but a very crude example, as the users might wish to consider other X-ray bands or hardness ratios coming from their preferred observatories, or might consider the exact models used to infer fluxes (de-reddened or not, power law vs. black body, etc). There are many other possibilities of use for this catalogue depending on the user's goal.

\section{Conclusion}\label{sect:conclusion}

After more than 15 years of multi-wavelength observation campaigns, the landscape of Galactic HMXBs has changed significantly. Much information about them is available throughout the literature. We present an updated catalogue of HMXBs in the Milky Way containing not only basic information such as identifiers, subclasses, and positions, but also multi-wavelength counterparts and orbital binary parameters. These are available  from an in-depth both automatised and manual survey performed across published papers and catalogues of high-energy sources.

\begin{figure}
    \centering
    \includegraphics[width=\columnwidth]{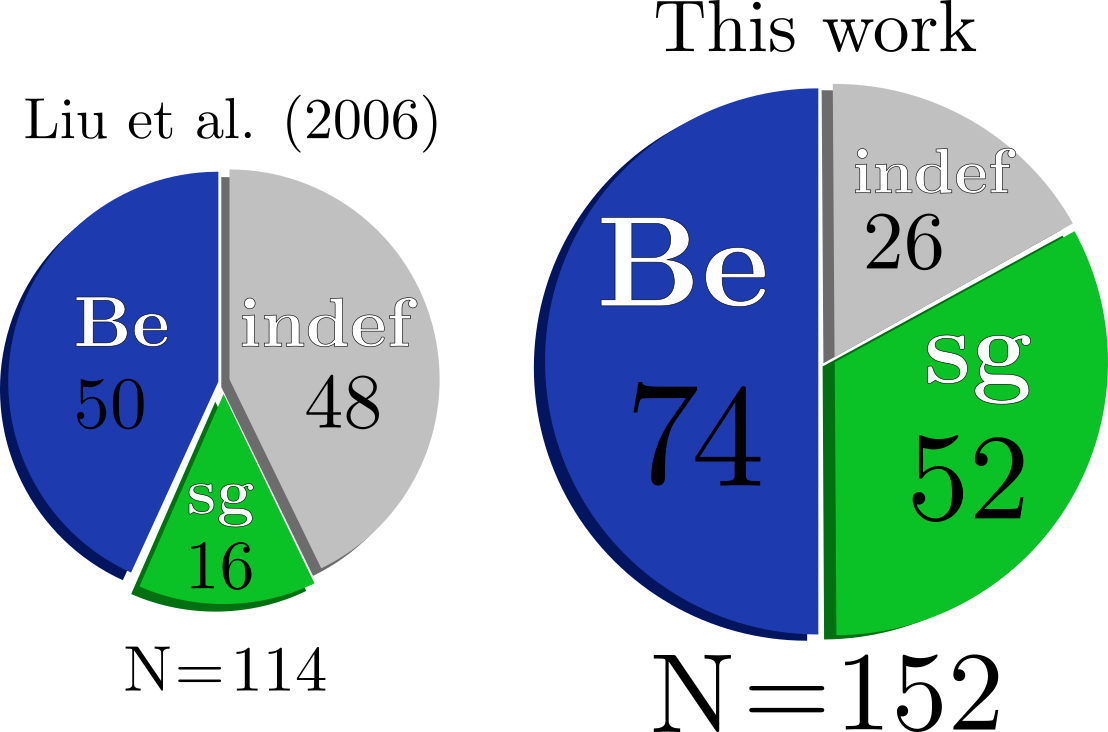}
    \caption{Evolution of the number and nature of HMXBs in the Galaxy with an identified spectral type from 2006 (left) to 2022 (right).}
    \label{fig:camembert}
\end{figure}

Compared to the last published catalogue of HMXBs by \citex{2006AA...455.1165L}, the total number of HMXBs known in the Galaxy has increased by roughly 33\% (see Figure\,\ref{fig:camembert}), by a factor of two when considering confirmed systems, and by a factor of three in the particular case of sgHMXBs. The latter most definitely benefited from the capabilities of \textit{INTEGRAL} and HMXBs in general, through many focused optical/nIR identification campaigns, as well as multiple follow-up efforts in the soft X-ray band, which are essential in the process of constraining the exact position of hard X-ray sources in the sky. In addition, the data collected by the \textit{Gaia} satellite since 2015 offer unrivalled estimates of positions and velocities, including distances to HMXBs across two-thirds of the catalogue, which it was not possible to achieve at this scale before.

The search for new X-ray binaries and information on them is still active, and the arrival of new observing facilities will ensure continued interest in this field. The eROSITA, SVOM, and LSST observatories will not only contribute to studying currently known or new persistent systems, but will also provide much more information on transient sources and therefore provide insight into other stages of binary evolution such as supernova explosions or merger events. The addition of the gravitational messenger by the LIGO/Virgo/KAGRA observatories will work in synergy with electromagnetic transient sky facilities to constrain the endpoint of binary evolution; we will soon, if we do not already, have access to observational data on phases spanning the entire life of massive binary stars. The coming years will thus provide many opportunities for studying the evolution of massive stars in binaries, which contribute to the Galactic ecology by their X-ray emission, heavy nucleus formation, and possible retro-action on the interstellar medium.

\begin{acknowledgements}
We thank the anonymous Referee for their insightful remarks that helped us improving both this paper and the online catalogue.
The authors were supported by the LabEx UnivEarthS: Interface project I10 "Binary rEvolution: from binary evolution towards merging of compact objects".
SC is grateful to the CNES (Centre National d'\'Etudes Spatiales) for the funding of MINE (Multi-wavelength \textit{INTEGRAL} Network).
FG is CONICET researcher. FG acknowledges support by PIP 0113 (CONICET), PICT-2017-2865 (ANPCyT) and PIBAA 1275 (CONICET).
This work made use of NASA's Astrophysics Data System (ADS) web services, and of the services associated to the Centre de Données Astronomiques de Strasbourg (CDS) Simbad and Vizier.
This work has made use of data from the European Space Agency (ESA) mission
{\it Gaia} (\url{https://www.cosmos.esa.int/gaia}), processed by the {\it Gaia}
Data Processing and Analysis Consortium (DPAC,
\url{https://www.cosmos.esa.int/web/gaia/dpac/consortium}). Funding for the DPAC
has been provided by national institutions, in particular, the institutions
participating in the {\it Gaia} Multilateral Agreement.

{\em Software:} Topcat \citexp{2005ASPC..347...29T}, {\sc matplotlib} \citexp{hunter_matplotlib_2007}, {\sc NumPy} \citexp{van_der_walt_numpy_2011}, {\sc scipy} \citexp{jones_scipy_2001} and {\sc Python} from \url{python.org}


\end{acknowledgements}

\bibliographystyle{aa}
\bibliography{references}

\begin{thebibliography}{359}
\expandafter\ifx\csname natexlab\endcsname\relax\def\natexlab#1{#1}\fi

\bibitem[{{Abdollahi} {et~al.}(2022){Abdollahi}, {Acero}, {Baldini}, {Ballet},
  {Bastieri}, {Bellazzini}, {Berenji}, {Berretta}, {Bissaldi}, {Blandford},
  {Bloom}, {Bonino}, {Brill}, {Britto}, {Bruel}, {Burnett}, {Buson}, {Cameron},
  {Caputo}, {Caraveo}, {Castro}, {Chaty}, {Cheung}, {Chiaro}, {Cibrario},
  {Ciprini}, {Coronado-Bl{\'a}zquez}, {Crnogorcevic}, {Cutini}, {D'Ammando},
  {De Gaetano}, {Digel}, {Di Lalla}, {Dirirsa}, {Di Venere}, {Dom{\'\i}nguez},
  {Fallah Ramazani}, {Fegan}, {Ferrara}, {Fiori}, {Fleischhack}, {Franckowiak},
  {Fukazawa}, {Funk}, {Fusco}, {Galanti}, {Gammaldi}, {Gargano}, {Garrappa},
  {Gasparrini}, {Giacchino}, {Giglietto}, {Giordano}, {Giroletti}, {Glanzman},
  {Green}, {Grenier}, {Grondin}, {Guillemot}, {Guiriec}, {Gustafsson},
  {Harding}, {Hays}, {Hewitt}, {Horan}, {Hou}, {J{\'o}hannesson}, {Karwin},
  {Kayanoki}, {Kerr}, {Kuss}, {Landriu}, {Larsson}, {Latronico},
  {Lemoine-Goumard}, {Li}, {Liodakis}, {Longo}, {Loparco}, {Lott}, {Lubrano},
  {Maldera}, {Malyshev}, {Manfreda}, {Mart{\'\i}-Devesa}, {Mazziotta}, {Mereu},
  {Meyer}, {Michelson}, {Mirabal}, {Mitthumsiri}, {Mizuno}, {Moiseev},
  {Monzani}, {Morselli}, {Moskalenko}, {Negro}, {Nuss}, {Omodei}, {Orienti},
  {Orlando}, {Paneque}, {Pei}, {Perkins}, {Persic}, {Pesce-Rollins},
  {Petrosian}, {Pillera}, {Poon}, {Porter}, {Principe}, {Rain{\`o}}, {Rando},
  {Rani}, {Razzano}, {Razzaque}, {Reimer}, {Reimer}, {Reposeur},
  {S{\'a}nchez-Conde}, {Saz Parkinson}, {Scotton}, {Serini}, {Sgr{\`o}},
  {Siskind}, {Smith}, {Spandre}, {Spinelli}, {Sueoka}, {Suson}, {Tajima},
  {Tak}, {Thayer}, {Thompson}, {Torres}, {Troja}, {Valverde}, {Wood}, \&
  {Zaharijas}}]{2022ApJS..260...53A}
{Abdollahi}, S., {Acero}, F., {Baldini}, L., {et~al.} 2022, \apjs, 260, 53

\bibitem[{{Abubekerov} {et~al.}(2004){Abubekerov}, {Antokhina}, \&
  {Cherepashchuk}}]{2004ARep...48...89A}
{Abubekerov}, M.~K., {Antokhina}, {\'E}.~A., \& {Cherepashchuk}, A.~M. 2004,
  Astronomy Reports, 48, 89

\bibitem[{{Adams} {et~al.}(2021){Adams}, {Benbow}, {Brill}, {Buckley},
  {Capasso}, {Chromey}, {Errando}, {Falcone}, {A Farrell}, {Feng}, \&
  et~al.}]{2021ApJ...923..241A}
{Adams}, C.~B., {Benbow}, W., {Brill}, A., {et~al.} 2021, \apj, 923, 241

\bibitem[{{Annala} \& {Poutanen}(2010)}]{2010AA...520A..76A}
{Annala}, M. \& {Poutanen}, J. 2010, \aap, 520, A76

\bibitem[{{Antokhin} {et~al.}(2022){Antokhin}, {Cherepashchuk}, {Antokhina}, \&
  {Tatarnikov}}]{2022ApJ...926..123A}
{Antokhin}, I.~I., {Cherepashchuk}, A.~M., {Antokhina}, E.~A., \& {Tatarnikov},
  A.~M. 2022, \apj, 926, 123

\bibitem[{{Aragona} {et~al.}(2010){Aragona}, {McSwain}, \& {De
  Becker}}]{2010ApJ...724..306A}
{Aragona}, C., {McSwain}, M.~V., \& {De Becker}, M. 2010, \apj, 724, 306

\bibitem[{{Aragona} {et~al.}(2009){Aragona}, {McSwain}, {Grundstrom}, {Marsh},
  {Roettenbacher}, {Hessler}, {Boyajian}, \& {Ray}}]{2009ApJ...698..514A}
{Aragona}, C., {McSwain}, M.~V., {Grundstrom}, E.~D., {et~al.} 2009, \apj, 698,
  514

\bibitem[{{Aret} {et~al.}(2016){Aret}, {Kraus}, \&
  {{\v{S}}lechta}}]{2016MNRAS.456.1424A}
{Aret}, A., {Kraus}, M., \& {{\v{S}}lechta}, M. 2016, \mnras, 456, 1424

\bibitem[{{Ash} {et~al.}(1999){Ash}, {Reynolds}, {Roche}, {Norton}, {Still}, \&
  {Morales-Rueda}}]{1999MNRAS.307..357A}
{Ash}, T.~D.~C., {Reynolds}, A.~P., {Roche}, P., {et~al.} 1999, \mnras, 307,
  357

\bibitem[{{Bailer-Jones} {et~al.}(2021){Bailer-Jones}, {Rybizki}, {Fouesneau},
  {Demleitner}, \& {Andrae}}]{2021AJ....161..147B}
{Bailer-Jones}, C.~A.~L., {Rybizki}, J., {Fouesneau}, M., {Demleitner}, M., \&
  {Andrae}, R. 2021, \aj, 161, 147

\bibitem[{{Bamba} {et~al.}(2001){Bamba}, {Yokogawa}, {Ueno}, {Koyama}, \&
  {Yamauchi}}]{2001PASJ...53.1179B}
{Bamba}, A., {Yokogawa}, J., {Ueno}, M., {Koyama}, K., \& {Yamauchi}, S. 2001,
  \pasj, 53, 1179

\bibitem[{{Barnstedt} {et~al.}(2008){Barnstedt}, {Staubert}, {Santangelo},
  {Ferrigno}, {Horns}, {Klochkov}, {Kretschmar}, {Kreykenbohm}, {Segreto}, \&
  {Wilms}}]{2008AA...486..293B}
{Barnstedt}, J., {Staubert}, R., {Santangelo}, A., {et~al.} 2008, \aap, 486,
  293

\bibitem[{{Barsukova} {et~al.}(2006){Barsukova}, {Borisov}, {Burenkov},
  {Klochkova}, {Goranskij}, \& {Metlova}}]{2006ASPC..355..305B}
{Barsukova}, E.~A., {Borisov}, N.~V., {Burenkov}, A.~N., {et~al.} 2006, in
  Astronomical Society of the Pacific Conference Series, Vol. 355, Stars with
  the B[e] Phenomenon, ed. M.~{Kraus} \& A.~S. {Miroshnichenko}, 305

\bibitem[{{Baykal} {et~al.}(2010){Baykal}, {G{\"o}{\v{g}}{\"u}{\c{s}}},
  {{\c{C}}a{\v{g}}da{\c{s}} {\.I}nam}, \& {Belloni}}]{2010ApJ...711.1306B}
{Baykal}, A., {G{\"o}{\v{g}}{\"u}{\c{s}}}, E., {{\c{C}}a{\v{g}}da{\c{s}}
  {\.I}nam}, S., \& {Belloni}, T. 2010, \apj, 711, 1306

\bibitem[{{Baykal} {et~al.}(2007){Baykal}, {Inam}, {Stark}, {Heffner},
  {Erkoca}, \& {Swank}}]{2007MNRAS.374.1108B}
{Baykal}, A., {Inam}, S.~{\c{C}}., {Stark}, M.~J., {et~al.} 2007, \mnras, 374,
  1108

\bibitem[{{Belczynski} \& {Ziolkowski}(2009)}]{2009ApJ...707..870B}
{Belczynski}, K. \& {Ziolkowski}, J. 2009, \apj, 707, 870

\bibitem[{{Belloni} {et~al.}(1993){Belloni}, {Hasinger}, {Pietsch},
  {Mereghetti}, {Bignami}, \& {Caraveo}}]{1993AA...271..487B}
{Belloni}, T., {Hasinger}, G., {Pietsch}, W., {et~al.} 1993, \aap, 271, 487

\bibitem[{{Bhargava} {et~al.}(2017){Bhargava}, {Rao}, {Singh}, {Choudhury},
  {Bhattacharyya}, {Chandra}, {Dewangan}, {Mukerjee}, {Stewart},
  {Bhattacharya}, {Mithun}, \& {Vadawale}}]{2017ApJ...849..141B}
{Bhargava}, Y., {Rao}, A.~R., {Singh}, K.~P., {et~al.} 2017, \apj, 849, 141

\bibitem[{{Bikmaev} {et~al.}(2017){Bikmaev}, {Nikolaeva}, {Shimansky},
  {Galeev}, {Zhuchkov}, {Irtuganov}, {Melnikov}, {Sakhibullin}, {Grebenev}, \&
  {Sharipova}}]{2017AstL...43..664B}
{Bikmaev}, I.~F., {Nikolaeva}, E.~A., {Shimansky}, V.~V., {et~al.} 2017,
  Astronomy Letters, 43, 664

\bibitem[{{Bildsten} {et~al.}(1997){Bildsten}, {Chakrabarty}, {Chiu}, {Finger},
  {Koh}, {Nelson}, {Prince}, {Rubin}, {Scott}, {Stollberg}, {Vaughan},
  {Wilson}, \& {Wilson}}]{1997ApJS..113..367B}
{Bildsten}, L., {Chakrabarty}, D., {Chiu}, J., {et~al.} 1997, \apjs, 113, 367

\bibitem[{{Bird} {et~al.}(2016){Bird}, {Bazzano}, {Malizia}, {Fiocchi},
  {Sguera}, {Bassani}, {Hill}, {Ubertini}, \& {Winkler}}]{2016ApJS..223...15B}
{Bird}, A.~J., {Bazzano}, A., {Malizia}, A., {et~al.} 2016, \apjs, 223, 15

\bibitem[{{Bissinger}(2016)}]{2016PhDT.......406B}
{Bissinger}, M. 2016, PhD thesis, Friedrich Alexander University of
  Erlangen-Nuremberg, Germany

\bibitem[{{Blair} \& {Candy}(1985)}]{1985MNRAS.212..219B}
{Blair}, D.~G. \& {Candy}, B.~N. 1985, \mnras, 212, 219

\bibitem[{{Blay} {et~al.}(2006){Blay}, {Negueruela}, {Reig}, {Coe}, {Corbet},
  {Fabregat}, \& {Tarasov}}]{2006AA...446.1095B}
{Blay}, P., {Negueruela}, I., {Reig}, P., {et~al.} 2006, \aap, 446, 1095

\bibitem[{{Blundell} {et~al.}(2007){Blundell}, {Bowler}, \&
  {Schmidtobreick}}]{2007AA...474..903B}
{Blundell}, K.~M., {Bowler}, M.~G., \& {Schmidtobreick}, L. 2007, \aap, 474,
  903

\bibitem[{{Bodaghee} {et~al.}(2012){Bodaghee}, {Tomsick}, \&
  {Rodriguez}}]{2012ApJ...753....3B}
{Bodaghee}, A., {Tomsick}, J.~A., \& {Rodriguez}, J. 2012, \apj, 753, 3

\bibitem[{{Bonnet-Bidaud} \& {Mouchet}(1998)}]{1998AA...332L...9B}
{Bonnet-Bidaud}, J.~M. \& {Mouchet}, M. 1998, \aap, 332, L9

\bibitem[{{Bradt} \& {McClintock}(1983)}]{1983ARAA..21...13B}
{Bradt}, H.~V.~D. \& {McClintock}, J.~E. 1983, \araa, 21, 13

\bibitem[{{Brodskaya}(1960)}]{1960IzKry..24..160B}
{Brodskaya}, E.~S. 1960, Izvestiya Ordena Trudovogo Krasnogo Znameni Krymskoj
  Astrofizicheskoj Observatorii, 24, 160

\bibitem[{{Butler} {et~al.}(2009){Butler}, {Tomsick}, {Chaty}, {Heras},
  {Rodriguez}, {Walter}, {Kaaret}, {Kalemci}, \&
  {{\"O}zbey}}]{2009ApJ...698..502B}
{Butler}, S.~C., {Tomsick}, J.~A., {Chaty}, S., {et~al.} 2009, \apj, 698, 502

\bibitem[{{Capitanio} {et~al.}(2011){Capitanio}, {Bird}, {Fiocchi}, {Scaringi},
  \& {Ubertini}}]{2011ApJS..195....9C}
{Capitanio}, F., {Bird}, A.~J., {Fiocchi}, M., {Scaringi}, S., \& {Ubertini},
  P. 2011, \apjs, 195, 9

\bibitem[{{Casares} {et~al.}(2011){Casares}, {Corral-Santana}, {Herrero},
  {Morales}, {Mu{\~n}oz-Darias}, {Negueruela}, {Paredes}, {Ribas}, {Rib{\'o}},
  {Steeghs}, {van Spaandonk}, \& {Vilardell}}]{2011ASSP...21..559C}
{Casares}, J., {Corral-Santana}, J.~M., {Herrero}, A., {et~al.} 2011, in
  Astrophysics and Space Science Proceedings, Vol.~21, High-Energy Emission
  from Pulsars and their Systems, 559--562

\bibitem[{{Casares} {et~al.}(2014){Casares}, {Negueruela}, {Rib{\'o}}, {Ribas},
  {Paredes}, {Herrero}, \& {Sim{\'o}n-D{\'\i}az}}]{2014Natur.505..378C}
{Casares}, J., {Negueruela}, I., {Rib{\'o}}, M., {et~al.} 2014, \nat, 505, 378

\bibitem[{{Casares} {et~al.}(2005{\natexlab{a}}){Casares}, {Ribas}, {Paredes},
  {Mart{\'\i}}, \& {Allende Prieto}}]{2005MNRAS.360.1105C}
{Casares}, J., {Ribas}, I., {Paredes}, J.~M., {Mart{\'\i}}, J., \& {Allende
  Prieto}, C. 2005{\natexlab{a}}, \mnras, 360, 1105

\bibitem[{{Casares} {et~al.}(2005{\natexlab{b}}){Casares}, {Rib{\'o}}, {Ribas},
  {Paredes}, {Mart{\'\i}}, \& {Herrero}}]{2005MNRAS.364..899C}
{Casares}, J., {Rib{\'o}}, M., {Ribas}, I., {et~al.} 2005{\natexlab{b}},
  \mnras, 364, 899

\bibitem[{{Chakrabarty} {et~al.}(1995){Chakrabarty}, {Koh}, {Bildsten},
  {Prince}, {Finger}, {Wilson}, {Pendleton}, \& {Rubin}}]{1995ApJ...446..826C}
{Chakrabarty}, D., {Koh}, T., {Bildsten}, L., {et~al.} 1995, \apj, 446, 826

\bibitem[{{Chaty}(2013)}]{2013AdSpR..52.2132C}
{Chaty}, S. 2013, Advances in Space Research, 52, 2132

\bibitem[{{Chaty}(2022)}]{2022abn..book.....C}
{Chaty}, S. 2022, {Accreting Binaries; Nature, formation, and evolution},
  AAS-IOP Astronomy (Institute of Physics Publishing)

\bibitem[{{Chaty} {et~al.}(2008){Chaty}, {Rahoui}, {Foellmi}, {Tomsick},
  {Rodriguez}, \& {Walter}}]{2008AA...484..783C}
{Chaty}, S., {Rahoui}, F., {Foellmi}, C., {et~al.} 2008, \aap, 484, 783

\bibitem[{{Chen} {et~al.}(2019){Chen}, {Davis}, {Doe}, {Evans}, {Fabbiano},
  {Galle}, {Gibbs}, {Grier}, {Hain}, {Hall}, {Harbo}, {He}, {Houck},
  {Karovska}, {Kashyap}, {Lauer}, {McCollough}, {McDowell}, {Miller},
  {Mitschang}, {Morgan}, {Mossman}, {Nichols}, {Nowak}, {Plummer}, {Refsdal},
  {Rots}, {Siemiginowska}, {Sundheim}, {Tibbetts}, {van}, {Winkelman}, \&
  {Zografou}}]{2019yCat.9057....0E}
{Chen}, J.~C., {Davis}, J.~E., {Doe}, S.~M., {et~al.} 2019, VizieR Online Data
  Catalog, IX/57

\bibitem[{{Cheng} {et~al.}(2014){Cheng}, {Shao}, \& {Li}}]{2014ApJ...786..128C}
{Cheng}, Z.~Q., {Shao}, Y., \& {Li}, X.~D. 2014, \apj, 786, 128

\bibitem[{{Cherepashchuk} {et~al.}(2021){Cherepashchuk}, {Belinski}, {Dodin},
  \& {Postnov}}]{2021MNRAS.507L..19C}
{Cherepashchuk}, A.~M., {Belinski}, A.~A., {Dodin}, A.~V., \& {Postnov}, K.~A.
  2021, \mnras, 507, L19

\bibitem[{{Chernyakova} {et~al.}(2005){Chernyakova}, {Lutovinov},
  {Rodr{\'\i}guez}, \& {Revnivtsev}}]{2005MNRAS.364..455C}
{Chernyakova}, M., {Lutovinov}, A., {Rodr{\'\i}guez}, J., \& {Revnivtsev}, M.
  2005, \mnras, 364, 455

\bibitem[{{Chojnowski} {et~al.}(2017){Chojnowski}, {Wisniewski}, {Whelan},
  {Labadie-Bartz}, {Borges Fernandes}, {Lin}, {Majewski}, {Stringfellow},
  {Mennickent}, {Roman-Lopes}, {Tang}, {Hearty}, {Holtzman}, {Pepper}, \&
  {Zasowski}}]{2017AJ....153..174C}
{Chojnowski}, S.~D., {Wisniewski}, J.~P., {Whelan}, D.~G., {et~al.} 2017, \aj,
  153, 174

\bibitem[{{Coe} {et~al.}(2007){Coe}, {Bird}, {Hill}, {McBride}, {Schurch},
  {Galache}, {Wilson}, {Finger}, {Buckley}, \&
  {Romero-Colmenero}}]{2007MNRAS.378.1427C}
{Coe}, M.~J., {Bird}, A.~J., {Hill}, A.~B., {et~al.} 2007, \mnras, 378, 1427

\bibitem[{{Coe} {et~al.}(1996){Coe}, {Fabregat}, {Negueruela}, {Roche}, \&
  {Steele}}]{1996MNRAS.281..333C}
{Coe}, M.~J., {Fabregat}, J., {Negueruela}, I., {Roche}, P., \& {Steele}, I.~A.
  1996, \mnras, 281, 333

\bibitem[{{Coe} {et~al.}(1994){Coe}, {Roche}, {Everall}, {Fabregat}, {Buckley},
  {Smith}, {Reynolds}, {Jupp}, \& {MacGillivray}}]{1994MNRAS.270L..57C}
{Coe}, M.~J., {Roche}, P., {Everall}, C., {et~al.} 1994, \mnras, 270, L57

\bibitem[{{Coleiro} {et~al.}(2013){Coleiro}, {Chaty}, {Zurita Heras}, {Rahoui},
  \& {Tomsick}}]{2013AA...560A.108C}
{Coleiro}, A., {Chaty}, S., {Zurita Heras}, J.~A., {Rahoui}, F., \& {Tomsick},
  J.~A. 2013, \aap, 560, A108

\bibitem[{{Coley} {et~al.}(2019){Coley}, {Corbet}, {F{\"u}rst}, {Huxtable},
  {Krimm}, {Pearlman}, \& {Pottschmidt}}]{2019ApJ...879...34C}
{Coley}, J.~B., {Corbet}, R. H.~D., {F{\"u}rst}, F., {et~al.} 2019, \apj, 879,
  34

\bibitem[{{Coley} {et~al.}(2014){Coley}, {Corbet}, {Mukai}, \&
  {Pottschmidt}}]{2014ApJ...793...77C}
{Coley}, J.~B., {Corbet}, R. H.~D., {Mukai}, K., \& {Pottschmidt}, K. 2014,
  \apj, 793, 77

\bibitem[{{Cominsky} {et~al.}(1978){Cominsky}, {Li}, {Bradt}, {Clark},
  {Rappaport}, {Johnston}, {Doxsey}, {Gursky}, {Schwartz}, \&
  {Schwarz}}]{1978IAUC.3163....1C}
{Cominsky}, L., {Li}, F., {Bradt}, H., {et~al.} 1978, \iaucirc, 3163, 1

\bibitem[{{Cook} \& {Warwick}(1987)}]{1987MNRAS.225..369C}
{Cook}, M.~C. \& {Warwick}, R.~S. 1987, \mnras, 225, 369

\bibitem[{{Corbet} {et~al.}(2006){Corbet}, {Barbier}, {Barthelmy}, {Cummings},
  {Fenimore}, {Gehrels}, {Hullinger}, {Krimm}, {Markwardt}, {Palmer},
  {Parsons}, {Sakamoto}, {Sato}, {Tueller}, \&
  {Remillard}}]{2006ATel..779....1C}
{Corbet}, R., {Barbier}, L., {Barthelmy}, S., {et~al.} 2006, The Astronomer's
  Telegram, 779, 1

\bibitem[{{Corbet} {et~al.}(2005){Corbet}, {Barbier}, {Barthelmy}, {Cummings},
  {Fenimore}, {Gehrels}, {Hullinger}, {Krimm}, {Markwardt}, {Palmer},
  {Parsons}, {Sakamoto}, {Sato}, {Tueller}, \& {Swift-Survey
  Team}}]{2005ATel..649....1C}
{Corbet}, R., {Barbier}, L., {Barthelmy}, S., {et~al.} 2005, The Astronomer's
  Telegram, 649, 1

\bibitem[{{Corbet}(1984)}]{1984AA...141...91C}
{Corbet}, R.~H.~D. 1984, \aap, 141, 91

\bibitem[{{Corbet} {et~al.}(2019){Corbet}, {Chomiuk}, {Coe}, {Coley}, {Dubus},
  {Edwards}, {Martin}, {McBride}, {Stevens}, {Strader}, \&
  {Townsend}}]{2019ApJ...884...93C}
{Corbet}, R.~H.~D., {Chomiuk}, L., {Coe}, M.~J., {et~al.} 2019, \apj, 884, 93

\bibitem[{{Corbet} {et~al.}(2022){Corbet}, {Coley}, {Gendreau}, {Guillot},
  {Islam}, {Jaisawal}, {Malacaria}, {Ng}, {Pottschmidt}, {Pradhan}, {Ray},
  {Sanna}, {Wilms}, \& {Wolff}}]{2022ATel15614....1C}
{Corbet}, R. H.~D., {Coley}, J.~B., {Gendreau}, K.~C., {et~al.} 2022, The
  Astronomer's Telegram, 15614, 1

\bibitem[{{Corbet} {et~al.}(2016){Corbet}, {Coley}, \&
  {Krimm}}]{2016ATel.9823....1C}
{Corbet}, R. H.~D., {Coley}, J.~B., \& {Krimm}, H.~A. 2016, The Astronomer's
  Telegram, 9823, 1

\bibitem[{{Corbet} {et~al.}(2017){Corbet}, {Coley}, \&
  {Krimm}}]{2017ApJ...846..161C}
{Corbet}, R. H.~D., {Coley}, J.~B., \& {Krimm}, H.~A. 2017, \apj, 846, 161

\bibitem[{{Corbet} {et~al.}(2021){Corbet}, {Coley}, {Krimm}, {Pottschmidt}, \&
  {Roche}}]{2021ApJ...906...13C}
{Corbet}, R. H.~D., {Coley}, J.~B., {Krimm}, H.~A., {Pottschmidt}, K., \&
  {Roche}, P. 2021, \apj, 906, 13

\bibitem[{{Corbet} {et~al.}(2004){Corbet}, {Hannikainen}, \&
  {Remillard}}]{2004ATel..269....1C}
{Corbet}, R.~H.~D., {Hannikainen}, D.~C., \& {Remillard}, R. 2004, The
  Astronomer's Telegram, 269, 1

\bibitem[{{Corbet} \& {Krimm}(2009)}]{2009ATel.2008....1C}
{Corbet}, R.~H.~D. \& {Krimm}, H.~A. 2009, The Astronomer's Telegram, 2008, 1

\bibitem[{{Corbet} \& {Krimm}(2010)}]{2010ATel.3079....1C}
{Corbet}, R.~H.~D. \& {Krimm}, H.~A. 2010, The Astronomer's Telegram, 3079, 1

\bibitem[{{Corbet} \& {Krimm}(2013)}]{2013ApJ...778...45C}
{Corbet}, R. H.~D. \& {Krimm}, H.~A. 2013, \apj, 778, 45

\bibitem[{{Corbet} {et~al.}(1999){Corbet}, {Marshall}, {Peele}, \&
  {Takeshima}}]{1999ApJ...517..956C}
{Corbet}, R.~H.~D., {Marshall}, F.~E., {Peele}, A.~G., \& {Takeshima}, T. 1999,
  \apj, 517, 956

\bibitem[{{Corbet} \& {Remillard}(2005)}]{2005ATel..377....1C}
{Corbet}, R.~H.~D. \& {Remillard}, R. 2005, The Astronomer's Telegram, 377, 1

\bibitem[{{Crampton} {et~al.}(1985){Crampton}, {Hutchings}, \&
  {Cowley}}]{1985ApJ...299..839C}
{Crampton}, D., {Hutchings}, J.~B., \& {Cowley}, A.~P. 1985, \apj, 299, 839

\bibitem[{{Cusumano} {et~al.}(2020){Cusumano}, {D'A{\`\i}}, {Segreto}, {La
  Parola}, \& {Del Santo}}]{2020MNRAS.498.2750C}
{Cusumano}, G., {D'A{\`\i}}, A., {Segreto}, A., {La Parola}, V., \& {Del
  Santo}, M. 2020, \mnras, 498, 2750

\bibitem[{{Cusumano} {et~al.}(2000){Cusumano}, {Maccarone}, {Nicastro},
  {Sacco}, \& {Kaaret}}]{2000ApJ...528L..25C}
{Cusumano}, G., {Maccarone}, M.~C., {Nicastro}, L., {Sacco}, B., \& {Kaaret},
  P. 2000, \apjl, 528, L25

\bibitem[{{Cusumano} {et~al.}(2013){Cusumano}, {Segreto}, {La Parola},
  {D'A{\`\i}}, {Masetti}, \& {Tagliaferri}}]{2013ApJ...775L..25C}
{Cusumano}, G., {Segreto}, A., {La Parola}, V., {et~al.} 2013, \apjl, 775, L25

\bibitem[{{Cusumano} {et~al.}(2015){Cusumano}, {Segreto}, {La Parola},
  {Masetti}, {D'A{\`\i}}, \& {Tagliaferri}}]{2015MNRAS.446.1041C}
{Cusumano}, G., {Segreto}, A., {La Parola}, V., {et~al.} 2015, \mnras, 446,
  1041

\bibitem[{{Cutri} {et~al.}(2003){Cutri}, {Skrutskie}, {van Dyk}, {Beichman},
  {Carpenter}, {Chester}, {Cambresy}, {Evans}, {Fowler}, {Gizis}, {Howard},
  {Huchra}, {Jarrett}, {Kopan}, {Kirkpatrick}, {Light}, {Marsh}, {McCallon},
  {Schneider}, {Stiening}, {Sykes}, {Weinberg}, {Wheaton}, {Wheelock}, \&
  {Zacarias}}]{2003yCat.2246....0C}
{Cutri}, R.~M., {Skrutskie}, M.~F., {van Dyk}, S., {et~al.} 2003, VizieR Online
  Data Catalog, II/246

\bibitem[{{D'A{\`\i}} {et~al.}(2015){D'A{\`\i}}, {Cusumano}, {La Parola}, \&
  {Segreto}}]{2015MNRAS.451.2835D}
{D'A{\`\i}}, A., {Cusumano}, G., {La Parola}, V., \& {Segreto}, A. 2015,
  \mnras, 451, 2835

\bibitem[{{Delgado-Mart{\'\i}} {et~al.}(2001){Delgado-Mart{\'\i}}, {Levine},
  {Pfahl}, \& {Rappaport}}]{2001ApJ...546..455D}
{Delgado-Mart{\'\i}}, H., {Levine}, A.~M., {Pfahl}, E., \& {Rappaport}, S.~A.
  2001, \apj, 546, 455

\bibitem[{{Densham} \& {Charles}(1982)}]{1982MNRAS.201..171D}
{Densham}, R.~H. \& {Charles}, P.~A. 1982, \mnras, 201, 171

\bibitem[{{D{\"o}nmez} {et~al.}(2020){D{\"o}nmez}, {Serim}, {{\.I}nam},
  {{\c{S}}ahiner}, {Serim}, \& {Baykal}}]{2020MNRAS.496.1768D}
{D{\"o}nmez}, {\c{C}}.~K., {Serim}, M.~M., {{\.I}nam}, S.~{\c{C}}., {et~al.}
  2020, \mnras, 496, 1768

\bibitem[{{Doroshenko} {et~al.}(2021){Doroshenko}, {Santangelo}, {Tsygankov},
  \& {Ji}}]{2021AA...647A.165D}
{Doroshenko}, V., {Santangelo}, A., {Tsygankov}, S.~S., \& {Ji}, L. 2021, \aap,
  647, A165

\bibitem[{{Doroshenko} {et~al.}(2018){Doroshenko}, {Tsygankov}, \&
  {Santangelo}}]{2018AA...613A..19D}
{Doroshenko}, V., {Tsygankov}, S., \& {Santangelo}, A. 2018, \aap, 613, A19

\bibitem[{{Drave} {et~al.}(2012){Drave}, {Bird}, {Townsend}, {Hill}, {McBride},
  {Sguera}, {Bazzano}, \& {Clark}}]{2012AA...539A..21D}
{Drave}, S.~P., {Bird}, A.~J., {Townsend}, L.~J., {et~al.} 2012, \aap, 539, A21

\bibitem[{{Dubus}(2013)}]{2013AARv..21...64D}
{Dubus}, G. 2013, \aapr, 21, 64

\bibitem[{{Ducci} {et~al.}(2019){Ducci}, {Romano}, {Ji}, \&
  {Santangelo}}]{2019A&A...631A.135D}
{Ducci}, L., {Romano}, P., {Ji}, L., \& {Santangelo}, A. 2019, \aap, 631, A135

\bibitem[{{Evans} {et~al.}(2020){Evans}, {Page}, {Osborne}, {Beardmore},
  {Willingale}, {Burrows}, {Kennea}, {Perri}, {Capalbi}, {Tagliaferri}, \&
  {Cenko}}]{2020ApJS..247...54E}
{Evans}, P.~A., {Page}, K.~L., {Osborne}, J.~P., {et~al.} 2020, \apjs, 247, 54

\bibitem[{{Fabrika}(1997)}]{1997ApSS.252..439F}
{Fabrika}, S.~N. 1997, \apss, 252, 439

\bibitem[{{Fairlamb} {et~al.}(2015){Fairlamb}, {Oudmaijer}, {Mendigut{\'\i}a},
  {Ilee}, \& {van den Ancker}}]{2015MNRAS.453..976F}
{Fairlamb}, J.~R., {Oudmaijer}, R.~D., {Mendigut{\'\i}a}, I., {Ilee}, J.~D., \&
  {van den Ancker}, M.~E. 2015, \mnras, 453, 976

\bibitem[{{Falanga} {et~al.}(2015){Falanga}, {Bozzo}, {Lutovinov},
  {Bonnet-Bidaud}, {Fetisova}, \& {Puls}}]{2015AA...577A.130F}
{Falanga}, M., {Bozzo}, E., {Lutovinov}, A., {et~al.} 2015, \aap, 577, A130

\bibitem[{{Ferrigno} {et~al.}(2022){Ferrigno}, {Bozzo}, \&
  {Romano}}]{2022AA...664A..99F}
{Ferrigno}, C., {Bozzo}, E., \& {Romano}, P. 2022, \aap, 664, A99

\bibitem[{{Ferrigno} {et~al.}(2013){Ferrigno}, {Farinelli}, {Bozzo},
  {Pottschmidt}, {Klochkov}, \& {Kretschmar}}]{2013AA...553A.103F}
{Ferrigno}, C., {Farinelli}, R., {Bozzo}, E., {et~al.} 2013, \aap, 553, A103

\bibitem[{{Filliatre} \& {Chaty}(2004)}]{2004ApJ...616..469F}
{Filliatre}, P. \& {Chaty}, S. 2004, \apj, 616, 469

\bibitem[{{Finger} {et~al.}(1999){Finger}, {Bildsten}, {Chakrabarty}, {Prince},
  {Scott}, {Wilson}, {Wilson}, \& {Zhang}}]{1999ApJ...517..449F}
{Finger}, M.~H., {Bildsten}, L., {Chakrabarty}, D., {et~al.} 1999, \apj, 517,
  449

\bibitem[{{Finger} {et~al.}(1996{\natexlab{a}}){Finger}, {Wilson}, \&
  {Chakrabarty}}]{1996AAS..120C.209F}
{Finger}, M.~H., {Wilson}, R.~B., \& {Chakrabarty}, D. 1996{\natexlab{a}},
  \aaps, 120, 209

\bibitem[{{Finger} {et~al.}(1996{\natexlab{b}}){Finger}, {Wilson}, \&
  {Harmon}}]{1996ApJ...459..288F}
{Finger}, M.~H., {Wilson}, R.~B., \& {Harmon}, B.~A. 1996{\natexlab{b}}, \apj,
  459, 288

\bibitem[{{Finley} {et~al.}(1992){Finley}, {Belloni}, \&
  {Cassinelli}}]{1992AA...262L..25F}
{Finley}, J.~P., {Belloni}, T., \& {Cassinelli}, J.~P. 1992, \aap, 262, L25

\bibitem[{{Fiocchi} {et~al.}(2013){Fiocchi}, {Bazzano}, {Bird}, {Drave},
  {Natalucci}, {Persi}, {Piro}, \& {Ubertini}}]{2013ApJ...762...19F}
{Fiocchi}, M., {Bazzano}, A., {Bird}, A.~J., {et~al.} 2013, \apj, 762, 19

\bibitem[{{Forman} {et~al.}(1978){Forman}, {Jones}, {Cominsky}, {Julien},
  {Murray}, {Peters}, {Tananbaum}, \& {Giacconi}}]{1978ApJS...38..357F}
{Forman}, W., {Jones}, C., {Cominsky}, L., {et~al.} 1978, \apjs, 38, 357

\bibitem[{{Fortin} {et~al.}(2018){Fortin}, {Chaty}, {Coleiro}, {Tomsick}, \&
  {Nitschelm}}]{2018AA...618A.150F}
{Fortin}, F., {Chaty}, S., {Coleiro}, A., {Tomsick}, J.~A., \& {Nitschelm},
  C.~H.~R. 2018, \aap, 618, A150

\bibitem[{{Fortin} {et~al.}(2020){Fortin}, {Chaty}, \&
  {Sander}}]{2020ApJ...894...86F}
{Fortin}, F., {Chaty}, S., \& {Sander}, A. 2020, \apj, 894, 86

\bibitem[{{Fortin} {et~al.}(2022{\natexlab{a}}){Fortin}, {Garc{\'\i}a}, \&
  {Chaty}}]{2022AA...665A..69F}
{Fortin}, F., {Garc{\'\i}a}, F., \& {Chaty}, S. 2022{\natexlab{a}}, \aap, 665,
  A69

\bibitem[{{Fortin} {et~al.}(2022{\natexlab{b}}){Fortin}, {Garc{\'\i}a},
  {Chaty}, {Chassande-Mottin}, \& {Simaz Bunzel}}]{2022AA...665A..31F}
{Fortin}, F., {Garc{\'\i}a}, F., {Chaty}, S., {Chassande-Mottin}, E., \& {Simaz
  Bunzel}, A. 2022{\natexlab{b}}, \aap, 665, A31

\bibitem[{{Gaia Collaboration}(2022)}]{2022yCat.1355....0G}
{Gaia Collaboration}. 2022, VizieR Online Data Catalog, I/355

\bibitem[{{Galloway} {et~al.}(2004){Galloway}, {Morgan}, \&
  {Levine}}]{2004ApJ...613.1164G}
{Galloway}, D.~K., {Morgan}, E.~H., \& {Levine}, A.~M. 2004, \apj, 613, 1164

\bibitem[{{Galloway} {et~al.}(2005){Galloway}, {Wang}, \&
  {Morgan}}]{2005ApJ...635.1217G}
{Galloway}, D.~K., {Wang}, Z., \& {Morgan}, E.~H. 2005, \apj, 635, 1217

\bibitem[{{Gamen} {et~al.}(2015){Gamen}, {Barb{\`a}}, {Walborn}, {Morrell},
  {Arias}, {Ma{\'\i}z Apell{\'a}niz}, {Sota}, \& {Alfaro}}]{2015AA...583L...4G}
{Gamen}, R., {Barb{\`a}}, R.~H., {Walborn}, N.~R., {et~al.} 2015, \aap, 583, L4

\bibitem[{{Garcia}(1993)}]{1993ApJS...87..197G}
{Garcia}, B. 1993, \apjs, 87, 197

\bibitem[{{Garc{\'\i}a} {et~al.}(2018){Garc{\'\i}a}, {Fogantini}, {Chaty}, \&
  {Combi}}]{2018AA...618A..61G}
{Garc{\'\i}a}, F., {Fogantini}, F.~A., {Chaty}, S., \& {Combi}, J.~A. 2018,
  \aap, 618, A61

\bibitem[{{Garrison} {et~al.}(1977){Garrison}, {Hiltner}, \&
  {Schild}}]{1977ApJS...35..111G}
{Garrison}, R.~F., {Hiltner}, W.~A., \& {Schild}, R.~E. 1977, \apjs, 35, 111

\bibitem[{{Gies} \& {Bolton}(1986)}]{1986ApJS...61..419G}
{Gies}, D.~R. \& {Bolton}, C.~T. 1986, \apjs, 61, 419

\bibitem[{{Gies} {et~al.}(2003){Gies}, {Bolton}, {Thomson}, {Huang}, {McSwain},
  {Riddle}, {Wang}, {Wiita}, {Wingert}, {Cs{\'a}k}, \&
  {Kiss}}]{2003ApJ...583..424G}
{Gies}, D.~R., {Bolton}, C.~T., {Thomson}, J.~R., {et~al.} 2003, \apj, 583, 424

\bibitem[{{Gomez} \& {Grindlay}(2021)}]{2021ApJ...913...48G}
{Gomez}, S. \& {Grindlay}, J.~E. 2021, \apj, 913, 48

\bibitem[{{Gontcharov}(2006)}]{2006AstL...32..759G}
{Gontcharov}, G.~A. 2006, Astronomy Letters, 32, 759

\bibitem[{{Gonz{\'a}lez-Gal{\'a}n}(2015)}]{2015arXiv150301087G}
{Gonz{\'a}lez-Gal{\'a}n}, A. 2015, arXiv e-prints, arXiv:1503.01087

\bibitem[{{Gonz{\'a}lez-Gal{\'a}n} {et~al.}(2014){Gonz{\'a}lez-Gal{\'a}n},
  {Negueruela}, {Castro}, {Sim{\'o}n-D{\'\i}az}, {Lorenzo}, \&
  {Vilardell}}]{2014AA...566A.131G}
{Gonz{\'a}lez-Gal{\'a}n}, A., {Negueruela}, I., {Castro}, N., {et~al.} 2014,
  \aap, 566, A131

\bibitem[{{Gotthelf} {et~al.}(2008){Gotthelf}, {Halpern}, {Camilo},
  {Markwardt}, \& {Swank}}]{2008ATel.1392....1G}
{Gotthelf}, E.~V., {Halpern}, J.~P., {Camilo}, F., {Markwardt}, C., \& {Swank},
  J. 2008, The Astronomer's Telegram, 1392, 1

\bibitem[{{G{\"o}{\v{g}}{\"u}{\c{s}}}
  {et~al.}(2005){G{\"o}{\v{g}}{\"u}{\c{s}}}, {Patel}, {Wilson}, {Woods},
  {Finger}, \& {Kouveliotou}}]{2005ApJ...632.1069G}
{G{\"o}{\v{g}}{\"u}{\c{s}}}, E., {Patel}, S.~K., {Wilson}, C.~A., {et~al.}
  2005, \apj, 632, 1069

\bibitem[{{Gregory}(2002)}]{2002ApJ...575..427G}
{Gregory}, P.~C. 2002, \apj, 575, 427

\bibitem[{{Grindlay} {et~al.}(1984){Grindlay}, {Petro}, \&
  {McClintock}}]{1984ApJ...276..621G}
{Grindlay}, J.~E., {Petro}, L.~D., \& {McClintock}, J.~E. 1984, \apj, 276, 621

\bibitem[{{Grundstrom} {et~al.}(2007){Grundstrom}, {Boyajian}, {Finch}, {Gies},
  {Huang}, {McSwain}, {O'Brien}, {Riddle}, {Trippe}, {Williams}, {Wingert}, \&
  {Zaballa}}]{2007ApJ...660.1398G}
{Grundstrom}, E.~D., {Boyajian}, T.~S., {Finch}, C., {et~al.} 2007, \apj, 660,
  1398

\bibitem[{{Grunhut} {et~al.}(2014){Grunhut}, {Bolton}, \&
  {McSwain}}]{2014AA...563A...1G}
{Grunhut}, J.~H., {Bolton}, C.~T., \& {McSwain}, M.~V. 2014, \aap, 563, A1

\bibitem[{{Haberl} {et~al.}(1998){Haberl}, {Angelini}, {Motch}, \&
  {White}}]{1998AA...330..189H}
{Haberl}, F., {Angelini}, L., {Motch}, C., \& {White}, N.~E. 1998, \aap, 330,
  189

\bibitem[{{Hardorp} {et~al.}(1964){Hardorp}, {Theile}, \&
  {Voigt}}]{1964LS....C03....0H}
{Hardorp}, J., {Theile}, I., \& {Voigt}, H.~H. 1964, Hamburger Sternw. Warner
  \& Swasey Obs., C03, 0

\bibitem[{{Hare} {et~al.}(2019){Hare}, {Halpern}, {Clavel}, {Grindlay},
  {Rahoui}, \& {Tomsick}}]{2019ApJ...878...15H}
{Hare}, J., {Halpern}, J.~P., {Clavel}, M., {et~al.} 2019, \apj, 878, 15

\bibitem[{{Harmanec} {et~al.}(2000){Harmanec}, {Habuda}, {{\v{S}}tefl},
  {Hadrava}, {Kor{\v{c}}{\'a}kov{\'a}}, {Koubsk{\'y}}, {Krti{\v{c}}ka},
  {Kub{\'a}t}, {{\v{S}}koda}, {{\v{S}}lechta}, \& {Wolf}}]{2000AA...364L..85H}
{Harmanec}, P., {Habuda}, P., {{\v{S}}tefl}, S., {et~al.} 2000, \aap, 364, L85

\bibitem[{{Harris} {et~al.}(1990){Harris}, {Forman}, {Gioia}, {Hale},
  {Harnden}, {Jones}, {Karakashian}, {Maccacaro}, {McSweeney}, \&
  {Primini}}]{1990EObsC...2.....H}
{Harris}, D.~E., {Forman}, W., {Gioia}, I.~M., {et~al.} 1990, Einstein
  Observatory Catalog of IPC X-ray Sources, 2

\bibitem[{{Harvey} {et~al.}(2022){Harvey}, {Rulten}, \&
  {Chadwick}}]{2022MNRAS.512.1141H}
{Harvey}, M., {Rulten}, C.~B., \& {Chadwick}, P.~M. 2022, \mnras, 512, 1141

\bibitem[{{Hemphill} {et~al.}(2019{\natexlab{a}}){Hemphill}, {Coley}, {Fuerst},
  {Kretschmar}, {Kuehnel}, {Malacaria}, \& {Pottschmidt}}]{2019ATel12556....1H}
{Hemphill}, P., {Coley}, J., {Fuerst}, F., {et~al.} 2019{\natexlab{a}}, The
  Astronomer's Telegram, 12556, 1

\bibitem[{{Hemphill} {et~al.}(2019{\natexlab{b}}){Hemphill}, {Rothschild},
  {Cheatham}, {F{\"u}rst}, {Kretschmar}, {K{\"u}hnel}, {Pottschmidt},
  {Staubert}, {Wilms}, \& {Wolff}}]{2019ApJ...873...62H}
{Hemphill}, P.~B., {Rothschild}, R.~E., {Cheatham}, D.~M., {et~al.}
  2019{\natexlab{b}}, \apj, 873, 62

\bibitem[{{Hill} {et~al.}(2005){Hill}, {Walter}, {Knigge}, {Bazzano},
  {B{\'e}langer}, {Bird}, {Dean}, {Galache}, {Malizia}, {Renaud}, {Stephen}, \&
  {Ubertini}}]{2005AA...439..255H}
{Hill}, A.~B., {Walter}, R., {Knigge}, C., {et~al.} 2005, \aap, 439, 255

\bibitem[{{Hillwig} {et~al.}(2004){Hillwig}, {Gies}, {Huang}, {McSwain},
  {Stark}, {van der Meer}, \& {Kaper}}]{2004ApJ...615..422H}
{Hillwig}, T.~C., {Gies}, D.~R., {Huang}, W., {et~al.} 2004, \apj, 615, 422

\bibitem[{{Hinkle} {et~al.}(2020){Hinkle}, {Lebzelter}, {Fekel}, {Straniero},
  {Joyce}, {Prato}, {Karnath}, \& {Habel}}]{2020ApJ...904..143H}
{Hinkle}, K.~H., {Lebzelter}, T., {Fekel}, F.~C., {et~al.} 2020, \apj, 904, 143

\bibitem[{{Houk}(1978)}]{1978mcts.book.....H}
{Houk}, N. 1978, {Michigan catalogue of two-dimensional spectral types for the
  HD stars}

\bibitem[{{Hu} {et~al.}(2017){Hu}, {Chou}, {Ng}, {Lin}, \&
  {Yen}}]{2017ApJ...844...16H}
{Hu}, C.-P., {Chou}, Y., {Ng}, C.~Y., {Lin}, L. C.-C., \& {Yen}, D. C.-C. 2017,
  \apj, 844, 16

\bibitem[{{Hulleman} {et~al.}(1998){Hulleman}, {in 't Zand}, \&
  {Heise}}]{1998AA...337L..25H}
{Hulleman}, F., {in 't Zand}, J.~J.~M., \& {Heise}, J. 1998, \aap, 337, L25

\bibitem[{Hunter(2007)}]{hunter_matplotlib_2007}
Hunter, J.~D. 2007, Computing in Science \& Engineering, 9, 90

\bibitem[{{Hutchings}(1984)}]{1984PASP...96..312H}
{Hutchings}, J.~B. 1984, \pasp, 96, 312

\bibitem[{{Hutchings} {et~al.}(1981){Hutchings}, {Cowley}, {Crampton}, \&
  {Williams}}]{1981PASP...93..741H}
{Hutchings}, J.~B., {Cowley}, A.~P., {Crampton}, D., \& {Williams}, G. 1981,
  \pasp, 93, 741

\bibitem[{{Hutchings} {et~al.}(1987){Hutchings}, {Crampton}, {Cowley}, \&
  {Thompson}}]{1987PASP...99..420H}
{Hutchings}, J.~B., {Crampton}, D., {Cowley}, A.~P., \& {Thompson}, I.~B. 1987,
  \pasp, 99, 420

\bibitem[{{Hutchings} {et~al.}(1982){Hutchings}, {Crampton}, {Cowley},
  {Cowley}, \& {Bord}}]{1982PASP...94..541H}
{Hutchings}, J.~B., {Crampton}, D., {Cowley}, D., {Cowley}, A.~P., \& {Bord},
  D.~J. 1982, \pasp, 94, 541

\bibitem[{{Hynes} {et~al.}(2002){Hynes}, {Clark}, {Barsukova}, {Callanan},
  {Charles}, {Collier Cameron}, {Fabrika}, {Garcia}, {Haswell}, {Horne},
  {Miroshnichenko}, {Negueruela}, {Reig}, {Welsh}, \&
  {Witherick}}]{2002AA...392..991H}
{Hynes}, R.~I., {Clark}, J.~S., {Barsukova}, E.~A., {et~al.} 2002, \aap, 392,
  991

\bibitem[{{in 't Zand} {et~al.}(1998){in 't Zand}, {Baykal}, \&
  {Strohmayer}}]{1998ApJ...496..386I}
{in 't Zand}, J.~J.~M., {Baykal}, A., \& {Strohmayer}, T.~E. 1998, \apj, 496,
  386

\bibitem[{{in't Zand} \& {Heise}(2004)}]{2004ATel..362....1I}
{in't Zand}, J. \& {Heise}, J. 2004, The Astronomer's Telegram, 362, 1

\bibitem[{{in't Zand} {et~al.}(2001{\natexlab{a}}){in't Zand}, {Corbet}, \&
  {Marshall}}]{2001ApJ...553L.165I}
{in't Zand}, J.~J.~M., {Corbet}, R.~H.~D., \& {Marshall}, F.~E.
  2001{\natexlab{a}}, \apjl, 553, L165

\bibitem[{{in't Zand} {et~al.}(2000){in't Zand}, {Halpern}, {Eracleous},
  {McCollough}, {Augusteijn}, {Remillard}, \& {Heise}}]{2000AA...361...85I}
{in't Zand}, J.~J.~M., {Halpern}, J., {Eracleous}, M., {et~al.} 2000, \aap,
  361, 85

\bibitem[{{in't Zand} {et~al.}(2001{\natexlab{b}}){in't Zand}, {Swank},
  {Corbet}, \& {Markwardt}}]{2001AA...380L..26I}
{in't Zand}, J.~J.~M., {Swank}, J., {Corbet}, R.~H.~D., \& {Markwardt}, C.~B.
  2001{\natexlab{b}}, \aap, 380, L26

\bibitem[{{Islam} {et~al.}(2015){Islam}, {Maitra}, {Pradhan}, \&
  {Paul}}]{2015MNRAS.446.4148I}
{Islam}, N., {Maitra}, C., {Pradhan}, P., \& {Paul}, B. 2015, \mnras, 446, 4148

\bibitem[{{Islam} \& {Paul}(2016)}]{2016MNRAS.461..816I}
{Islam}, N. \& {Paul}, B. 2016, \mnras, 461, 816

\bibitem[{{Israel} {et~al.}(2000){Israel}, {Covino}, {Campana}, {Polcaro},
  {Roche}, {Stella}, {Di Paola}, {Lazzati}, {Mereghetti}, {Giallongo},
  {Fontana}, \& {Verrecchia}}]{2000MNRAS.314...87I}
{Israel}, G.~L., {Covino}, S., {Campana}, S., {et~al.} 2000, \mnras, 314, 87

\bibitem[{{Israel} {et~al.}(2001){Israel}, {Negueruela}, {Campana}, {Covino},
  {Di Paola}, {Maxwell}, {Norton}, {Speziali}, {Verrecchia}, \&
  {Stella}}]{2001AA...371.1018I}
{Israel}, G.~L., {Negueruela}, I., {Campana}, S., {et~al.} 2001, \aap, 371,
  1018

\bibitem[{{Ives} {et~al.}(1975){Ives}, {Sanford}, \& {Bell
  Burnell}}]{1975Natur.254..578I}
{Ives}, J.~C., {Sanford}, P.~W., \& {Bell Burnell}, S.~J. 1975, \nat, 254, 578

\bibitem[{{Iyer} \& {Paul}(2017)}]{2017MNRAS.471..355I}
{Iyer}, N. \& {Paul}, B. 2017, \mnras, 471, 355

\bibitem[{{Jain} {et~al.}(2009){Jain}, {Paul}, \&
  {Dutta}}]{2009MNRAS.397L..11J}
{Jain}, C., {Paul}, B., \& {Dutta}, A. 2009, \mnras, 397, L11

\bibitem[{{Jain} {et~al.}(2011){Jain}, {Paul}, \&
  {Maitra}}]{2011ATel.3785....1J}
{Jain}, C., {Paul}, B., \& {Maitra}, C. 2011, The Astronomer's Telegram, 3785,
  1

\bibitem[{{Jaisawal} {et~al.}(2021){Jaisawal}, {Naik}, {Gupta}, {Agrawal},
  {Jana}, {Chhotaray}, \& {Epili}}]{2021JApA...42...33J}
{Jaisawal}, G.~K., {Naik}, S., {Gupta}, S., {et~al.} 2021, Journal of
  Astrophysics and Astronomy, 42, 33

\bibitem[{{Jaisawal} {et~al.}(2020){Jaisawal}, {Naik}, {Ho}, {Kumari}, {Epili},
  \& {Vasilopoulos}}]{2020MNRAS.498.4830J}
{Jaisawal}, G.~K., {Naik}, S., {Ho}, W. C.~G., {et~al.} 2020, \mnras, 498, 4830

\bibitem[{{Janot-Pacheco} {et~al.}(1981){Janot-Pacheco}, {Ilovaisky}, \&
  {Chevalier}}]{1981AA....99..274J}
{Janot-Pacheco}, E., {Ilovaisky}, S.~A., \& {Chevalier}, C. 1981, \aap, 99, 274

\bibitem[{{Jaschek} \& {Egret}(1982)}]{1982IAUS...98..261J}
{Jaschek}, M. \& {Egret}, D. 1982, in Be Stars, ed. M.~{Jaschek} \& H.~G.
  {Groth}, Vol.~98, 261

\bibitem[{{Jenke} \& {Wilson-Hodge}(2017)}]{2017ATel10812....1J}
{Jenke}, P. \& {Wilson-Hodge}, C.~A. 2017, The Astronomer's Telegram, 10812, 1

\bibitem[{{Jenke} {et~al.}(2011){Jenke}, {Finger}, {Wilson-Hodge}, \&
  {Camero-Arranz}}]{2011AIPC.1379..212J}
{Jenke}, P.~A., {Finger}, M.~H., {Wilson-Hodge}, C.~A., \& {Camero-Arranz}, A.
  2011, in American Institute of Physics Conference Series, Vol. 1379,
  AstroPhysics of Neutron Stars 2010: A Conference in Honor of M. Ali Alpar,
  ed. E.~{G{\"o}{\u{g}}{\"u}{\c{s}}}, T.~{Belloni}, \& {\"U}.~{Ertan}, 212--213

\bibitem[{{Johnston} {et~al.}(1992){Johnston}, {Manchester}, {Lyne}, {Bailes},
  {Kaspi}, {Qiao}, \& {D'Amico}}]{1992ApJ...387L..37J}
{Johnston}, S., {Manchester}, R.~N., {Lyne}, A.~G., {et~al.} 1992, \apjl, 387,
  L37

\bibitem[{{Johnston} {et~al.}(1994){Johnston}, {Manchester}, {Lyne},
  {Nicastro}, \& {Spyromilio}}]{1994MNRAS.268..430J}
{Johnston}, S., {Manchester}, R.~N., {Lyne}, A.~G., {Nicastro}, L., \&
  {Spyromilio}, J. 1994, \mnras, 268, 430

\bibitem[{Jones {et~al.}(2001)Jones, Oliphant, \& Peterson}]{jones_scipy_2001}
Jones, E., Oliphant, T., \& Peterson, P. 2001

\bibitem[{{Jonker} {et~al.}(2007){Jonker}, {Nelemans}, \&
  {Bassa}}]{2007MNRAS.374..999J}
{Jonker}, P.~G., {Nelemans}, G., \& {Bassa}, C.~G. 2007, \mnras, 374, 999

\bibitem[{{Kaaret} {et~al.}(2000){Kaaret}, {Cusumano}, \&
  {Sacco}}]{2000ApJ...542L..41K}
{Kaaret}, P., {Cusumano}, G., \& {Sacco}, B. 2000, \apjl, 542, L41

\bibitem[{{Kaaret} {et~al.}(1999){Kaaret}, {Piraino}, {Halpern}, \&
  {Eracleous}}]{1999ApJ...523..197K}
{Kaaret}, P., {Piraino}, S., {Halpern}, J., \& {Eracleous}, M. 1999, \apj, 523,
  197

\bibitem[{{Kabiraj} \& {Paul}(2020)}]{2020MNRAS.497.1059K}
{Kabiraj}, S. \& {Paul}, B. 2020, \mnras, 497, 1059

\bibitem[{{Kaper} {et~al.}(2006){Kaper}, {van der Meer}, \&
  {Najarro}}]{2006AA...457..595K}
{Kaper}, L., {van der Meer}, A., \& {Najarro}, F. 2006, \aap, 457, 595

\bibitem[{{Karasev} {et~al.}(2010){Karasev}, {Lutovinov}, \&
  {Burenin}}]{2010MNRAS.409L..69K}
{Karasev}, D.~I., {Lutovinov}, A.~A., \& {Burenin}, R.~A. 2010, \mnras, 409,
  L69

\bibitem[{{Karasev} {et~al.}(2012){Karasev}, {Lutovinov}, {Revnivtsev}, \&
  {Krivonos}}]{2012AstL...38..629K}
{Karasev}, D.~I., {Lutovinov}, A.~A., {Revnivtsev}, M.~G., \& {Krivonos}, R.~A.
  2012, Astronomy Letters, 38, 629

\bibitem[{{Kaur} {et~al.}(2008){Kaur}, {Paul}, {Kumar}, \&
  {Sagar}}]{2008MNRAS.386.2253K}
{Kaur}, R., {Paul}, B., {Kumar}, B., \& {Sagar}, R. 2008, \mnras, 386, 2253

\bibitem[{{Kelley} {et~al.}(1981){Kelley}, {Apparao}, {Doxsey}, {Jernigan},
  {Naranan}, \& {Rappaport}}]{1981ApJ...243..251K}
{Kelley}, R.~L., {Apparao}, K.~M.~V., {Doxsey}, R.~E., {et~al.} 1981, \apj,
  243, 251

\bibitem[{{Kelley} {et~al.}(1983){Kelley}, {Rappaport}, \&
  {Ayasli}}]{1983ApJ...274..765K}
{Kelley}, R.~L., {Rappaport}, S., \& {Ayasli}, S. 1983, \apj, 274, 765

\bibitem[{{Kennea} {et~al.}(2010){Kennea}, {Curran}, {Krimm}, {Romano},
  {Mangano}, {Evans}, {Yamaoka}, \& {Burrows}}]{2010ATel.3060....1K}
{Kennea}, J.~A., {Curran}, P., {Krimm}, H., {et~al.} 2010, The Astronomer's
  Telegram, 3060, 1

\bibitem[{{Kharchenko} {et~al.}(2007){Kharchenko}, {Scholz}, {Piskunov},
  {R{\"o}ser}, \& {Schilbach}}]{2007AN....328..889K}
{Kharchenko}, N.~V., {Scholz}, R.~D., {Piskunov}, A.~E., {R{\"o}ser}, S., \&
  {Schilbach}, E. 2007, Astronomische Nachrichten, 328, 889

\bibitem[{{Kinugasa} {et~al.}(1998){Kinugasa}, {Torii}, {Hashimoto}, {Tsunemi},
  {Hayashida}, {Kitamoto}, {Kamata}, {Dotani}, {Nagase}, {Sugizaki}, {Ueda},
  {Kawai}, {Makishima}, \& {Yamauchi}}]{1998ApJ...495..435K}
{Kinugasa}, K., {Torii}, K., {Hashimoto}, Y., {et~al.} 1998, \apj, 495, 435

\bibitem[{{Koenigsberger} {et~al.}(2003){Koenigsberger}, {Canalizo}, {Arrieta},
  {Richer}, \& {Georgiev}}]{2003RMxAA..39...17K}
{Koenigsberger}, G., {Canalizo}, G., {Arrieta}, A., {Richer}, M.~G., \&
  {Georgiev}, L. 2003, \rmxaa, 39, 17

\bibitem[{{Koenigsberger} {et~al.}(2006){Koenigsberger}, {Georgiev}, {Moreno},
  {Richer}, {Toledano}, {Canalizo}, \& {Arrieta}}]{2006AA...458..513K}
{Koenigsberger}, G., {Georgiev}, L., {Moreno}, E., {et~al.} 2006, \aap, 458,
  513

\bibitem[{{Krti{\v{c}}ka} {et~al.}(2015){Krti{\v{c}}ka}, {Kub{\'a}t}, \&
  {Krti{\v{c}}kov{\'a}}}]{2015AA...579A.111K}
{Krti{\v{c}}ka}, J., {Kub{\'a}t}, J., \& {Krti{\v{c}}kov{\'a}}, I. 2015, \aap,
  579, A111

\bibitem[{{La Palombara} {et~al.}(2021){La Palombara}, {Sidoli}, {Esposito},
  {Israel}, \& {Rodr{\'\i}guez Castillo}}]{2021AA...649A.118L}
{La Palombara}, N., {Sidoli}, L., {Esposito}, P., {Israel}, G.~L., \&
  {Rodr{\'\i}guez Castillo}, G.~A. 2021, \aap, 649, A118

\bibitem[{{La Parola} {et~al.}(2013{\natexlab{a}}){La Parola}, {Cusumano},
  {Segreto}, {D'A{\`\i}}, {Masetti}, \& {D'Elia}}]{2013ApJ...775L..24L}
{La Parola}, V., {Cusumano}, G., {Segreto}, A., {et~al.} 2013{\natexlab{a}},
  \apjl, 775, L24

\bibitem[{{La Parola} {et~al.}(2013{\natexlab{b}}){La Parola}, {D'A{\`\i}},
  {Cusumano}, {Segreto}, {Masetti}, \& {Melandri}}]{2013arXiv1305.3916L}
{La Parola}, V., {D'A{\`\i}}, A., {Cusumano}, G., {et~al.} 2013{\natexlab{b}},
  arXiv e-prints, arXiv:1305.3916

\bibitem[{{Lamb} {et~al.}(1980){Lamb}, {Markert}, {Hartman}, {Thompson}, \&
  {Bignami}}]{1980ApJ...239..651L}
{Lamb}, R.~C., {Markert}, T.~H., {Hartman}, R.~C., {Thompson}, D.~J., \&
  {Bignami}, G.~F. 1980, \apj, 239, 651

\bibitem[{{LaSala} {et~al.}(1998){LaSala}, {Charles}, {Smith},
  {Balucinska-Church}, \& {Church}}]{1998MNRAS.301..285L}
{LaSala}, J., {Charles}, P.~A., {Smith}, R.~A.~D., {Balucinska-Church}, M., \&
  {Church}, M.~J. 1998, \mnras, 301, 285

\bibitem[{{Levenhagen} \& {Leister}(2006)}]{2006MNRAS.371..252L}
{Levenhagen}, R.~S. \& {Leister}, N.~V. 2006, \mnras, 371, 252

\bibitem[{{Levine} {et~al.}(2011){Levine}, {Bradt}, {Chakrabarty}, {Corbet}, \&
  {Harris}}]{2011ApJS..196....6L}
{Levine}, A.~M., {Bradt}, H.~V., {Chakrabarty}, D., {Corbet}, R. H.~D., \&
  {Harris}, R.~J. 2011, \apjs, 196, 6

\bibitem[{{Levine} {et~al.}(2004){Levine}, {Rappaport}, {Remillard}, \&
  {Savcheva}}]{2004ApJ...617.1284L}
{Levine}, A.~M., {Rappaport}, S., {Remillard}, R., \& {Savcheva}, A. 2004,
  \apj, 617, 1284

\bibitem[{{Lin} {et~al.}(2002){Lin}, {Church}, {Nagase}, \&
  {Ba{\l}uci{\'n}ska-Church}}]{2002MNRAS.337.1245L}
{Lin}, X.~B., {Church}, M.~J., {Nagase}, F., \& {Ba{\l}uci{\'n}ska-Church}, M.
  2002, \mnras, 337, 1245

\bibitem[{{Lindstr{\o}m} {et~al.}(2005){Lindstr{\o}m}, {Griffin}, {Kiss},
  {Uemura}, {Derekas}, {M{\'e}sz{\'a}ros}, \&
  {Sz{\'e}kely}}]{2005MNRAS.363..882L}
{Lindstr{\o}m}, C., {Griffin}, J., {Kiss}, L.~L., {et~al.} 2005, \mnras, 363,
  882

\bibitem[{{Liu} {et~al.}(2000){Liu}, {van Paradijs}, \& {van den
  Heuvel}}]{2000AAS..147...25L}
{Liu}, Q.~Z., {van Paradijs}, J., \& {van den Heuvel}, E.~P.~J. 2000, \aaps,
  147, 25

\bibitem[{{Liu} {et~al.}(2005){Liu}, {van Paradijs}, \& {van den
  Heuvel}}]{2005AA...442.1135L}
{Liu}, Q.~Z., {van Paradijs}, J., \& {van den Heuvel}, E.~P.~J. 2005, \aap,
  442, 1135

\bibitem[{{Liu} {et~al.}(2006){Liu}, {van Paradijs}, \& {van den
  Heuvel}}]{2006AA...455.1165L}
{Liu}, Q.~Z., {van Paradijs}, J., \& {van den Heuvel}, E.~P.~J. 2006, \aap,
  455, 1165

\bibitem[{{Lopes de Oliveira} {et~al.}(2006){Lopes de Oliveira}, {Motch},
  {Haberl}, {Negueruela}, \& {Janot-Pacheco}}]{2006AA...454..265L}
{Lopes de Oliveira}, R., {Motch}, C., {Haberl}, F., {Negueruela}, I., \&
  {Janot-Pacheco}, E. 2006, \aap, 454, 265

\bibitem[{{Lutovinov} {et~al.}(2005){Lutovinov}, {Rodriguez}, {Revnivtsev}, \&
  {Shtykovskiy}}]{2005AA...433L..41L}
{Lutovinov}, A., {Rodriguez}, J., {Revnivtsev}, M., \& {Shtykovskiy}, P. 2005,
  \aap, 433, L41

\bibitem[{{Lutovinov} {et~al.}(2016){Lutovinov}, {Buckley}, {Townsend},
  {Tsygankov}, \& {Kennea}}]{2016MNRAS.462.3823L}
{Lutovinov}, A.~A., {Buckley}, D. A.~H., {Townsend}, L.~J., {Tsygankov}, S.~S.,
  \& {Kennea}, J. 2016, \mnras, 462, 3823

\bibitem[{{Lyubimkov} {et~al.}(1997){Lyubimkov}, {Rostopchin}, {Roche}, \&
  {Tarasov}}]{1997MNRAS.286..549L}
{Lyubimkov}, L.~S., {Rostopchin}, S.~I., {Roche}, P., \& {Tarasov}, A.~E. 1997,
  \mnras, 286, 549

\bibitem[{{Maitra} {et~al.}(2023){Maitra}, {Kaltenbrunner}, {Haberl},
  {Buckley}, {Monageng}, {Udalski}, {Carpano}, {Coley}, {Doroshenko}, {Ducci},
  {Malacaria}, {K{\"o}nig}, {Santangelo}, {Vasilopoulos}, \&
  {Wilms}}]{2023A&A...669A..30M}
{Maitra}, C., {Kaltenbrunner}, D., {Haberl}, F., {et~al.} 2023, \aap, 669, A30

\bibitem[{{Ma{\'\i}z Apell{\'a}niz} {et~al.}(2016){Ma{\'\i}z Apell{\'a}niz},
  {Sota}, {Arias}, {Barb{\'a}}, {Walborn}, {Sim{\'o}n-D{\'\i}az}, {Negueruela},
  {Marco}, {Le{\~a}o}, {Herrero}, {Gamen}, \& {Alfaro}}]{2016ApJS..224....4M}
{Ma{\'\i}z Apell{\'a}niz}, J., {Sota}, A., {Arias}, J.~I., {et~al.} 2016,
  \apjs, 224, 4

\bibitem[{{Makishima} {et~al.}(1984){Makishima}, {Kawai}, {Koyama},
  {Shibazaki}, {Nagase}, \& {Nakagawa}}]{1984PASJ...36..679M}
{Makishima}, K., {Kawai}, N., {Koyama}, K., {et~al.} 1984, \pasj, 36, 679

\bibitem[{{Malacaria} {et~al.}(2022){Malacaria}, {Bhargava}, {Coley}, {Ducci},
  {Pradhan}, {Ballhausen}, {Fuerst}, {Islam}, {Jaisawal}, {Jenke},
  {Kretschmar}, {Kreykenbohm}, {Pottschmidt}, {Sokolova-Lapa}, {Staubert},
  {Wilms}, {Wilson-Hodge}, \& {Wolff}}]{2022ApJ...927..194M}
{Malacaria}, C., {Bhargava}, Y., {Coley}, J.~B., {et~al.} 2022, \apj, 927, 194

\bibitem[{{Malacaria} {et~al.}(2021){Malacaria}, {Kretschmar}, {Madsen},
  {Wilson-Hodge}, {Coley}, {Jenke}, {Lutovinov}, {Pottschmidt}, {Tsygankov}, \&
  {Wilms}}]{2021ApJ...909..153M}
{Malacaria}, C., {Kretschmar}, P., {Madsen}, K.~K., {et~al.} 2021, \apj, 909,
  153

\bibitem[{{Marcu-Cheatham} {et~al.}(2015){Marcu-Cheatham}, {Pottschmidt},
  {K{\"u}hnel}, {M{\"u}ller}, {Falkner}, {Caballero}, {Finger}, {Jenke},
  {Wilson-Hodge}, {F{\"u}rst}, {Grinberg}, {Hemphill}, {Kreykenbohm},
  {Klochkov}, {Rothschild}, {Terada}, {Enoto}, {Iwakiri}, {Wolff}, {Becker},
  {Wood}, \& {Wilms}}]{2015ApJ...815...44M}
{Marcu-Cheatham}, D.~M., {Pottschmidt}, K., {K{\"u}hnel}, M., {et~al.} 2015,
  \apj, 815, 44

\bibitem[{{Markwardt} {et~al.}(2010){Markwardt}, {Baumgartner}, {Skinner}, \&
  {Corbet}}]{2010ATel.2564....1M}
{Markwardt}, C.~B., {Baumgartner}, W.~H., {Skinner}, G.~K., \& {Corbet}, R.
  H.~D. 2010, The Astronomer's Telegram, 2564, 1

\bibitem[{{Markwardt} {et~al.}(2008){Markwardt}, {Pereira}, {Ray}, {Smith}, \&
  {Swank}}]{2008ATel.1679....1M}
{Markwardt}, C.~B., {Pereira}, D., {Ray}, P.~S., {Smith}, E., \& {Swank}, J.~H.
  2008, The Astronomer's Telegram, 1679, 1

\bibitem[{{Marsden} {et~al.}(1998){Marsden}, {Gruber}, {Heindl}, {Pelling}, \&
  {Rothschild}}]{1998ApJ...502L.129M}
{Marsden}, D., {Gruber}, D.~E., {Heindl}, W.~A., {Pelling}, M.~R., \&
  {Rothschild}, R.~E. 1998, \apjl, 502, L129

\bibitem[{{Mart{\'\i}} {et~al.}(2016){Mart{\'\i}}, {Luque-Escamilla}, \&
  {Mu{\~n}oz-Arjonilla}}]{2016AA...596A..46M}
{Mart{\'\i}}, J., {Luque-Escamilla}, P.~L., \& {Mu{\~n}oz-Arjonilla}, {\'A}.~J.
  2016, \aap, 596, A46

\bibitem[{{Martins} {et~al.}(2005){Martins}, {Schaerer}, \&
  {Hillier}}]{2005AA...436.1049M}
{Martins}, F., {Schaerer}, D., \& {Hillier}, D.~J. 2005, \aap, 436, 1049

\bibitem[{{Masetti} {et~al.}(2018){Masetti}, {Ferreira}, {Saito}, {Kammers}, \&
  {Minniti}}]{2018ATel11992....1M}
{Masetti}, N., {Ferreira}, T.~S., {Saito}, R.~K., {Kammers}, R., \& {Minniti},
  D. 2018, The Astronomer's Telegram, 11992, 1

\bibitem[{{Masetti} {et~al.}(2012){Masetti}, {Landi}, {Parisi}, {Bazzano}, \&
  {Bird}}]{2012ATel.4209....1M}
{Masetti}, N., {Landi}, R., {Parisi}, P., {Bazzano}, A., \& {Bird}, A.~J. 2012,
  The Astronomer's Telegram, 4209, 1

\bibitem[{{Masetti} {et~al.}(2008){Masetti}, {Mason}, {Morelli}, {Cellone},
  {McBride}, {Palazzi}, {Bassani}, {Bazzano}, {Bird}, {Charles}, {Dean},
  {Galaz}, {Gehrels}, {Landi}, {Malizia}, {Minniti}, {Panessa}, {Romero},
  {Stephen}, {Ubertini}, \& {Walter}}]{2008AA...482..113M}
{Masetti}, N., {Mason}, E., {Morelli}, L., {et~al.} 2008, \aap, 482, 113

\bibitem[{{Masetti} {et~al.}(2010){Masetti}, {Parisi}, {Palazzi},
  {Jim{\'e}nez-Bail{\'o}n}, {Chavushyan}, {Bassani}, {Bazzano}, {Bird}, {Dean},
  {Charles}, {Galaz}, {Landi}, {Malizia}, {Mason}, {McBride}, {Minniti},
  {Morelli}, {Schiavone}, {Stephen}, \& {Ubertini}}]{2010AA...519A..96M}
{Masetti}, N., {Parisi}, P., {Palazzi}, E., {et~al.} 2010, \aap, 519, A96

\bibitem[{{Masetti} {et~al.}(2013){Masetti}, {Parisi}, {Palazzi},
  {Jim{\'e}nez-Bail{\'o}n}, {Chavushyan}, {McBride}, {Rojas}, {Steward},
  {Bassani}, {Bazzano}, {Bird}, {Charles}, {Galaz}, {Landi}, {Malizia},
  {Mason}, {Minniti}, {Morelli}, {Schiavone}, {Stephen}, \&
  {Ubertini}}]{2013AA...556A.120M}
{Masetti}, N., {Parisi}, P., {Palazzi}, E., {et~al.} 2013, \aap, 556, A120

\bibitem[{{Masetti} {et~al.}(2009){Masetti}, {Parisi}, {Palazzi},
  {Jim{\'e}nez-Bail{\'o}n}, {Morelli}, {Chavushyan}, {Mason}, {McBride},
  {Bassani}, {Bazzano}, {Bird}, {Dean}, {Galaz}, {Gehrels}, {Landi}, {Malizia},
  {Minniti}, {Schiavone}, {Stephen}, \& {Ubertini}}]{2009AA...495..121M}
{Masetti}, N., {Parisi}, P., {Palazzi}, E., {et~al.} 2009, \aap, 495, 121

\bibitem[{{Mason} {et~al.}(2012){Mason}, {Clark}, {Norton}, {Crowther},
  {Tauris}, {Langer}, {Negueruela}, \& {Roche}}]{2012MNRAS.422..199M}
{Mason}, A.~B., {Clark}, J.~S., {Norton}, A.~J., {et~al.} 2012, \mnras, 422,
  199

\bibitem[{{Mason} {et~al.}(2009){Mason}, {Clark}, {Norton}, {Negueruela}, \&
  {Roche}}]{2009AA...505..281M}
{Mason}, A.~B., {Clark}, J.~S., {Norton}, A.~J., {Negueruela}, I., \& {Roche},
  P. 2009, \aap, 505, 281

\bibitem[{{Mason} {et~al.}(2010){Mason}, {Norton}, {Clark}, {Negueruela}, \&
  {Roche}}]{2010AA...509A..79M}
{Mason}, A.~B., {Norton}, A.~J., {Clark}, J.~S., {Negueruela}, I., \& {Roche},
  P. 2010, \aap, 509, A79

\bibitem[{{Mason} {et~al.}(2011){Mason}, {Norton}, {Clark}, {Negueruela}, \&
  {Roche}}]{2011AA...532A.124M}
{Mason}, A.~B., {Norton}, A.~J., {Clark}, J.~S., {Negueruela}, I., \& {Roche},
  P. 2011, \aap, 532, A124

\bibitem[{{Massi} \& {Torricelli-Ciamponi}(2016)}]{2016AA...585A.123M}
{Massi}, M. \& {Torricelli-Ciamponi}, G. 2016, \aap, 585, A123

\bibitem[{{Mathew} \& {Subramaniam}(2011)}]{2011BASI...39..517M}
{Mathew}, B. \& {Subramaniam}, A. 2011, Bulletin of the Astronomical Society of
  India, 39, 517

\bibitem[{{Mattana} {et~al.}(2006){Mattana}, {G{\"o}tz}, {Falanga}, {Senziani},
  {de Luca}, {Esposito}, \& {Caraveo}}]{2006AA...460L...1M}
{Mattana}, F., {G{\"o}tz}, D., {Falanga}, M., {et~al.} 2006, \aap, 460, L1

\bibitem[{{McBride} {et~al.}(2006){McBride}, {Wilms}, {Coe}, {Kreykenbohm},
  {Rothschild}, {Coburn}, {Galache}, {Kretschmar}, {Edge}, \&
  {Staubert}}]{2006AA...451..267M}
{McBride}, V.~A., {Wilms}, J., {Coe}, M.~J., {et~al.} 2006, \aap, 451, 267

\bibitem[{{McBride} {et~al.}(2007){McBride}, {Wilms}, {Kreykenbohm}, {Coe},
  {Rothschild}, {Kretschmar}, {Pottschmidt}, {Fisher}, \&
  {Hamson}}]{2007AA...470.1065M}
{McBride}, V.~A., {Wilms}, J., {Kreykenbohm}, I., {et~al.} 2007, \aap, 470,
  1065

\bibitem[{{McClintock} {et~al.}(1976){McClintock}, {Rappaport}, {Joss},
  {Bradt}, {Buff}, {Clark}, {Hearn}, {Lewin}, {Matilsky}, {Mayer}, \&
  {Primini}}]{1976ApJ...206L..99M}
{McClintock}, J.~E., {Rappaport}, S., {Joss}, P.~C., {et~al.} 1976, \apjl, 206,
  L99

\bibitem[{{McClintock} {et~al.}(1977){McClintock}, {Rappaport}, {Nugent}, \&
  {Li}}]{1977ApJ...216L..15M}
{McClintock}, J.~E., {Rappaport}, S.~A., {Nugent}, J.~J., \& {Li}, F.~K. 1977,
  \apjl, 216, L15

\bibitem[{{McCollum} \& {Laine}(2019)}]{2019ATel12560....1M}
{McCollum}, B. \& {Laine}, S. 2019, The Astronomer's Telegram, 12560, 1

\bibitem[{{Mereghetti} {et~al.}(2008){Mereghetti}, {Romano}, \&
  {Sidoli}}]{2008AA...483..249M}
{Mereghetti}, S., {Romano}, P., \& {Sidoli}, L. 2008, \aap, 483, 249

\bibitem[{{Miller-Jones} {et~al.}(2021){Miller-Jones}, {Bahramian}, {Orosz},
  {Mandel}, {Gou}, {Maccarone}, {Neijssel}, {Zhao}, {Zi{\'o}{\l}kowski},
  {Reid}, {Uttley}, {Zheng}, {Byun}, {Dodson}, {Grinberg}, {Jung}, {Kim},
  {Marcote}, {Markoff}, {Rioja}, {Rushton}, {Russell}, {Sivakoff}, {Tetarenko},
  {Tudose}, \& {Wilms}}]{2021Sci...371.1046M}
{Miller-Jones}, J. C.~A., {Bahramian}, A., {Orosz}, J.~A., {et~al.} 2021,
  Science, 371, 1046

\bibitem[{{Miller-Jones} {et~al.}(2018){Miller-Jones}, {Deller}, {Shannon},
  {Dodson}, {Mold{\'o}n}, {Rib{\'o}}, {Dubus}, {Johnston}, {Paredes}, {Ransom},
  \& {Tomsick}}]{2018MNRAS.479.4849M}
{Miller-Jones}, J.~C.~A., {Deller}, A.~T., {Shannon}, R.~M., {et~al.} 2018,
  \mnras, 479, 4849

\bibitem[{{Morel} \& {Grosdidier}(2005)}]{2005MNRAS.356..665M}
{Morel}, T. \& {Grosdidier}, Y. 2005, \mnras, 356, 665

\bibitem[{{Moritani} {et~al.}(2018){Moritani}, {Kawano}, {Chimasu}, {Kawachi},
  {Takahashi}, {Takata}, \& {Carciofi}}]{2018PASJ...70...61M}
{Moritani}, Y., {Kawano}, T., {Chimasu}, S., {et~al.} 2018, \pasj, 70, 61

\bibitem[{{Motch} {et~al.}(1997){Motch}, {Haberl}, {Dennerl}, {Pakull}, \&
  {Janot-Pacheco}}]{1997AA...323..853M}
{Motch}, C., {Haberl}, F., {Dennerl}, K., {Pakull}, M., \& {Janot-Pacheco}, E.
  1997, \aap, 323, 853

\bibitem[{{Motch} {et~al.}(2003){Motch}, {Herent}, \&
  {Guillout}}]{2003AN....324...61M}
{Motch}, C., {Herent}, O., \& {Guillout}, P. 2003, Astronomische Nachrichten,
  324, 61

\bibitem[{{Motch} \& {Janot-Pacheco}(1987)}]{1987AA...182L..55M}
{Motch}, C. \& {Janot-Pacheco}, E. 1987, \aap, 182, L55

\bibitem[{{Mukerjee} \& {Antia}(2021)}]{2021ApJ...920..139M}
{Mukerjee}, K. \& {Antia}, H.~M. 2021, \apj, 920, 139

\bibitem[{{Nabizadeh} {et~al.}(2019){Nabizadeh}, {Tsygankov}, {Karasev},
  {M{\"o}nkk{\"o}nen}, {Lutovinov}, {Nagirner}, \&
  {Poutanen}}]{2019AA...622A.198N}
{Nabizadeh}, A., {Tsygankov}, S.~S., {Karasev}, D.~I., {et~al.} 2019, \aap,
  622, A198

\bibitem[{{Nabizadeh} {et~al.}(2022){Nabizadeh}, {Tsygankov}, {Molkov},
  {Karasev}, {Ji}, {Lutovinov}, \& {Poutanen}}]{2022AA...657A..58N}
{Nabizadeh}, A., {Tsygankov}, S.~S., {Molkov}, S.~V., {et~al.} 2022, \aap, 657,
  A58

\bibitem[{{Naz{\'e}} {et~al.}(2022){Naz{\'e}}, {Rauw}, {Czesla}, {Smith}, \&
  {Robrade}}]{2022MNRAS.510.2286N}
{Naz{\'e}}, Y., {Rauw}, G., {Czesla}, S., {Smith}, M.~A., \& {Robrade}, J.
  2022, \mnras, 510, 2286

\bibitem[{{Negueruela} {et~al.}(2008){Negueruela}, {Casares}, {Verrecchia},
  {Blay}, {Israel}, \& {Covino}}]{2008ATel.1876....1N}
{Negueruela}, I., {Casares}, J., {Verrecchia}, F., {et~al.} 2008, The
  Astronomer's Telegram, 1876, 1

\bibitem[{{Negueruela} {et~al.}(2003){Negueruela}, {Israel}, {Marco}, {Norton},
  \& {Speziali}}]{2003AA...397..739N}
{Negueruela}, I., {Israel}, G.~L., {Marco}, A., {Norton}, A.~J., \& {Speziali},
  R. 2003, \aap, 397, 739

\bibitem[{{Negueruela} \& {Okazaki}(2001)}]{2001AA...369..108N}
{Negueruela}, I. \& {Okazaki}, A.~T. 2001, \aap, 369, 108

\bibitem[{{Negueruela} {et~al.}(2011){Negueruela}, {Rib{\'o}}, {Herrero},
  {Lorenzo}, {Khangulyan}, \& {Aharonian}}]{2011ApJ...732L..11N}
{Negueruela}, I., {Rib{\'o}}, M., {Herrero}, A., {et~al.} 2011, \apjl, 732, L11

\bibitem[{{Negueruela} {et~al.}(1999){Negueruela}, {Roche}, {Fabregat}, \&
  {Coe}}]{1999MNRAS.307..695N}
{Negueruela}, I., {Roche}, P., {Fabregat}, J., \& {Coe}, M.~J. 1999, \mnras,
  307, 695

\bibitem[{{Negueruela} \& {Schurch}(2007)}]{2007AA...461..631N}
{Negueruela}, I. \& {Schurch}, M.~P.~E. 2007, \aap, 461, 631

\bibitem[{{Negueruela} {et~al.}(2006{\natexlab{a}}){Negueruela}, {Smith},
  {Harrison}, \& {Torrej{\'o}n}}]{2006ApJ...638..982N}
{Negueruela}, I., {Smith}, D.~M., {Harrison}, T.~E., \& {Torrej{\'o}n}, J.~M.
  2006{\natexlab{a}}, \apj, 638, 982

\bibitem[{{Negueruela} {et~al.}(2006{\natexlab{b}}){Negueruela}, {Smith},
  {Reig}, {Chaty}, \& {Torrej{\'o}n}}]{2006ESASP.604..165N}
{Negueruela}, I., {Smith}, D.~M., {Reig}, P., {Chaty}, S., \& {Torrej{\'o}n},
  J.~M. 2006{\natexlab{b}}, in ESA Special Publication, Vol. 604, The X-ray
  Universe 2005, ed. A.~{Wilson}, 165

\bibitem[{{Nemravov{\'a}} {et~al.}(2012){Nemravov{\'a}}, {Harmanec},
  {Koubsk{\'y}}, {Miroshnichenko}, {Yang}, {{\v{S}}lechta}, {Buil},
  {Kor{\v{c}}{\'a}kov{\'a}}, \& {Votruba}}]{2012AA...537A..59N}
{Nemravov{\'a}}, J., {Harmanec}, P., {Koubsk{\'y}}, P., {et~al.} 2012, \aap,
  537, A59

\bibitem[{{Nespoli} {et~al.}(2008{\natexlab{a}}){Nespoli}, {Fabregat}, \&
  {Mennickent}}]{2008ATel.1396....1N}
{Nespoli}, E., {Fabregat}, J., \& {Mennickent}, R.~E. 2008{\natexlab{a}}, The
  Astronomer's Telegram, 1396, 1

\bibitem[{{Nespoli} {et~al.}(2008{\natexlab{b}}){Nespoli}, {Fabregat}, \&
  {Mennickent}}]{2008AA...486..911N}
{Nespoli}, E., {Fabregat}, J., \& {Mennickent}, R.~E. 2008{\natexlab{b}}, \aap,
  486, 911

\bibitem[{{Nikolaeva} {et~al.}(2013){Nikolaeva}, {Bikmaev}, {Melnikov},
  {Galeev}, {Zhuchkov}, \& {Irtuganov}}]{2013BCrAO.109...27N}
{Nikolaeva}, E.~A., {Bikmaev}, I.~F., {Melnikov}, S.~S., {et~al.} 2013,
  Bulletin Crimean Astrophysical Observatory, 109, 27

\bibitem[{{Ochsenbein} {et~al.}(2000){Ochsenbein}, {Bauer}, \&
  {Marcout}}]{2000AAS..143...23O}
{Ochsenbein}, F., {Bauer}, P., \& {Marcout}, J. 2000, \aaps, 143, 23

\bibitem[{{O'Connor} {et~al.}(2022){O'Connor}, {G{\"o}{\u{g}}{\"u}{\c{s}}},
  {Huppenkothen}, {Kouveliotou}, {Gorgone}, {Townsend}, {Calamida}, {Fruchter},
  {Buckley}, {Baring}, {Kennea}, {Younes}, {Arzoumanian}, {Bellm}, {Cenko},
  {Gendreau}, {Granot}, {Hailey}, {Harrison}, {Hartmann}, {Kaper}, {Kutyrev},
  {Slane}, {Stern}, {Troja}, {van der Horst}, {Wijers}, \&
  {Woudt}}]{2022ApJ...927..139O}
{O'Connor}, B., {G{\"o}{\u{g}}{\"u}{\c{s}}}, E., {Huppenkothen}, D., {et~al.}
  2022, \apj, 927, 139

\bibitem[{{Okazaki} \& {Negueruela}(2001)}]{2001AA...377..161O}
{Okazaki}, A.~T. \& {Negueruela}, I. 2001, \aap, 377, 161

\bibitem[{{Orosz} {et~al.}(2001){Orosz}, {Kuulkers}, {van der Klis},
  {McClintock}, {Garcia}, {Callanan}, {Bailyn}, {Jain}, \&
  {Remillard}}]{2001ApJ...555..489O}
{Orosz}, J.~A., {Kuulkers}, E., {van der Klis}, M., {et~al.} 2001, \apj, 555,
  489

\bibitem[{{Pacheco} {et~al.}(1982){Pacheco}, {Chevalier}, \&
  {Ilovaisky}}]{1982IAUS...98..151P}
{Pacheco}, E.~J., {Chevalier}, C., \& {Ilovaisky}, S.~A. 1982, in Be Stars, ed.
  M.~{Jaschek} \& H.~G. {Groth}, Vol.~98, 151--154

\bibitem[{{Pakull} {et~al.}(2003){Pakull}, {Motch}, \&
  {Negueruela}}]{2003ATel..202....1P}
{Pakull}, M.~W., {Motch}, C., \& {Negueruela}, I. 2003, The Astronomer's
  Telegram, 202, 1

\bibitem[{{Parkes} {et~al.}(1978){Parkes}, {Murdin}, \&
  {Mason}}]{1978MNRAS.184P..73P}
{Parkes}, G.~E., {Murdin}, P.~G., \& {Mason}, K.~O. 1978, \mnras, 184, 73P

\bibitem[{{Parkes} {et~al.}(1980){Parkes}, {Murdin}, \&
  {Mason}}]{1980MNRAS.190..537P}
{Parkes}, G.~E., {Murdin}, P.~G., \& {Mason}, K.~O. 1980, \mnras, 190, 537

\bibitem[{{Paul} {et~al.}(2001){Paul}, {Agrawal}, {Mukerjee}, {Rao}, {Seetha},
  \& {Kasturirangan}}]{2001AA...370..529P}
{Paul}, B., {Agrawal}, P.~C., {Mukerjee}, K., {et~al.} 2001, \aap, 370, 529

\bibitem[{{Pearlman} {et~al.}(2019){Pearlman}, {Coley}, {Corbet}, \&
  {Pottschmidt}}]{2019ApJ...873...86P}
{Pearlman}, A.~B., {Coley}, J.~B., {Corbet}, R. H.~D., \& {Pottschmidt}, K.
  2019, \apj, 873, 86

\bibitem[{{Pellizza} {et~al.}(2011){Pellizza}, {Chaty}, \&
  {Chisari}}]{2011AA...526A..15P}
{Pellizza}, L.~J., {Chaty}, S., \& {Chisari}, N.~E. 2011, \aap, 526, A15

\bibitem[{{Pellizza} {et~al.}(2006){Pellizza}, {Chaty}, \&
  {Negueruela}}]{2006AA...455..653P}
{Pellizza}, L.~J., {Chaty}, S., \& {Negueruela}, I. 2006, \aap, 455, 653

\bibitem[{{Picchi} {et~al.}(2020){Picchi}, {Shore}, {Harvey}, \&
  {Berdyugin}}]{2020AA...640A..96P}
{Picchi}, P., {Shore}, S.~N., {Harvey}, E.~J., \& {Berdyugin}, A. 2020, \aap,
  640, A96

\bibitem[{{Pike} \& {Harrison}(2020)}]{2020ATel14291....1P}
{Pike}, S.~N. \& {Harrison}, F.~A. 2020, The Astronomer's Telegram, 14291, 1

\bibitem[{{Piraino} {et~al.}(1999){Piraino}, {Santangelo}, {Giarrusso},
  {Segreto}, {Cusumano}, {del Sordo}, {dal Fiume}, {Orlandini}, {Robba},
  {Burderi}, {Oosterbroek}, \& {Parmar}}]{1999NuPhS..69..220P}
{Piraino}, S., {Santangelo}, A., {Giarrusso}, S., {et~al.} 1999, Nuclear
  Physics B Proceedings Supplements, 69, 220

\bibitem[{{Polcaro} {et~al.}(1990){Polcaro}, {Rossi}, {Giovannelli},
  {Ferrari-Toniolo}, {La Padula}, {Persi}, {Manchanda}, {Golinskaya}, {Kurt},
  {Misykima}, {Shafer}, {Shamolin}, {Smirnov}, {Sheffer}, {Boyarchuck}, \&
  {Gershberg}}]{1990AA...231..354P}
{Polcaro}, V.~F., {Rossi}, C., {Giovannelli}, F., {et~al.} 1990, \aap, 231, 354

\bibitem[{{Popper}(1950)}]{1950ApJ...111..495P}
{Popper}, D.~M. 1950, \apj, 111, 495

\bibitem[{{Porter}(1996)}]{1996MNRAS.280L..31P}
{Porter}, J.~M. 1996, \mnras, 280, L31

\bibitem[{{Pradhan} {et~al.}(2013){Pradhan}, {Maitra}, {Paul}, \&
  {Paul}}]{2013MNRAS.436..945P}
{Pradhan}, P., {Maitra}, C., {Paul}, B., \& {Paul}, B.~C. 2013, \mnras, 436,
  945

\bibitem[{{Raguzova} \& {Popov}(2005)}]{2005AAT...24..151R}
{Raguzova}, N.~V. \& {Popov}, S.~B. 2005, Astronomical and Astrophysical
  Transactions, 24, 151

\bibitem[{{Rahoui} \& {Chaty}(2008)}]{2008AA...492..163R}
{Rahoui}, F. \& {Chaty}, S. 2008, \aap, 492, 163

\bibitem[{{Rahoui} {et~al.}(2008){Rahoui}, {Chaty}, {Lagage}, \&
  {Pantin}}]{2008AA...484..801R}
{Rahoui}, F., {Chaty}, S., {Lagage}, P.~O., \& {Pantin}, E. 2008, \aap, 484,
  801

\bibitem[{{Raichur} \& {Paul}(2010{\natexlab{a}})}]{2010MNRAS.406.2663R}
{Raichur}, H. \& {Paul}, B. 2010{\natexlab{a}}, \mnras, 406, 2663

\bibitem[{{Raichur} \& {Paul}(2010{\natexlab{b}})}]{2010MNRAS.401.1532R}
{Raichur}, H. \& {Paul}, B. 2010{\natexlab{b}}, \mnras, 401, 1532

\bibitem[{{Ratti} {et~al.}(2010){Ratti}, {Bassa}, {Torres}, {Kuiper},
  {Miller-Jones}, \& {Jonker}}]{2010MNRAS.408.1866R}
{Ratti}, E.~M., {Bassa}, C.~G., {Torres}, M.~A.~P., {et~al.} 2010, \mnras, 408,
  1866

\bibitem[{{Ray} \& {Chakrabarty}(2002)}]{2002ApJ...581.1293R}
{Ray}, P.~S. \& {Chakrabarty}, D. 2002, \apj, 581, 1293

\bibitem[{{Reed}(2003)}]{2003AJ....125.2531R}
{Reed}, B.~C. 2003, \aj, 125, 2531

\bibitem[{{Reig} {et~al.}(2017){Reig}, {Blay}, \&
  {Blinov}}]{2017AA...598A..16R}
{Reig}, P., {Blay}, P., \& {Blinov}, D. 2017, \aap, 598, A16

\bibitem[{{Reig} {et~al.}(2020){Reig}, {Fabregat}, \&
  {Alfonso-Garz{\'o}n}}]{2020AA...640A..35R}
{Reig}, P., {Fabregat}, J., \& {Alfonso-Garz{\'o}n}, J. 2020, \aap, 640, A35

\bibitem[{{Reig} {et~al.}(2001){Reig}, {Negueruela}, {Buckley}, {Coe},
  {Fabregat}, \& {Haigh}}]{2001AA...367..266R}
{Reig}, P., {Negueruela}, I., {Buckley}, D.~A.~H., {et~al.} 2001, \aap, 367,
  266

\bibitem[{{Reig} {et~al.}(2004){Reig}, {Negueruela}, {Fabregat}, {Chato},
  {Blay}, \& {Mavromatakis}}]{2004AA...421..673R}
{Reig}, P., {Negueruela}, I., {Fabregat}, J., {et~al.} 2004, \aap, 421, 673

\bibitem[{{Reig} {et~al.}(2005{\natexlab{a}}){Reig}, {Negueruela}, {Fabregat},
  {Chato}, \& {Coe}}]{2005AA...440.1079R}
{Reig}, P., {Negueruela}, I., {Fabregat}, J., {Chato}, R., \& {Coe}, M.~J.
  2005{\natexlab{a}}, \aap, 440, 1079

\bibitem[{{Reig} {et~al.}(2005{\natexlab{b}}){Reig}, {Negueruela},
  {Papamastorakis}, {Manousakis}, \& {Kougentakis}}]{2005AA...440..637R}
{Reig}, P., {Negueruela}, I., {Papamastorakis}, G., {Manousakis}, A., \&
  {Kougentakis}, T. 2005{\natexlab{b}}, \aap, 440, 637

\bibitem[{{Reig} {et~al.}(2011){Reig}, {Nespoli}, {Fabregat}, \&
  {Mennickent}}]{2011AA...533A..23R}
{Reig}, P., {Nespoli}, E., {Fabregat}, J., \& {Mennickent}, R.~E. 2011, \aap,
  533, A23

\bibitem[{{Reig} \& {Roche}(1999)}]{1999MNRAS.306..100R}
{Reig}, P. \& {Roche}, P. 1999, \mnras, 306, 100

\bibitem[{{Reig} \& {Zezas}(2018)}]{2018AA...613A..52R}
{Reig}, P. \& {Zezas}, A. 2018, \aap, 613, A52

\bibitem[{{Reig} {et~al.}(2010){Reig}, {Zezas}, \&
  {Gkouvelis}}]{2010AA...522A.107R}
{Reig}, P., {Zezas}, A., \& {Gkouvelis}, L. 2010, \aap, 522, A107

\bibitem[{{Rho} {et~al.}(2004){Rho}, {Moon}, {Gotthelf}, {Pannuti}, \&
  {Corbet}}]{2004HEAD....8.1730R}
{Rho}, J., {Moon}, D.~S., {Gotthelf}, E., {Pannuti}, T., \& {Corbet}, R. 2004,
  in AAS/High Energy Astrophysics Division, Vol.~8, AAS/High Energy
  Astrophysics Division \#8, 17.30

\bibitem[{{Rivinius} {et~al.}(2013){Rivinius}, {Carciofi}, \&
  {Martayan}}]{2013AARv..21...69R}
{Rivinius}, T., {Carciofi}, A.~C., \& {Martayan}, C. 2013, \aapr, 21, 69

\bibitem[{{Rodes-Roca} {et~al.}(2018){Rodes-Roca}, {Bernabeu}, {Magazz{\`u}},
  {Torrej{\'o}n}, \& {Solano}}]{2018MNRAS.476.2110R}
{Rodes-Roca}, J.~J., {Bernabeu}, G., {Magazz{\`u}}, A., {Torrej{\'o}n}, J.~M.,
  \& {Solano}, E. 2018, \mnras, 476, 2110

\bibitem[{{Rodes-Roca} {et~al.}(2013){Rodes-Roca}, {Torrej{\'o}n},
  {Mart{\'\i}nez-N{\'u}{\~n}ez}, {Bernab{\'e}u}, \&
  {Magazz{\'u}}}]{2013AA...555A.115R}
{Rodes-Roca}, J.~J., {Torrej{\'o}n}, J.~M., {Mart{\'\i}nez-N{\'u}{\~n}ez}, S.,
  {Bernab{\'e}u}, G., \& {Magazz{\'u}}, A. 2013, \aap, 555, A115

\bibitem[{{Romano} {et~al.}(2010){Romano}, {Sidoli}, {Ducci}, {Cusumano}, {La
  Parola}, {Pagani}, {Page}, {Kennea}, {Burrows}, {Gehrels}, {Sguera}, \&
  {Bazzano}}]{2010MNRAS.401.1564R}
{Romano}, P., {Sidoli}, L., {Ducci}, L., {et~al.} 2010, \mnras, 401, 1564

\bibitem[{{Roy} {et~al.}(2020){Roy}, {Agrawal}, {Singari}, \&
  {Misra}}]{2020RAA....20..155R}
{Roy}, J., {Agrawal}, P.~C., {Singari}, B., \& {Misra}, R. 2020, Research in
  Astronomy and Astrophysics, 20, 155

\bibitem[{{Safi-Harb} {et~al.}(2007){Safi-Harb}, {Rib{\'o}}, {Butt},
  {Matheson}, {Negueruela}, {Lu}, {Jia}, \& {Chen}}]{2007ApJ...659..407S}
{Safi-Harb}, S., {Rib{\'o}}, M., {Butt}, Y., {et~al.} 2007, \apj, 659, 407

\bibitem[{{Salganik} {et~al.}(2022){Salganik}, {Tsygankov}, {Djupvik},
  {Karasev}, {Lutovinov}, {Buckley}, {Gromadzki}, \&
  {Poutanen}}]{2022MNRAS.509.5955S}
{Salganik}, A., {Tsygankov}, S.~S., {Djupvik}, A.~A., {et~al.} 2022, \mnras,
  509, 5955

\bibitem[{{Saraswat} \& {Apparao}(1992)}]{1992ApJ...401..678S}
{Saraswat}, P. \& {Apparao}, K. M.~V. 1992, \apj, 401, 678

\bibitem[{{Segreto} {et~al.}(2013){Segreto}, {La Parola}, {Cusumano},
  {D'A{\`\i}}, {Masetti}, \& {Campana}}]{2013AA...558A..99S}
{Segreto}, A., {La Parola}, V., {Cusumano}, G., {et~al.} 2013, \aap, 558, A99

\bibitem[{{Sguera} {et~al.}(2011){Sguera}, {Drave}, {Bird}, {Bazzano}, {Landi},
  \& {Ubertini}}]{2011MNRAS.417..573S}
{Sguera}, V., {Drave}, S.~P., {Bird}, A.~J., {et~al.} 2011, \mnras, 417, 573

\bibitem[{{Sguera} {et~al.}(2013){Sguera}, {Drave}, {Sidoli}, {Masetti},
  {Landi}, {Bird}, \& {Bazzano}}]{2013AA...556A..27S}
{Sguera}, V., {Drave}, S.~P., {Sidoli}, L., {et~al.} 2013, \aap, 556, A27

\bibitem[{{Sguera} {et~al.}(2007){Sguera}, {Hill}, {Bird}, {Dean}, {Bazzano},
  {Ubertini}, {Masetti}, {Landi}, {Malizia}, {Clark}, \&
  {Molina}}]{2007AA...467..249S}
{Sguera}, V., {Hill}, A.~B., {Bird}, A.~J., {et~al.} 2007, \aap, 467, 249

\bibitem[{{Sguera} {et~al.}(2020){Sguera}, {Sidoli}, {Bird}, {Paizis}, \&
  {Bazzano}}]{2020MNRAS.491.4543S}
{Sguera}, V., {Sidoli}, L., {Bird}, A.~J., {Paizis}, A., \& {Bazzano}, A. 2020,
  \mnras, 491, 4543

\bibitem[{{Sharma} {et~al.}(2022){Sharma}, {Sharma}, {Jain}, \&
  {Dutta}}]{2022MNRAS.509.5747S}
{Sharma}, P., {Sharma}, R., {Jain}, C., \& {Dutta}, A. 2022, \mnras, 509, 5747

\bibitem[{{Shaw} {et~al.}(2009){Shaw}, {Hill}, {Kuulkers}, {Brandt},
  {Chenevez}, \& {Kretschmar}}]{2009MNRAS.393..419S}
{Shaw}, S.~E., {Hill}, A.~B., {Kuulkers}, E., {et~al.} 2009, \mnras, 393, 419

\bibitem[{{Shenavrin} {et~al.}(2011){Shenavrin}, {Taranova}, \&
  {Nadzhip}}]{2011ARep...55...31S}
{Shenavrin}, V.~I., {Taranova}, O.~G., \& {Nadzhip}, A.~E. 2011, Astronomy
  Reports, 55, 31

\bibitem[{{Shirke} {et~al.}(2021){Shirke}, {Bala}, {Roy}, \&
  {Bhattacharya}}]{2021JApA...42...58S}
{Shirke}, P., {Bala}, S., {Roy}, J., \& {Bhattacharya}, D. 2021, Journal of
  Astrophysics and Astronomy, 42, 58

\bibitem[{{Sidoli} {et~al.}(2016){Sidoli}, {Esposito}, {Motta}, {Israel}, \&
  {Rodr{\'\i}guez Castillo}}]{2016MNRAS.460.3637S}
{Sidoli}, L., {Esposito}, P., {Motta}, S.~E., {Israel}, G.~L., \&
  {Rodr{\'\i}guez Castillo}, G.~A. 2016, \mnras, 460, 3637

\bibitem[{{Sidoli} {et~al.}(2017){Sidoli}, {Israel}, {Esposito},
  {Rodr{\'\i}guez Castillo}, \& {Postnov}}]{2017MNRAS.469.3056S}
{Sidoli}, L., {Israel}, G.~L., {Esposito}, P., {Rodr{\'\i}guez Castillo},
  G.~A., \& {Postnov}, K. 2017, \mnras, 469, 3056

\bibitem[{{Sidoli} \& {Paizis}(2018)}]{2018MNRAS.481.2779S}
{Sidoli}, L. \& {Paizis}, A. 2018, \mnras, 481, 2779

\bibitem[{{Sidoli} {et~al.}(2020){Sidoli}, {Postnov}, {Tiengo}, {Esposito},
  {Sguera}, {Paizis}, \& {Rodr{\'\i}guez Castillo}}]{2020AA...638A..71S}
{Sidoli}, L., {Postnov}, K., {Tiengo}, A., {et~al.} 2020, \aap, 638, A71

\bibitem[{{Sidoli} {et~al.}(2022){Sidoli}, {Sguera}, {Esposito}, {Oskinova}, \&
  {Polletta}}]{2022MNRAS.512.2929S}
{Sidoli}, L., {Sguera}, V., {Esposito}, P., {Oskinova}, L., \& {Polletta}, M.
  2022, \mnras, 512, 2929

\bibitem[{{Smith} {et~al.}(2005){Smith}, {Hazelton}, {Coburn}, {Boggs},
  {Fivian}, {Hurford}, {Hudson}, {Grefenstette}, \&
  {Gilmore}}]{2005ATel..557....1S}
{Smith}, D.~M., {Hazelton}, B., {Coburn}, W., {et~al.} 2005, The Astronomer's
  Telegram, 557, 1

\bibitem[{{Smith} {et~al.}(2002){Smith}, {Heindl}, \&
  {Swank}}]{2002ApJ...578L.129S}
{Smith}, D.~M., {Heindl}, W.~A., \& {Swank}, J.~H. 2002, \apjl, 578, L129

\bibitem[{{Smith} {et~al.}(2012){Smith}, {Lopes de Oliveira}, \&
  {Motch}}]{2012ApJ...755...64S}
{Smith}, M.~A., {Lopes de Oliveira}, R., \& {Motch}, C. 2012, \apj, 755, 64

\bibitem[{{Sota} {et~al.}(2014){Sota}, {Ma{\'\i}z Apell{\'a}niz}, {Morrell},
  {Barb{\'a}}, {Walborn}, {Gamen}, {Arias}, \& {Alfaro}}]{2014ApJS..211...10S}
{Sota}, A., {Ma{\'\i}z Apell{\'a}niz}, J., {Morrell}, N.~I., {et~al.} 2014,
  \apjs, 211, 10

\bibitem[{{Sota} {et~al.}(2011){Sota}, {Ma{\'\i}z Apell{\'a}niz}, {Walborn},
  {Alfaro}, {Barb{\'a}}, {Morrell}, {Gamen}, \& {Arias}}]{2011ApJS..193...24S}
{Sota}, A., {Ma{\'\i}z Apell{\'a}niz}, J., {Walborn}, N.~R., {et~al.} 2011,
  \apjs, 193, 24

\bibitem[{{Stecchini} {et~al.}(2017){Stecchini}, {Castro}, {Jablonski},
  {D'Amico}, \& {Braga}}]{2017ApJ...843L..10S}
{Stecchini}, P.~E., {Castro}, M., {Jablonski}, F., {D'Amico}, F., \& {Braga},
  J. 2017, \apjl, 843, L10

\bibitem[{{Stecchini} {et~al.}(2020){Stecchini}, {D'Amico}, {Jablonski},
  {Castro}, \& {Braga}}]{2020MNRAS.493.2694S}
{Stecchini}, P.~E., {D'Amico}, F., {Jablonski}, F., {Castro}, M., \& {Braga},
  J. 2020, \mnras, 493, 2694

\bibitem[{{Stella} {et~al.}(1985){Stella}, {White}, {Davelaar}, {Parmar},
  {Blissett}, \& {van der Klis}}]{1985ApJ...288L..45S}
{Stella}, L., {White}, N.~E., {Davelaar}, J., {et~al.} 1985, \apjl, 288, L45

\bibitem[{{Stickland} {et~al.}(1997){Stickland}, {Lloyd}, \&
  {Radziun-Woodham}}]{1997MNRAS.286L..21S}
{Stickland}, D., {Lloyd}, C., \& {Radziun-Woodham}, A. 1997, \mnras, 286, L21

\bibitem[{{Stollberg} {et~al.}(1993){Stollberg}, {Finger}, {Wilson}, {Harmon},
  {Rubin}, {Zhang}, \& {Fishman}}]{1993IAUC.5836....1S}
{Stollberg}, M.~T., {Finger}, M.~H., {Wilson}, R.~B., {et~al.} 1993, \iaucirc,
  5836, 1

\bibitem[{{Stoyanov} {et~al.}(2014){Stoyanov}, {Zamanov}, {Latev}, {Abedin}, \&
  {Tomov}}]{2014AN....335.1060S}
{Stoyanov}, K.~A., {Zamanov}, R.~K., {Latev}, G.~Y., {Abedin}, A.~Y., \&
  {Tomov}, N.~A. 2014, Astronomische Nachrichten, 335, 1060

\bibitem[{{Strader} {et~al.}(2015){Strader}, {Chomiuk}, {Cheung}, {Salinas}, \&
  {Peacock}}]{2015ApJ...813L..26S}
{Strader}, J., {Chomiuk}, L., {Cheung}, C.~C., {Salinas}, R., \& {Peacock}, M.
  2015, \apjl, 813, L26

\bibitem[{{Strohmayer} {et~al.}(2009){Strohmayer}, {Rodriquez}, {Markwardt},
  {Tomsick}, {Bodaghee}, {Chaty}, {Corbel}, \& {Paizis}}]{2009ATel.2002....1S}
{Strohmayer}, T., {Rodriquez}, J., {Markwardt}, C., {et~al.} 2009, The
  Astronomer's Telegram, 2002, 1

\bibitem[{{Taylor}(2005)}]{2005ASPC..347...29T}
{Taylor}, M.~B. 2005, in Astronomical Society of the Pacific Conference Series,
  Vol. 347, Astronomical Data Analysis Software and Systems XIV, ed.
  P.~{Shopbell}, M.~{Britton}, \& R.~{Ebert}, 29

\bibitem[{{Thompson} {et~al.}(2007){Thompson}, {Tomsick}, {in 't Zand},
  {Rothschild}, \& {Walter}}]{2007ApJ...661..447T}
{Thompson}, T. W.~J., {Tomsick}, J.~A., {in 't Zand}, J.~J.~M., {Rothschild},
  R.~E., \& {Walter}, R. 2007, \apj, 661, 447

\bibitem[{{Torii} {et~al.}(1998){Torii}, {Kinugasa}, {Katayama}, {Kohmura},
  {Tsunemi}, {Sakano}, {Nishiuchi}, {Koyama}, {Yamauchi}, \&
  {Shigeo}}]{1998ApJ...508..854T}
{Torii}, K., {Kinugasa}, K., {Katayama}, K., {et~al.} 1998, \apj, 508, 854

\bibitem[{{Torii} {et~al.}(1999){Torii}, {Sugizaki}, {Kohmura}, {Endo}, \&
  {Nagase}}]{1999ApJ...523L..65T}
{Torii}, K., {Sugizaki}, M., {Kohmura}, T., {Endo}, T., \& {Nagase}, F. 1999,
  \apjl, 523, L65

\bibitem[{{Torrej{\'o}n} {et~al.}(2010){Torrej{\'o}n}, {Negueruela}, {Smith},
  \& {Harrison}}]{2010AA...510A..61T}
{Torrej{\'o}n}, J.~M., {Negueruela}, I., {Smith}, D.~M., \& {Harrison}, T.~E.
  2010, \aap, 510, A61

\bibitem[{{Torrej{\'o}n} \& {Orr}(2001)}]{2001AA...377..148T}
{Torrej{\'o}n}, J.~M. \& {Orr}, A. 2001, \aap, 377, 148

\bibitem[{{Townsend} {et~al.}(2011){Townsend}, {Coe}, {Corbet}, \&
  {Hill}}]{2011MNRAS.416.1556T}
{Townsend}, L.~J., {Coe}, M.~J., {Corbet}, R.~H.~D., \& {Hill}, A.~B. 2011,
  \mnras, 416, 1556

\bibitem[{{Tsygankov} {et~al.}(2021){Tsygankov}, {Lutovinov}, {Molkov},
  {Djupvik}, {Karasev}, {Doroshenko}, {Mushtukov}, {Malacaria}, {Kretschmar},
  \& {Poutanen}}]{2021ApJ...909..154T}
{Tsygankov}, S.~S., {Lutovinov}, A.~A., {Molkov}, S.~V., {et~al.} 2021, \apj,
  909, 154

\bibitem[{{Tsygankov} {et~al.}(2017){Tsygankov}, {Wijnands}, {Lutovinov},
  {Degenaar}, \& {Poutanen}}]{2017MNRAS.470..126T}
{Tsygankov}, S.~S., {Wijnands}, R., {Lutovinov}, A.~A., {Degenaar}, N., \&
  {Poutanen}, J. 2017, \mnras, 470, 126

\bibitem[{{Uchida} {et~al.}(2021){Uchida}, {Takahashi}, {Fukazawa}, \&
  {Makishima}}]{2021PASJ...73.1389U}
{Uchida}, N., {Takahashi}, H., {Fukazawa}, Y., \& {Makishima}, K. 2021, \pasj,
  73, 1389

\bibitem[{{van den Eijnden} {et~al.}(2021){van den Eijnden}, {Degenaar},
  {Russell}, {Wijnands}, {Bahramian}, {Miller-Jones}, {Hern{\'a}ndez
  Santisteban}, {Gallo}, {Atri}, {Plotkin}, {Maccarone}, {Sivakoff}, {Miller},
  {Reynolds}, {Russell}, {Maitra}, {Heinke}, {Armas Padilla}, \&
  {Shaw}}]{2021MNRAS.507.3899V}
{van den Eijnden}, J., {Degenaar}, N., {Russell}, T.~D., {et~al.} 2021, \mnras,
  507, 3899

\bibitem[{{van den Heuvel}(2019)}]{2019IAUS..346....1V}
{van den Heuvel}, E. P.~J. 2019, IAU Symposium, 346, 1

\bibitem[{{van der Meer} {et~al.}(2007){van der Meer}, {Kaper}, {van Kerkwijk},
  {Heemskerk}, \& {van den Heuvel}}]{2007AA...473..523V}
{van der Meer}, A., {Kaper}, L., {van Kerkwijk}, M.~H., {Heemskerk}, M.~H.~M.,
  \& {van den Heuvel}, E.~P.~J. 2007, \aap, 473, 523

\bibitem[{van~der Walt {et~al.}(2011)van~der Walt, Colbert, \&
  Varoquaux}]{van_der_walt_numpy_2011}
van~der Walt, S., Colbert, C., \& Varoquaux, G. 2011, CSE, 13

\bibitem[{{van Kerkwijk} {et~al.}(1996){van Kerkwijk}, {Geballe}, {King}, {van
  der Klis}, \& {van Paradijs}}]{1996AA...314..521V}
{van Kerkwijk}, M.~H., {Geballe}, T.~R., {King}, D.~L., {van der Klis}, M., \&
  {van Paradijs}, J. 1996, \aap, 314, 521

\bibitem[{{van Paradijs}(1995)}]{1995xrbi.nasa..536V}
{van Paradijs}, J. 1995, in X-ray Binaries, 536--577

\bibitem[{{van Soelen} {et~al.}(2022){van Soelen}, {Mc Keague}, {Malyshev},
  {Chernyakova}, {Komin}, {Matchett}, \& {Monageng}}]{2022MNRAS.515.1078V}
{van Soelen}, B., {Mc Keague}, S., {Malyshev}, D., {et~al.} 2022, \mnras, 515,
  1078

\bibitem[{{Verrecchia} {et~al.}(2002){Verrecchia}, {Israel}, {Negueruela},
  {Covino}, {Polcaro}, {Clark}, {Steele}, {Gualandi}, {Speziali}, \&
  {Stella}}]{2002AA...393..983V}
{Verrecchia}, F., {Israel}, G.~L., {Negueruela}, I., {et~al.} 2002, \aap, 393,
  983

\bibitem[{{Vieira} {et~al.}(2003){Vieira}, {Corradi}, {Alencar}, {Mendes},
  {Torres}, {Quast}, {Guimar{\~a}es}, \& {da Silva}}]{2003AJ....126.2971V}
{Vieira}, S.~L.~A., {Corradi}, W.~J.~B., {Alencar}, S.~H.~P., {et~al.} 2003,
  \aj, 126, 2971

\bibitem[{{Vijapurkar} \& {Drilling}(1993)}]{1993ApJS...89..293V}
{Vijapurkar}, J. \& {Drilling}, J.~S. 1993, \apjs, 89, 293

\bibitem[{{Waisberg} \& {Romani}(2015)}]{2015ApJ...805...18W}
{Waisberg}, I.~R. \& {Romani}, R.~W. 2015, \apj, 805, 18

\bibitem[{{Walter} {et~al.}(2015){Walter}, {Lutovinov}, {Bozzo}, \&
  {Tsygankov}}]{2015AARv..23....2W}
{Walter}, R., {Lutovinov}, A.~A., {Bozzo}, E., \& {Tsygankov}, S.~S. 2015,
  \aapr, 23, 2

\bibitem[{{Walter} {et~al.}(2006){Walter}, {Zurita Heras}, {Bassani},
  {Bazzano}, {Bodaghee}, {Dean}, {Dubath}, {Parmar}, {Renaud}, \&
  {Ubertini}}]{2006AA...453..133W}
{Walter}, R., {Zurita Heras}, J., {Bassani}, L., {et~al.} 2006, \aap, 453, 133

\bibitem[{{Wang} \& {Gies}(1998)}]{1998PASP..110.1310W}
{Wang}, Z.~X. \& {Gies}, D.~R. 1998, \pasp, 110, 1310

\bibitem[{{Warwick} {et~al.}(1981){Warwick}, {Marshall}, {Fraser}, {Watson},
  {Lawrence}, {Page}, {Pounds}, {Ricketts}, {Sims}, \&
  {Smith}}]{1981MNRAS.197..865W}
{Warwick}, R.~S., {Marshall}, N., {Fraser}, G.~W., {et~al.} 1981, \mnras, 197,
  865

\bibitem[{{Webb} {et~al.}(2020){Webb}, {Coriat}, {Traulsen}, {Ballet}, {Motch},
  {Carrera}, {Koliopanos}, {Authier}, {de la Calle}, {Ceballos}, {Colomo},
  {Chuard}, {Freyberg}, {Garcia}, {Kolehmainen}, {Lamer}, {Lin}, {Maggi},
  {Michel}, {Page}, {Page}, {Perea-Calderon}, {Pineau}, {Rodriguez}, {Rosen},
  {Santos Lleo}, {Saxton}, {Schwope}, {Tom{\'a}s}, {Watson}, \&
  {Zakardjian}}]{2020AA...641A.136W}
{Webb}, N.~A., {Coriat}, M., {Traulsen}, I., {et~al.} 2020, \aap, 641, A136

\bibitem[{{Wen} {et~al.}(2006){Wen}, {Levine}, {Corbet}, \&
  {Bradt}}]{2006ApJS..163..372W}
{Wen}, L., {Levine}, A.~M., {Corbet}, R. H.~D., \& {Bradt}, H.~V. 2006, \apjs,
  163, 372

\bibitem[{{Wenger} {et~al.}(2000){Wenger}, {Ochsenbein}, {Egret}, {Dubois},
  {Bonnarel}, {Borde}, {Genova}, {Jasniewicz}, {Lalo{\"e}}, {Lesteven}, \&
  {Monier}}]{2000AAS..143....9W}
{Wenger}, M., {Ochsenbein}, F., {Egret}, D., {et~al.} 2000, \aaps, 143, 9

\bibitem[{{White} {et~al.}(2000){White}, {Giommi}, \&
  {Angelini}}]{2000yCat.9031....0W}
{White}, N.~E., {Giommi}, P., \& {Angelini}, L. 2000, VizieR Online Data
  Catalog, IX/31

\bibitem[{{White} {et~al.}(1976){White}, {Mason}, {Huckle}, {Charles}, \&
  {Sanford}}]{1976ApJ...209L.119W}
{White}, N.~E., {Mason}, K.~O., {Huckle}, H.~E., {Charles}, P.~A., \&
  {Sanford}, P.~W. 1976, \apjl, 209, L119

\bibitem[{{Williams} {et~al.}(2010){Williams}, {Gies}, {Matson}, {Touhami},
  {Grundstrom}, {Huang}, \& {McSwain}}]{2010ApJ...723L..93W}
{Williams}, S.~J., {Gies}, D.~R., {Matson}, R.~A., {et~al.} 2010, \apjl, 723,
  L93

\bibitem[{{Wilson} {et~al.}(2002){Wilson}, {Finger}, {Coe}, {Laycock}, \&
  {Fabregat}}]{2002ApJ...570..287W}
{Wilson}, C.~A., {Finger}, M.~H., {Coe}, M.~J., {Laycock}, S., \& {Fabregat},
  J. 2002, \apj, 570, 287

\bibitem[{{Wilson} {et~al.}(1998){Wilson}, {Finger}, {Harmon}, {Chakrabarty},
  \& {Strohmayer}}]{1998ApJ...499..820W}
{Wilson}, C.~A., {Finger}, M.~H., {Harmon}, B.~A., {Chakrabarty}, D., \&
  {Strohmayer}, T. 1998, \apj, 499, 820

\bibitem[{{Wilson} {et~al.}(1997){Wilson}, {Finger}, {Harmon}, {Scott},
  {Wilson}, {Bildsten}, {Chakrabarty}, \& {Prince}}]{1997ApJ...479..388W}
{Wilson}, C.~A., {Finger}, M.~H., {Harmon}, B.~A., {et~al.} 1997, \apj, 479,
  388

\bibitem[{{Wolff} {et~al.}(2022){Wolff}, {Ray}, {Ng}, {Pradhan}, {Pottschmidt},
  {Corbet}, {Gendreau}, {Jaisawal}, {Sanna}, {Coley}, {Guillot}, {Malacaria},
  \& {Wilms}}]{2022ATel15556....1W}
{Wolff}, M.~T., {Ray}, P.~S., {Ng}, M., {et~al.} 2022, The Astronomer's
  Telegram, 15556, 1

\bibitem[{{Wood} {et~al.}(1984){Wood}, {Meekins}, {Yentis}, {Smathers},
  {McNutt}, {Bleach}, {Byram}, {Chupp}, {Friedman}, \&
  {Meidav}}]{1984ApJS...56..507W}
{Wood}, K.~S., {Meekins}, J.~F., {Yentis}, D.~J., {et~al.} 1984, \apjs, 56, 507

\bibitem[{{Zamanov} {et~al.}(2019){Zamanov}, {Stoyanov}, {Wolter}, {Marchev},
  \& {Petrov}}]{2019AA...622A.173Z}
{Zamanov}, R., {Stoyanov}, K.~A., {Wolter}, U., {Marchev}, D., \& {Petrov},
  N.~I. 2019, \aap, 622, A173

\bibitem[{{Zhao} {et~al.}(2019){Zhao}, {Heinke}, {Tsygankov}, {Ho}, {Potekhin},
  \& {Shaw}}]{2019MNRAS.488.4427Z}
{Zhao}, Y., {Heinke}, C.~O., {Tsygankov}, S.~S., {et~al.} 2019, \mnras, 488,
  4427

\bibitem[{{Zorec} {et~al.}(2005){Zorec}, {Fr{\'e}mat}, \&
  {Cidale}}]{2005AA...441..235Z}
{Zorec}, J., {Fr{\'e}mat}, Y., \& {Cidale}, L. 2005, \aap, 441, 235

\bibitem[{{Zurita Heras} \& {Chaty}(2008)}]{2008AA...489..657Z}
{Zurita Heras}, J.~A. \& {Chaty}, S. 2008, \aap, 489, 657

\bibitem[{{Zurita Heras} {et~al.}(2006){Zurita Heras}, {De Cesare}, {Walter},
  {Bodaghee}, {B{\'e}langer}, {Courvoisier}, {Shaw}, \&
  {Stephen}}]{2006AA...448..261Z}
{Zurita Heras}, J.~A., {De Cesare}, G., {Walter}, R., {et~al.} 2006, \aap, 448,
  261

\end{thebibliography}

\appendix
\onecolumn
\section{Catalogue of Galactic HMXBs}

\tiny
\begin{longtable}{lllllllll}\caption{Catalogue of Galactic HMXBs: General information. Spectype refers to the spectral type of the donor star in the binaries, Class indicates the general category of HMXB ($\gamma$ indicates a HMGB), Right Ascension (RA) and Declination (Dec) are given in J2000 alongside the 90\%\,positional error radius, Distance is queried from \citex{2021AJ....161..147B}, $RV$ is the systemic radial velocity, and $Var$ is the variability flag as detailed in Section\,\ref{subsect:contents}.}\label{tab:cat:general}\\
\hline\hline\\[-2ex]
Main ID &Spectype &Class &RA [deg] &Dec [deg] &err [mas] &Distance [pc] &RV [km/s] &Var \\
\hline\\[-2ex]
\endfirsthead
\caption{continued.}\\
\hline\hline\\[-2ex]
Main ID &Spectype &Class &RA [deg] &Dec [deg] &err [mas] &Distance [pc] &RV [km/s] &Var \\
\hline\\[-2ex]
\endhead
\hline
\endfoot
IGR J00370+6122 & BN0.7Ib~[1] & sg & 9.29013 & 61.36013 & 0.008 & 3401$_{-171}^{+186}$ & -80.0$\pm$3.0~[2] & Y\\[3pt]
gam Cas & B0.5IVpe~[3] & Be & 14.17745 & 60.71672 & 1.834 &  & -0.018$\pm$0.075~[4] & Y\\[3pt]
EM* AS   14 & B2~[5] &  & 18.99604 & 59.15394 & 0.011 & 2592$_{-140}^{+156}$ &    & \\[3pt]
2S 0114+650 & B1Iae~[6] & sg & 19.51123 & 65.29162 & 0.007 & 4475$_{-183}^{+217}$ & -31.0$\pm$5.0~[7] & Y\\[3pt]
4U 0115+634 & B0.2Ve~[8] & Be & 19.63319 & 63.74252 & 0.011 & 5787$_{-453}^{+817}$ &    & Y\\[3pt]
IGR J01363+6610 & B1Ve~[9] & Be & 23.95772 & 66.21202 & 0.007 & 5816$_{-407}^{+413}$ &    & Y\\[3pt]
RX J0146.9+6121 & B1IIIe~[10] & Be & 26.75088 & 61.35657 & 0.012 & 2751$_{-138}^{+162}$ &    & Y\\[3pt]
IGR J01583+6713 & B2IVe+~[11] & Be & 29.57703 & 67.22318 & 0.009 & 6048$_{-472}^{+464}$ &    & Y\\[3pt]
LS I+61 303 & B0Ve~[12] & $\gamma$\,Be & 40.13193 & 61.22933 & 0.007 & 2504$_{-67}^{+72}$ & -41.41$\pm$0.6~[13] & Y\\[3pt]
Swift J0243.6+6124 & O9.5V~[14] & Be & 40.91843 & 61.43438 & 0.007 & 5189$_{-314}^{+291}$ &    & Y\\[3pt]
V 0332+53 & O8.5Ve~[15] & Be & 53.74963 & 53.17314 & 0.014 & 5584$_{-506}^{+730}$ &    & Y\\[3pt]
X Per & B1Ve~[16] & Be & 58.84615 & 31.04584 & 0.03 & 595$_{-13}^{+17}$ & 1.0$\pm$0.9~[17] & Y\\[3pt]
XTE J0421+560 & B0/2I[e]~[18] & sgB[e] & 64.92556 & 55.99936 & 0.013 & 4094$_{-206}^{+276}$ & -51.0$\pm$2.0~[19] & Y\\[3pt]
RX J0440.9+4431 & B0.2Ve~[20] & Be & 70.24721 & 44.53034 & 0.012 & 2444$_{-77}^{+60}$ &    & Y\\[3pt]
EXO 051910+3737.7 & B0III-IVe~[21] & Be & 80.6468 & 37.67599 & 0.025 & 1317$_{-52}^{+55}$ & -20.5$\pm$4.4~[22] & \\[3pt]
1A 0535+262 & O9.5III-Ve~[23] & Be & 84.72739 & 26.31578 & 0.02 & 1793$_{-71}^{+76}$ & -30.0$\pm$4.0~[24] & Y\\[3pt]
AAO+28  342 & B5ne~[25] & Be & 88.97934 & 28.78511 & 0.025 & 1600$_{-73}^{+95}$ &    & \\[3pt]
IGR J06074+2205 & B0.5Ve~[26] & Be & 91.86089 & 22.0966 & 0.016 & 5989$_{-603}^{+559}$ & 18.9$\pm$4.1~[27] & Y\\[3pt]
HD 259440 & B0pe~[28] & $\gamma$\,Be & 98.2469 & 5.80032 & 0.018 & 1772$_{-86}^{+93}$ & 36.9$\pm$0.8~[29] & \\[3pt]
SAX J0635.2+0533 & B2V-B1IIIe~[30] & Be & 98.82616 & 5.55174 & 0.012 & 6290$_{-492}^{+655}$ &    & \\[3pt]
3A 0656-072 & O9.7Ve~[31] & Be & 104.56044 & -7.21249 & 0.038 & 218$_{-2}^{+2}$ &    & Y\\[3pt]
3A 0726-260 & O5Ve~[32] & Be & 112.22324 & -26.10802 & 0.009 & 7869$_{-1082}^{+1199}$ &    & Y\\[3pt]
SGR 0755-2933 & O6-8~[33] &  & 118.92702 & -29.56486 & 0.009 & 3356$_{-154}^{+159}$ &    & Y\\[3pt]
RX J0812.4-3114 & B0.2IVe~[34] & Be & 123.11815 & -31.24781 & 0.007 & 6659$_{-422}^{+528}$ &    & Y\\[3pt]
IGR J08262-3736 & OBV~[35] & Be & 126.55688 & -37.61995 & 0.006 & 5124$_{-228}^{+279}$ &    & Y\\[3pt]
GS 0834-430 & B0-2III-Ve~[36] & Be & 128.98108 & -43.1856 & 60.0 &  &    & \\[3pt]
IGR J08408-4503 & O8.5Ib-II(f)p~[37] & SFXT & 130.19919 & -45.0584 & 60.0 &  & 15.3$\pm$0.5~[38] & Y\\[3pt]
Vela X-1 & B0.5Iae-1b~[39] & sg & 135.52856 & -40.55465 & 0.011 & 1960$_{-52}^{+57}$ & -3.2$\pm$0.9~[40] & Y\\[3pt]
GRO J1008-57 & B0e~[41] & Be & 152.44564 & -58.29321 & 0.012 & 3536$_{-152}^{+148}$ &    & Y\\[3pt]
IGR J10101-5654 & sgB[e]~[42] & sgB[e] & 152.54942 & -56.92554 & 0.069 & 4474$_{-1171}^{+1094}$ &    & Y\\[3pt]
1FGL J1018.6-5856 & O6V~[43] & $\gamma$\,Be & 154.73156 & -58.94609 & 0.009 & 4324$_{-204}^{+209}$ & 55.3$\pm$13.1~[44] & Y\\[3pt]
4U 1036-56 & B0III-Ve~[10] & Be & 159.39717 & -56.79886 & 60.0 &  &    & Y\\[3pt]
HD 96670 & O7V(f)n~[45] & Be & 166.808 & -59.8731 & 0.023 & 3131$_{-253}^{+268}$ & -27.5$\pm$0.02~[46] & \\[3pt]
1A 1118-615 & O9.5III-Ve~[47] & Be & 170.23817 & -61.91671 & 0.009 & 2899$_{-76}^{+86}$ &    & Y\\[3pt]
Cen X-3 & O6-7II-III~[48] & sg & 170.31286 & -60.62378 & 0.012 & 6784$_{-572}^{+634}$ & 32.0$\pm$13.0~[49] & Y\\[3pt]
IGR J11215-5952 & B1Ia~[50] & SFXT & 170.44505 & -59.86331 & 0.01 & 7175$_{-563}^{+516}$ &    & Y\\[3pt]
IGR J11305-6256 & B0IIIe~[51] & Be & 172.77874 & -62.94692 & 0.036 & 1731$_{-111}^{+139}$ & -22.0$\pm$7.4~[52] & Y\\[3pt]
IGR J11435-6109 & B0.5Ve~[42] & Be & 176.00117 & -61.1268 & 0.024 & 7863$_{-1217}^{+1373}$ &    & Y\\[3pt]
1E 1145.1-6141 & B2Iae~[53] & sg & 176.86894 & -61.95372 & 0.008 & 8097$_{-566}^{+603}$ & -13.0$\pm$3.0~[54] & Y\\[3pt]
2E 1145.5-6155 & B1III-Ve~[55] & Be & 177.00003 & -62.20691 & 0.014 & 2063$_{-87}^{+80}$ & -17.0$\pm$7.4~[52] & Y\\[3pt]
EXMS B1210-645 & B2V~[56] & Be & 183.31157 & -64.87513 & 0.012 & 3352$_{-161}^{+180}$ &    & \\[3pt]
GX 301-2 & B1.5Iaeq~[57] & sg & 186.65645 & -62.77036 & 0.012 & 3604$_{-195}^{+204}$ & 4.1$\pm$2.4~[58] & \\[3pt]
IGR J12341-6143 & ~[59] & SFXT? & 188.467 & -61.796 & 238800.001 &  &    & Y\\[3pt]
1H 1238-599 &    &  & 190.42741 & -60.27217 & 4300.0 &  &    & \\[3pt]
1H 1249-637 & B0.5IVpe~[60] & Be & 190.70932 & -63.05864 & 0.053 & 439$_{-15}^{+15}$ & 22.0$\pm$7.0~[52] & Y\\[3pt]
1A 1244-604 &    &  & 190.82482 & -60.20152 & 70.0 &  &    & \\[3pt]
GX 304-1 & B2Vne~[61] & Be & 195.3212 & -61.60184 & 0.009 & 1856$_{-45}^{+36}$ &    & Y\\[3pt]
IGR J13020-6359 & B0-6Ve~[62] & Be & 195.49461 & -63.96912 & 0.06 & 5952$_{-1339}^{+2377}$ &    & Y\\[3pt]
PSR B1259-63 & O9.5Ve~[63] & $\gamma$\,Be & 195.69849 & -63.83573 & 0.009 & 2170$_{-58}^{+62}$ & 0.0$\pm$1.0~[64] & \\[3pt]
IGR J13186-6257 & B0-6Ve~[62] & Be & 199.60662 & -62.97081 & 0.056 & 3469$_{-901}^{+1102}$ &    & Y\\[3pt]
SAX J1324.4-6200 & Be?~[65] & Be? & 201.11105 & -62.02198 & 110.0 &  &    & Y\\[3pt]
HD 119682 & B0Ve~[60] & Be & 206.63567 & -62.92339 & 0.017 & 1597$_{-65}^{+80}$ &    & Y\\[3pt]
IGR J14059-6116 & O6III~[66] & $\gamma$ & 211.31006 & -61.30784 & 80.0 &  &    & Y\\[3pt]
MAXI J1409-619 &    &  & 212.01132 & -61.98389 & 70.0 &  &    & \\[3pt]
H 1417-624 & B1e~[67] & Be & 215.30061 & -62.69892 & 0.022 & 7428$_{-1811}^{+3095}$ &    & Y\\[3pt]
IGR J14331-6112 & BV-IIIe~[68] & Be & 218.28465 & -61.26107 & 0.064 & 2551$_{-565}^{+609}$ &    & \\[3pt]
IGR J14488-5942 & O-BVe~[42] & Be & 222.18009 & -59.70382 & 70.0 &  &    & Y\\[3pt]
Cir X-1 & B5-A0I~[69] & sg & 230.17018 & -57.16674 & 0.1 & 7469$_{-2103}^{+2320}$ & 26.0$\pm$3.0~[69] & Y\\[3pt]
4U 1538-522 & B0.2Ia~[70] & sg & 235.5973 & -52.38601 & 0.011 & 5614$_{-434}^{+489}$ & -158.0$\pm$11.0~[71] & Y\\[3pt]
XTE J1543-568 & Be?~[72] & Be? & 236.02206 & -56.76167 & 0.085 & 5388$_{-1925}^{+2675}$ &    & Y\\[3pt]
1H 1555-552 & B2IIIn~[73] &  & 238.59067 & -55.32899 & 0.013 & 1327$_{-35}^{+27}$ &    & \\[3pt]
H 1553-542 & B1-2V~[74] & Be & 239.45139 & -54.4151 & 2.229 &  &    & Y\\[3pt]
IGR J16195-4945 & ON9.7Iab~[42] & SFXT? & 244.8841 & -49.74183 & 0.042 & 2642$_{-258}^{+275}$ &    & Y\\[3pt]
IGR J16207-5129 & B1Ia~[75] & sg & 245.19273 & -51.50171 & 0.038 & 8582$_{-1718}^{+2016}$ &    & Y\\[3pt]
SWIFT J1626.6-5156 & B0Ve~[76] & Be & 246.65219 & -51.94182 & 60.0 &  &    & Y\\[3pt]
IGR J16283-4838 & OBI~[77] & sg & 247.04513 & -48.64891 & 110.0 &  &    & Y\\[3pt]
IGR J16318-4848 & sgB[e]~[78] & sgB[e] & 247.95126 & -48.81688 & 0.131 & 6663$_{-2034}^{+4052}$ &    & Y\\[3pt]
AX J1631.9-4752 & O8I~[79] & sg & 248.00733 & -47.87472 & 80.0 &  &    & Y\\[3pt]
IGR J16328-4726 & O8Iafpe~[42] & SFXT & 248.15829 & -47.39455 & 0.517 & 2131$_{-1012}^{+1769}$ &    & Y\\[3pt]
IGR J16327-4940 & OBIII~[35] &  & 248.16638 & -49.70383 & 0.041 & 13138$_{-3256}^{+4426}$ &    & Y\\[3pt]
IGR J16374-5043 & ~[59] & SFXT? & 249.30624 & -50.7246 & 0.493 & 4224$_{-1799}^{+3433}$ &    & Y\\[3pt]
AX J163904-4642 & BIV-V~[80] & Be & 249.77285 & -46.70354 & 527.513 &  &    & Y\\[3pt]
IGR J16418-4532 & BN0.5Ia~[42] & SFXT? & 250.46162 & -45.54038 & 60.0 &  &    & Y\\[3pt]
IGR J16465-4507 & O9.5Ia~[75] & SFXT & 251.6469 & -45.11796 & 0.016 & 2912$_{-148}^{+224}$ &    & Y\\[3pt]
IGR J16479-4514 & O9.5Iab~[75] & SFXT & 252.02734 & -45.20189 & 60.0 &  &    & Y\\[3pt]
IGR J16493-4348 & B0.5Ib~[81] & sg & 252.36231 & -43.81918 & 60.0 &  &    & Y\\[3pt]
AX J1700-419 &    &  & 255.01813 & -41.96822 & 0.296 & 3282$_{-1153}^{+1564}$ &    & Y\\[3pt]
AX J1700.2-4220 & B0.5IVe~[82] & Be & 255.08037 & -42.33864 & 727.403 &  &    & \\[3pt]
OAO 1657-415 & Ofpe/WN9~[83] & WR & 255.20368 & -41.65596 & 60.0 &  & -57.2$\pm$3.0~[84] & Y\\[3pt]
4U 1700-377 & O6Iafcp~[37] & sg & 255.98657 & -37.84412 & 0.021 & 1499$_{-57}^{+50}$ & -60.0$\pm$10.0~[85] & Y\\[3pt]
AX J1714.1-3912 & ~[86] & sgB[e]/SFXT? & 258.43297 & -39.20157 & 1.785 &  &    & Y\\[3pt]
XTE J1716-389 & sg?~[87] & sg? & 258.98525 & -38.86493 & 80.0 &  &    & Y\\[3pt]
EXO 1722-363 & B0-1Ia~[83] & sg & 261.29746 & -36.28265 & 60.0 &  & -6.5$\pm$3.8~[88] & Y\\[3pt]
IGR J17354-3255 & O9Iab~[42] & SFXT & 263.865 & -32.93182 & 0.329 & 3826$_{-1769}^{+1523}$ &    & Y\\[3pt]
IGR J17375-3022 & ~[59] & SFXT? & 264.39208 & -30.38806 & 900.0 &  &    & Y\\[3pt]
XTE J1739-302 & O8Iab(f)~[89] & SFXT & 264.79813 & -30.34381 & 0.04 & 1929$_{-165}^{+201}$ &    & Y\\[3pt]
GRS 1736-297 &    &  & 264.88797 & -29.72458 & 0.591 & 2271$_{-995}^{+1723}$ &    & Y\\[3pt]
XTE J1743-363 & O9I/GIII-I~[87] & sg & 265.75557 & -36.37283 & 60.0 &  &    & Y\\[3pt]
1E 1740.7-2942 &    &  & 265.97848 & -29.74528 & 218.139 &  &    & Y\\[3pt]
RX J1744.7-2713 & B0.5V-IIIe~[90] & Be & 266.19069 & -27.22903 & 0.02 & 1207$_{-29}^{+33}$ &    & Y\\[3pt]
AX J1749.1-2733 & B1-3V~[91] & Be & 267.2787 & -27.54251 & 717.583 &  &    & Y\\[3pt]
AX J1749.2-2725 & B3V~[91] & Be & 267.30167 & -27.42729 & 728.882 &  &    & Y\\[3pt]
GRO J1750-27 & Be?~[92] & Be? & 267.304 & -26.64406 & 900.0 &  &    & Y\\[3pt]
IGR J17503-2636 & sg?~[93] & SFXT? & 267.57526 & -26.60467 & 60.0 &  &    & \\[3pt]
IGR J17544-2619 & O9Ib~[94] & SFXT & 268.6053 & -26.33127 & 0.022 & 2425$_{-134}^{+156}$ & -46.8$\pm$4.0~[95] & Y\\[3pt]
GRS 1758-258 & AV~[96] & IMXB? & 270.30197 & -25.74328 & 0.919 & 4630$_{-1998}^{+2554}$ &    & Y\\[3pt]
IGR J18027-2016 & B1Ib~[97] & sg & 270.67474 & -20.28817 & 0.105 & 8526$_{-2438}^{+4119}$ & 51.7$\pm$2.4~[98] & Y\\[3pt]
2MASS J18081689-1919395 &    &  & 272.07039 & -19.32766 & 110.0 &  &    & \\[3pt]
SWIFT J1816.7-1613 & B0-2e~[99] & Be & 274.178 & -16.22283 & 765.639 &  &    & Y\\[3pt]
SAX J1818.6-1703 & B0Iab~[97] & SFXT & 274.6579 & -17.04668 & 0.151 & 2791$_{-762}^{+1145}$ &    & Y\\[3pt]
SAX J1819.3-2525 & B9III~[100] &  & 274.84014 & -25.40718 & 0.02 & 4738$_{-601}^{+765}$ & 72.7$\pm$3.3~[101] & Y\\[3pt]
AX J1820.5-1434 & B0-2IV-Ve~[102] & Be & 275.1254 & -14.57302 & 80.0 &  &    & Y\\[3pt]
IGR J18214-1318 & B0V-O9I~[103] & Be/sg & 275.33234 & -13.31089 & 0.23 & 4209$_{-1286}^{+1709}$ &    & Y\\[3pt]
IGR J18219-1347 & Be?~[104] & Be? & 275.47837 & -13.79074 & 745.065 &  &    & Y\\[3pt]
IGR J18246-1425 &    &  & 276.09842 & -14.41551 & 3600.0 &  &    & \\[3pt]
IGR J18256-1035 &    &  & 276.43262 & -10.58389 & 1.012 & 3775$_{-2274}^{+2759}$ &    & \\[3pt]
LS 5039 & ON6.5V(f)~[105] & $\gamma$\,Be & 276.56277 & -14.84844 & 0.013 & 1898$_{-54}^{+59}$ & 17.3$\pm$0.5~[106] & Y\\[3pt]
XTE J1829-098 & ~[59] & SFXT? & 277.43329 & -9.85645 & 738.219 &  &    & Y\\[3pt]
ATO J278.3657-10.5901 & B0.5Ve~[107] & Be & 278.3657 & -10.59012 & 0.154 & 1021$_{-144}^{+255}$ &    & Y\\[3pt]
SNR 021.5-00.9 & B0Ve~[108] & Be & 278.36792 & -10.40243 & 0.024 & 1996$_{-115}^{+112}$ &    & Y\\[3pt]
AX J1838.0-0655 &    &  & 279.51928 & -6.90243 & 8.636 &  &    & \\[3pt]
AX J1841.0-0536 & B1Ib~[75] & SFXT & 280.28017 & -5.58165 & 0.016 & 2828$_{-129}^{+143}$ & 74.4$\pm$2.1~[109] & Y\\[3pt]
GS 1839-06 &    &  & 280.39706 & -5.84236 & 24000.0 &  &    & \\[3pt]
IGR J18450-0435 & O9.5I~[110] & SFXT & 281.25662 & -4.56576 & 0.019 & 5436$_{-641}^{+645}$ & 61.0$\pm$1.4~[109] & Y\\[3pt]
GS 1843+009 & B0-2IV-Ve~[102] & Be & 281.40347 & 0.86316 & 0.168 & 5097$_{-1904}^{+4192}$ &    & Y\\[3pt]
IGR J18462-0223 & OBI?~[111] & SFXT & 281.55377 & -2.37469 & 0.236 & 6784$_{-2839}^{+4048}$ &    & Y\\[3pt]
IGR J18482+0049 &    &  & 282.06417 & 0.79257 & 120.0 &  &    & \\[3pt]
3A 1845-024 & OBI?~[112] & sg? & 282.07035 & -2.42361 & 425.859 &  &    & Y\\[3pt]
IGR J18483-0311 & B0.5Ia~[113] & SFXT & 282.07169 & -3.17137 & 0.137 & 2722$_{-761}^{+1473}$ &    & Y\\[3pt]
XTE J1855-026 & B0Iaep~[114] & sg & 283.87668 & -2.60468 & 0.014 & 7731$_{-735}^{+1090}$ &    & \\[3pt]
XTE J1858+034 & K-M?~[115] &  & 284.68191 & 3.43475 & 80.0 &  &    & Y\\[3pt]
XTE J1859+083 & B0-2Ve~[116] & Be & 284.75681 & 8.24567 & 60.0 &  &    & \\[3pt]
4U 1901+03 & B8-9IV~[117] & Be & 285.91413 & 3.20436 & 0.194 & 5575$_{-1787}^{+2302}$ &    & Y\\[3pt]
XTE J1906+090 & Be?~[118] & Be? & 286.19777 & 9.04491 & 0.559 & 6370$_{-1965}^{+3471}$ &    & Y\\[3pt]
4U 1907+097 & O9.5Iab~[75] & sg & 287.40853 & 9.82978 & 0.026 & 5351$_{-858}^{+866}$ &    & Y\\[3pt]
AX J1910.7+0917 & BI~[119] & sg & 287.68168 & 9.27476 & 70.0 &  &    & Y\\[3pt]
4U 1909+07 & O7.5-9.5If~[120] & sg & 287.70087 & 7.59766 & 0.663 & 4965$_{-2092}^{+3850}$ &    & \\[3pt]
IGR J19113+1533 & sgB[e]~[35] & sgB[e] & 287.82131 & 15.55377 & 0.573 & 3582$_{-1600}^{+2138}$ &    & Y\\[3pt]
SS 433 & A7Ib~[121] & sg & 287.95651 & 4.98271 & 0.018 & 7290$_{-857}^{+1186}$ & 69.0$\pm$4.7~[122] & Y\\[3pt]
IGR J19140+0951 & B1Iab~[75] & sg & 288.51761 & 9.88286 & 0.304 & 4020$_{-1681}^{+1826}$ &    & Y\\[3pt]
IGR J19149+1036 &    &  & 288.73635 & 10.61015 & 1.514 & 1101$_{-242}^{+270}$ &    & \\[3pt]
IGR J19294+1816 & B1Ve~[123] & Be & 292.48293 & 18.31061 & 0.649 & 3189$_{-1241}^{+1980}$ &    & Y\\[3pt]
1RXS J194211.9+255552 & OB~[124] &  & 295.54652 & 25.93489 & 0.032 & 9057$_{-2317}^{+3895}$ &    & \\[3pt]
XTE J1946+274 & B0-1IV-Ve~[125] & Be & 296.41397 & 27.3654 & 0.017 & 13139$_{-2332}^{+3380}$ &    & Y\\[3pt]
KS 1947+300 & B0Ve~[126] & Be & 297.39784 & 30.20881 & 0.009 & 15126$_{-2634}^{+3148}$ &    & Y\\[3pt]
IGR J19498+2534 & BIa~[127] & SFXT & 297.48095 & 25.56659 & 0.013 & 5991$_{-874}^{+1125}$ &    & Y\\[3pt]
4U 1954+319 & MI~[128] & sg & 298.9264 & 32.09693 & 0.016 & 3432$_{-249}^{+261}$ &    & Y\\[3pt]
Cyg X-1 & O9.7Iabpvar~[129] & sg & 299.59029 & 35.20158 & 0.011 & 2146$_{-53}^{+64}$ & -7.0$\pm$5.0~[130] & \\[3pt]
IGR J20006+3210 & BV-III~[68] & Be & 300.09105 & 32.18974 & 0.014 & 8356$_{-1199}^{+1766}$ &    & \\[3pt]
W63 X-1 & Be?~[131] & Be? & 304.79994 & 45.66718 & 0.196 & 767$_{-107}^{+157}$ &    & \\[3pt]
RX J2030.5+4751 & B0.5V-IIIe~[10] & Be & 307.6285 & 47.86407 & 0.014 & 2293$_{-85}^{+95}$ &    & \\[3pt]
EXO 2030+375 & B0Ve~[132] & Be & 308.06363 & 37.63743 & 0.061 & 2410$_{-391}^{+497}$ &    & Y\\[3pt]
Cyg X-3 & WN4/5-6/7~[133] & WR & 308.10742 & 40.95776 & 60.0 &  &    & Y\\[3pt]
GRO J2058+42 & O9.5-B0IV-Ve~[9] & Be & 314.69806 & 41.77698 & 0.011 & 8861$_{-927}^{+1065}$ &    & \\[3pt]
SAX J2103.5+4545 & B0Ve~[134] & Be & 315.89877 & 45.75153 & 0.011 & 6218$_{-448}^{+576}$ &    & Y\\[3pt]
IGR J21347+4737 & B3V~[56] & Be & 323.58487 & 47.63338 & 0.011 & 8300$_{-805}^{+851}$ &    & Y\\[3pt]
Cep X-4 & B1-B2Ve~[135] & Be & 324.87785 & 56.98622 & 0.012 & 7446$_{-492}^{+576}$ &    & \\[3pt]
1H 2202+501 & Be~[136] & Be & 330.40919 & 50.16795 & 0.011 & 1116$_{-16}^{+14}$ & -16.8$\pm$2.5~[27] & \\[3pt]
4U 2206+543 & O9.5Vep~[137] & Be & 331.98429 & 54.51843 & 0.012 & 3104$_{-135}^{+133}$ & -54.5$\pm$1.0~[138] & Y\\[3pt]
SAX J2239.3+6116 & B0Ve~[139] & Be & 339.83683 & 61.27405 & 0.012 & 7387$_{-675}^{+855}$ &    & \\[3pt]
MWC 656 & B1.5-B2IIIe~[140] & Be & 340.73874 & 44.72172 & 0.013 & 1984$_{-76}^{+80}$ & -14.1$\pm$2.1~[140] & \\[3pt]
2MASS J22535512+6243368 & BIII~[141] &  & 343.47967 & 62.72688 & 0.015 & 9748$_{-1166}^{+1676}$ &    & Y\\[3pt]
\hline
\end{longtable}
\tablebib{[1]\cite{2014AA...566A.131G}; [2]\cite{2014AA...563A...1G}; [3]\cite{2011ARep...55...31S}; [4]\cite{2012AA...537A..59N}; [5]\cite{1960IzKry..24..160B}; [6]\cite{2015AA...579A.111K}; [7]\cite{2003RMxAA..39...17K}; [8]\cite{2001AA...369..108N}; [9]\cite{2005AA...440..637R}; [10]\cite{1997AA...323..853M}; [11]\cite{2008MNRAS.386.2253K}; [12]\cite{1981PASP...93..741H}; [13]\cite{2009ApJ...698..514A}; [14]\cite{2020AA...640A..35R}; [15]\cite{1999MNRAS.307..695N}; [16]\cite{2019AA...622A.173Z}; [17]\cite{2007ApJ...660.1398G}; [18]\cite{2002AA...392..991H}; [19]\cite{2016MNRAS.456.1424A}; [20]\cite{2005AA...440.1079R}; [21]\cite{1990AA...231..354P}; [22]\cite{2006AstL...32..759G}; [23]\cite{1998PASP..110.1310W}; [24]\cite{1984PASP...96..312H}; [25]\cite{1950ApJ...111..495P}; [26]\cite{2010AA...522A.107R}; [27]\cite{2017AJ....153..174C}; [28]\cite{1982IAUS...98..261J}; [29]\cite{2018PASJ...70...61M}; [30]\cite{1999ApJ...523..197K}; [31]\cite{2003ATel..202....1P}; [32]\cite{2016ApJS..224....4M}; [33]\cite{2003AJ....125.2531R}; [34]\cite{2001AA...367..266R}; [35]\cite{2010AA...519A..96M}; [36]\cite{2000MNRAS.314...87I}; [37]\cite{2014ApJS..211...10S}; [38]\cite{2015AA...583L...4G}; [39]\cite{1978mcts.book.....H}; [40]\cite{1997MNRAS.286L..21S}; [41]\cite{1994MNRAS.270L..57C}; [42]\cite{2013AA...560A.108C}; [43]\cite{2015ApJ...813L..26S}; [44]\cite{2022MNRAS.515.1078V}; [45]\cite{1993ApJS...87..197G}; [46]\cite{2021ApJ...913...48G}; [47]\cite{1981AA....99..274J}; [48]\cite{1999MNRAS.307..357A}; [49]\cite{2007AA...473..523V}; [50]\cite{1993ApJS...89..293V}; [51]\cite{1977ApJS...35..111G}; [52]\cite{2007AN....328..889K}; [53]\cite{1982MNRAS.201..171D}; [54]\cite{1987PASP...99..420H}; [55]\cite{1982IAUS...98..151P}; [56]\cite{2009AA...495..121M}; [57]\cite{1982PASP...94..541H}; [58]\cite{2006AA...457..595K}; [59]\cite{2020MNRAS.491.4543S}; [60]\cite{2006MNRAS.371..252L}; [61]\cite{1980MNRAS.190..537P}; [62]\cite{2018AA...618A.150F}; [63]\cite{2011ApJ...732L..11N}; [64]\cite{1994MNRAS.268..430J}; [65]\cite{2008AA...483..249M}; [66]\cite{2019ApJ...884...93C}; [67]\cite{1984ApJ...276..621G}; [68]\cite{2008AA...482..113M}; [69]\cite{2007MNRAS.374..999J}; [70]\cite{1978MNRAS.184P..73P}; [71]\cite{2004ARep...48...89A}; [72]\cite{2001ApJ...553L.165I}; [73]\cite{2003AJ....126.2971V}; [74]\cite{2016MNRAS.462.3823L}; [75]\cite{2008AA...486..911N}; [76]\cite{2011AA...533A..23R}; [77]\cite{2011AA...526A..15P}; [78]\cite{2004ApJ...616..469F}; [79]\cite{2008AA...484..801R}; [80]\cite{2008AA...484..783C}; [81]\cite{2008ATel.1396....1N}; [82]\cite{2007AA...461..631N}; [83]\cite{2009AA...505..281M}; [84]\cite{2012MNRAS.422..199M}; [85]\cite{1986ApJS...61..419G}; [86]\cite{2022MNRAS.512.2929S}; [87]\cite{2010MNRAS.408.1866R}; [88]\cite{2010AA...509A..79M}; [89]\cite{2006ApJ...638..982N}; [90]\cite{2006AA...454..265L}; [91]\cite{2010MNRAS.409L..69K}; [92]\cite{1997ApJS..113..367B}; [93]\cite{2018ATel11992....1M}; [94]\cite{2006AA...455..653P}; [95]\cite{2013BCrAO.109...27N}; [96]\cite{2016AA...596A..46M}; [97]\cite{2010AA...510A..61T}; [98]\cite{2011AA...532A.124M}; [99]\cite{2019AA...622A.198N}; [100]\cite{2001ApJ...555..489O}; [101]\cite{2005MNRAS.363..882L}; [102]\cite{2001AA...371.1018I}; [103]\cite{2009ApJ...698..502B}; [104]\cite{2012AstL...38..629K}; [105]\cite{2011MNRAS.416.1556T}; [106]\cite{2011ASSP...21..559C}; [107]\cite{2003AN....324...61M}; [108]\cite{2011BASI...39..517M}; [109]\cite{2015arXiv150301087G}; [110]\cite{1996MNRAS.281..333C}; [111]\cite{2013AA...556A..27S}; [112]\cite{2022AA...657A..58N}; [113]\cite{2008AA...492..163R}; [114]\cite{2008ATel.1876....1N}; [115]\cite{2021ApJ...909..154T}; [116]\cite{2022MNRAS.509.5955S}; [117]\cite{2019ATel12560....1M}; [118]\cite{2005ApJ...632.1069G}; [119]\cite{2013AA...555A.115R}; [120]\cite{2005MNRAS.356..665M}; [121]\cite{2004ApJ...615..422H}; [122]\cite{2020AA...640A..96P}; [123]\cite{2018MNRAS.476.2110R}; [124]\cite{2012ATel.4209....1M}; [125]\cite{2002AA...393..983V}; [126]\cite{2003AA...397..739N}; [127]\cite{2019ApJ...878...15H}; [128]\cite{2020ApJ...904..143H}; [129]\cite{2011ApJS..193...24S}; [130]\cite{2003ApJ...583..424G}; [131]\cite{2004HEAD....8.1730R}; [132]\cite{1987AA...182L..55M}; [133]\cite{1996AA...314..521V}; [134]\cite{2004AA...421..673R}; [135]\cite{1998AA...332L...9B}; [136]\cite{1964LS....C03....0H}; [137]\cite{2006AA...446.1095B}; [138]\cite{2014AN....335.1060S}; [139]\cite{2017AA...598A..16R}; [140]\cite{2014Natur.505..378C}; [141]\cite{2013AA...556A.120M}; }\clearpage
\begin{longtable}{lllllll}\caption{Catalogue of Galactic HMXBs: Orbital data. Mx and Mo refer to the mass of the compact object and the companion star, respectively. Sup. orb. period is the super-orbital period of the system.}\label{tab:cat:orbital}\\
\hline\hline\\[-2ex]
Main ID &Mx [\Msun\,] &Mo [\Msun\,] &Period [d] &Sup. orbital period [d] &Eccentricity &Spin period [s] \\
\hline\\[-2ex]
\endfirsthead
\caption{continued.}\\
\hline\hline\\[-2ex]
Main ID &Mx [\Msun\,] &Mo [\Msun\,] &Orbital period [d] &Sup. orbital period [d] &Eccentricity &Spin period [s] \\
\hline\\[-2ex]
\endhead
\hline
\endfoot
IGR J00370+6122 &    & 22.0~[1] & 15.66$\pm$1e-3~[2] &    & 0.48$\pm$0.03~[1] & 674.0~[2]\\[3pt]
gam Cas &    & 13.0~[3] & 203.37$\pm$0.089~[4] &    & 0.26$\pm$0.035~[3] &   \\[3pt]
EM* AS   14 &    &    &    &    &    &   \\[3pt]
2S 0114+650 &    & 16.0$\pm$2.0~[5] & 11.6$\pm$6e-4~[6] & 30.76$\pm$0.03~[7] & 0.18$\pm$0.05~[8] & 10008.0$\pm$36.0~[9]\\[3pt]
4U 0115+634 &    & 17.5~[10]$\dagger$ & 24.32$\pm$4e-4~[11] &    & 0.34$\pm$5e-3~[11] & 3.6~[12]\\[3pt]
IGR J01363+6610 &    & 12.5~[10]$\dagger$ & 159.0$\pm$2.0~[13] &    &    &   \\[3pt]
RX J0146.9+6121 &    & 9.6~[10]$\dagger$ & 330.0~[14] &    &    & 1407.4$\pm$3.0~[15]\\[3pt]
IGR J01583+6713 &    & 12.5~[10]$\dagger$ &    &    &    & 469.2~[16]\\[3pt]
LS I+61 303 &    & 12.5~[17] & 26.5$\pm$3e-3~[18] & 1628.0$\pm$48.0~[19] & 0.54$\pm$0.034~[20] &   \\[3pt]
Swift J0243.6+6124 &    &    & 28.3$\pm$0.2~[21] &    & 0.09$\pm$7e-3~[21] & 9.8661$\pm$3e-4~[22]\\[3pt]
V 0332+53 &    & 18.8~[23]$\dagger$ & 36.5$\pm$0.29~[11] &    & 0.42$\pm$7e-3~[11] & 4.4~[24]\\[3pt]
X Per &    & 15.5~[25] & 250.3$\pm$0.6~[26] &    & 0.11$\pm$0.018~[26] & 837.6712$\pm$3e-4~[26]\\[3pt]
XTE J0421+560 &    &    & 19.41$\pm$0.02~[27] &    &    &   \\[3pt]
RX J0440.9+4431 &    & 17.5~[10]$\dagger$ & 150.0~[28] &    &    & 202.5$\pm$0.5~[29]\\[3pt]
EXO 051910+3737.7 &    & 17.5~[10]$\dagger$ &    &    &    &   \\[3pt]
1A 0535+262 &    & 20.0~[30] & 110.3$\pm$0.3~[30] &    & 0.47$\pm$0.02~[31] & 103.4$\pm$0.02~[32]\\[3pt]
AAO+28  342 &    &    &    &    &    &   \\[3pt]
IGR J06074+2205 &    & 14.6~[10]$\dagger$ &    &    &    & 373.226$\pm$0.013~[33]\\[3pt]
HD 259440 & 1.4~[34] & 15.7$\pm$2.5~[34] & 317.3$\pm$0.7~[35] &    & 0.62$\pm$0.16~[36] &   \\[3pt]
SAX J0635.2+0533 &    & 9.6~[10]$\dagger$ & 11.2$\pm$0.5~[37] &    & 0.29$\pm$0.09~[37] & 0.034~[38]\\[3pt]
3A 0656-072 &    & 15.6~[23]$\dagger$ & 101.2~[39] &    &    & 160.4$\pm$0.4~[40]\\[3pt]
3A 0726-260 &    & 38.1~[23]$\dagger$ & 34.55$\pm$1e-2~[41] &    &    & 103.144$\pm$1e-3~[42]\\[3pt]
SGR 0755-2933 &    & 25.29~[23]$\dagger$ & 260.0~[43] &    &    & 307.8$\pm$0.04~[43]\\[3pt]
RX J0812.4-3114 &    & 17.5~[10]$\dagger$ & 80.39$\pm$3.0~[44] &    &    & 31.908$\pm$9e-3~[44]\\[3pt]
IGR J08262-3736 &    &    &    &    &    &   \\[3pt]
GS 0834-430 &    & 13.5~[10]$\dagger$ & 105.8$\pm$0.4~[45] &    & 0.14$\pm$0.04~[45] & 12.3203$\pm$2e-3~[46]\\[3pt]
IGR J08408-4503 &    & 33.0~[47] & 9.54$\pm$2e-4~[47] &    & 0.63$\pm$0.03~[47] &   \\[3pt]
Vela X-1 & 2.12$\pm$0.16~[48] & 26.0$\pm$1.0~[48] & 8.96$\pm$4e-4~[49] &    & 0.11$\pm$0.079~[49] & 283.0~[50]\\[3pt]
GRO J1008-57 &    & 17.5~[10]$\dagger$ & 247.8$\pm$0.4~[51] &    & 0.68$\pm$0.02~[51] & 93.587$\pm$5e-3~[52]\\[3pt]
IGR J10101-5654 &    &    &    &    &    &   \\[3pt]
1FGL J1018.6-5856 & 1.4~[53] & 23.0$\pm$3.0~[53] & 16.55$\pm$4e-4~[54] &    & 0.53$\pm$0.033~[54] &   \\[3pt]
4U 1036-56 &    & 17.5~[10]$\dagger$ & 60.9~[14] &    &    & 860.0$\pm$2.0~[29]\\[3pt]
HD 96670 & 6.2$\pm$0.9~[55] & 22.7$\pm$5.2~[55] & 5.28$\pm$5e-4~[55] &    & 0.12$\pm$1e-2~[55] &   \\[3pt]
1A 1118-615 &    & 18.0~[23]$\dagger$ & 24.0~[39] &    &    & 405.3$\pm$0.6~[56]\\[3pt]
Cen X-3 & 1.34$\pm$0.16~[57] & 20.2$\pm$1.8~[57] & 2.03$\pm$0.029~[58] &    & 0.0~[59] & 4.80188$\pm$9e-5~[58]\\[3pt]
IGR J11215-5952 &    &    & 164.6$\pm$0.1~[14] &    & 0.8~[14] & 186.78$\pm$0.03~[60]\\[3pt]
IGR J11305-6256 &    & 17.5~[10]$\dagger$ & 120.83$\pm$0.34~[61] &    &    &   \\[3pt]
IGR J11435-6109 &    & 14.6~[10]$\dagger$ & 52.46$\pm$0.06~[62] &    &    & 161.76$\pm$1e-2~[63]\\[3pt]
1E 1145.1-6141 & 1.7$\pm$0.3~[64] & 14.0$\pm$4.0~[64] & 14.36$\pm$2e-3~[65] &    & 0.2$\pm$0.03~[65] & 298.0$\pm$4.0~[66]\\[3pt]
2E 1145.5-6155 &    & 12.5~[10]$\dagger$ & 187.5~[14] &    & 0.5~[14] & 292.274$\pm$1e-3~[67]\\[3pt]
EXMS B1210-645 &    & 9.6~[10]$\dagger$ & 6.7$\pm$5e-4~[68] &    &    &   \\[3pt]
GX 301-2 &    & 43.0$\pm$10.0~[69] & 41.5$\pm$2e-3~[69] &    & 0.46$\pm$0.014~[69] & 680.0~[70]\\[3pt]
IGR J12341-6143 &    &    &    &    &    &   \\[3pt]
1H 1238-599 &    &    &    &    &    & 191.0~[71]\\[3pt]
1H 1249-637 &    & 9.6~[72] & 226.0$\pm$6.0~[73] &    &    & 14200.0$\pm$1400.0~[74]\\[3pt]
1A 1244-604 &    &    &    &    &    &   \\[3pt]
GX 304-1 &    & 9.6~[10]$\dagger$ & 132.5~[39] &    &    & 272.0~[75]\\[3pt]
IGR J13020-6359 &    & 17.5~[10]$\dagger$ &    &    &    & 700.0~[76]\\[3pt]
PSR B1259-63 &    & 22.5$\pm$7.5~[77] & 1236.72$\pm$6e-6~[77] &    & 0.87$\pm$6e-8~[77] & 0.04776~[78]\\[3pt]
IGR J13186-6257 &    & 17.5~[10]$\dagger$ &    &    &    &   \\[3pt]
SAX J1324.4-6200 &    &    & 1.13~[79] &    &    & 172.84~[80]\\[3pt]
HD 119682 &    & 17.5~[10]$\dagger$ & 90.0~[81] &    &    & 1500.0$\pm$100.0~[82]\\[3pt]
IGR J14059-6116 &    & 34.53~[23]$\dagger$ & 13.71$\pm$2e-3~[83] &    &    &   \\[3pt]
MAXI J1409-619 &    &    & 14.7$\pm$0.4~[84] &    &    & 503.0$\pm$10.0~[85]\\[3pt]
H 1417-624 &    & 12.5~[10]$\dagger$ & 42.12~[86] &    & 0.45~[86] & 17.64~[87]\\[3pt]
IGR J14331-6112 &    &    &    &    &    &   \\[3pt]
IGR J14488-5942 &    &    & 49.63$\pm$0.05~[88] &    &    & 33.419$\pm$1e-3~[88]\\[3pt]
Cir X-1 &    &    & 16.68$\pm$0.15~[89] &    & 0.45$\pm$0.07~[89] &   \\[3pt]
4U 1538-522 & 1.18$\pm$0.29~[90] & 20.0~[90] & 3.73$\pm$2e-5~[91] & 14.91$\pm$3e-3~[92] & 0.18$\pm$1e-2~[48] & 526.42$\pm$0.07~[91]\\[3pt]
XTE J1543-568 &    &    & 75.56$\pm$0.25~[93] &    & 0.03~[93] & 27.12156$\pm$6e-4~[93]\\[3pt]
1H 1555-552 &    & 19.4$\pm$5.0~[94] &    &    &    &   \\[3pt]
H 1553-542 &    & 10.8~[10]$\dagger$ & 30.6$\pm$2.2~[95] &    &    & 9.282155$\pm$3e-6~[96]\\[3pt]
IGR J16195-4945 &    & 27.8~[23]$\dagger$ & 16.0~[39] &    &    &   \\[3pt]
IGR J16207-5129 &    &    & 9.73~[97] &    &    &   \\[3pt]
SWIFT J1626.6-5156 &    & 17.5~[10]$\dagger$ & 132.89$\pm$0.03~[98] &    & 0.08$\pm$1e-2~[98] & 15.346577$\pm$1e-6~[98]\\[3pt]
IGR J16283-4838 &    &    & 287.6$\pm$1.7~[99] &    &    &   \\[3pt]
IGR J16318-4848 &    &    & 80.09$\pm$0.012~[100] &    &    &   \\[3pt]
AX J1631.9-4752 &    & 33.7~[23]$\dagger$ & 8.96$\pm$1e-2~[101] &    & 0.2$\pm$1e-2~[102] & 1309.0$\pm$40.0~[103]\\[3pt]
IGR J16328-4726 &    & 33.7~[23]$\dagger$ & 10.07$\pm$2e-3~[104] &    &    &   \\[3pt]
IGR J16327-4940 &    &    &    &    &    &   \\[3pt]
IGR J16374-5043 &    &    &    &    &    &   \\[3pt]
AX J163904-4642 &    &    & 4.24$\pm$1e-5~[105] & 14.99$\pm$1e-2~[7] &    & 908.79$\pm$1e-2~[105]\\[3pt]
IGR J16418-4532 &    &    & 3.75$\pm$4e-3~[106] & 14.73$\pm$6e-3~[7] &    & 1246.0~[107]\\[3pt]
IGR J16465-4507 &    & 27.8~[23]$\dagger$ & 30.24~[14] &    &    & 228.0~[107]\\[3pt]
IGR J16479-4514 &    & 27.8~[23]$\dagger$ & 3.32$\pm$1e-3~[108] & 11.88$\pm$2e-3~[7] &    &   \\[3pt]
IGR J16493-4348 &    &    & 6.78$\pm$4e-4~[109] & 20.06$\pm$7e-3~[110] & 0.0~[109] & 1093.1036$\pm$4e-4~[109]\\[3pt]
AX J1700-419 &    &    &    &    &    & 714.5$\pm$0.3~[111]\\[3pt]
AX J1700.2-4220 &    & 14.6~[10]$\dagger$ & 44.03$\pm$0.03~[88] &    &    & 54.22$\pm$0.03~[112]\\[3pt]
OAO 1657-415 & 1.42$\pm$0.26~[113] & 14.3$\pm$0.8~[113] & 10.45$\pm$1e-4~[114] &    & 0.11$\pm$1e-3~[115] & 37.03322$\pm$1e-4~[116]\\[3pt]
4U 1700-377 & 1.96$\pm$0.19~[48] & 46.0$\pm$5.0~[48] & 3.41$\pm$4e-6~[117] &    & 0.03$\pm$0.02~[117] &   \\[3pt]
AX J1714.1-3912 &    &    &    &    &    &   \\[3pt]
XTE J1716-389 &    &    & 99.1$\pm$0.4~[118] &    &    &   \\[3pt]
EXO 1722-363 & 1.91$\pm$0.45~[48] & 18.0$\pm$2.0~[48] & 9.74$\pm$4e-4~[119] &    & $<$0.19~[119] & 414.8$\pm$0.5~[120]\\[3pt]
IGR J17354-3255 &    & 29.6~[23]$\dagger$ & 8.45$\pm$2e-3~[121] &    &    &   \\[3pt]
IGR J17375-3022 &    &    &    &    &    &   \\[3pt]
XTE J1739-302 &    & 33.7~[23]$\dagger$ & 51.47~[14] &    &    &   \\[3pt]
GRS 1736-297 &    &    &    &    &    &   \\[3pt]
XTE J1743-363 &    & 29.63~[23]$\dagger$ &    &    &    &   \\[3pt]
1E 1740.7-2942 & 5.0$\pm$1.1~[122] &    & 12.61$\pm$0.06~[123] &    &    &   \\[3pt]
RX J1744.7-2713 &    & 14.6~[10]$\dagger$ &    &    &    & 3245.0$\pm$350.0~[124]\\[3pt]
AX J1749.1-2733 &    & 9.6~[10]$\dagger$ & 185.5$\pm$1.1~[125] &    &    & 66.09$\pm$0.07~[125]\\[3pt]
AX J1749.2-2725 &    & 7.7~[10]$\dagger$ &    &    &    & 220.38$\pm$0.2~[126]\\[3pt]
GRO J1750-27 &    &    & 29.81$\pm$1e-3~[127] &    & 0.36$\pm$2e-3~[127] & 4.45349$\pm$2e-5~[127]\\[3pt]
IGR J17503-2636 &    &    &    &    &    &   \\[3pt]
IGR J17544-2619 & 1.4~[128] & 23.0$\pm$2.0~[128] & 12.17$\pm$7e-3~[129] &    & 0.44$\pm$0.14~[129] & 71.49$\pm$0.02~[130]\\[3pt]
GRS 1758-258 &    &    & 18.45$\pm$0.1~[131] &    &    &   \\[3pt]
IGR J18027-2016 & 1.5$\pm$0.4~[132] & 20.0$\pm$3.0~[132] & 4.57$\pm$9e-4~[132] &    & $<$0.2~[132] & 139.612$\pm$6e-3~[133]\\[3pt]
2MASS J18081689-1919395 &    &    &    &    &    &   \\[3pt]
SWIFT J1816.7-1613 &    & 12.5~[10]$\dagger$ & 151.1$\pm$0.5~[88] &    &    & 143.6863$\pm$2e-4~[134]\\[3pt]
SAX J1818.6-1703 &    &    & 30.0~[14] &    &    &   \\[3pt]
SAX J1819.3-2525 & 10.2$\pm$1.5~[135] & 6.8$\pm$1.4~[135] & 2.82$\pm$2e-3~[136] &    &    &   \\[3pt]
AX J1820.5-1434 &    & 12.5~[10]$\dagger$ & 54.0$\pm$0.4~[137] &    &    & 152.26$\pm$0.04~[138]\\[3pt]
IGR J18214-1318 &    &    & 5.42$\pm$4e-4~[139] &    & 0.17~[139] &   \\[3pt]
IGR J18219-1347 &    &    & 72.44$\pm$0.3~[140] &    &    & 56.468$\pm$3e-4~[141]\\[3pt]
IGR J18246-1425 &    &    &    &    &    & 120.0~[142]\\[3pt]
IGR J18256-1035 &    &    &    &    &    &   \\[3pt]
LS 5039 &    & 23.0~[143] & 3.91$\pm$8e-5~[144] &    & 0.35$\pm$0.03~[144] &   \\[3pt]
XTE J1829-098 &    &    & 244.2$\pm$0.2~[145] &    &    & 7.847089$\pm$2e-5~[146]\\[3pt]
ATO J278.3657-10.5901 &    & 14.6~[10]$\dagger$ &    &    &    &   \\[3pt]
SNR 021.5-00.9 &    & 17.5~[10]$\dagger$ &    &    &    &   \\[3pt]
AX J1838.0-0655 &    &    &    &    &    & 0.07049821$\pm$3e-5~[147]\\[3pt]
AX J1841.0-0536 &    &    & 6.45$\pm$2e-3~[148] &    & 0.16$\pm$0.11~[148] & 4.7394$\pm$8e-4~[149]\\[3pt]
GS 1839-06 &    &    &    &    &    &   \\[3pt]
IGR J18450-0435 &    & 29.6~[23]$\dagger$ & 4.74$\pm$3e-4~[148] &    & 0.34$\pm$0.11~[148] &   \\[3pt]
GS 1843+009 &    & 13.5~[10]$\dagger$ &    &    &    & 29.4764$\pm$8e-4~[150]\\[3pt]
IGR J18462-0223 &    &    & 2.14~[151] &    &    & 997.0$\pm$1.0~[152]\\[3pt]
IGR J18482+0049 &    &    &    &    &    &   \\[3pt]
3A 1845-024 &    &    & 242.18$\pm$1e-2~[153] &    & 0.88$\pm$1e-2~[153] & 94.7171$\pm$3e-4~[154]\\[3pt]
IGR J18483-0311 &    &    & 18.55$\pm$3e-3~[155] &    & 0.4~[156] & 21.0526$\pm$5e-4~[157]\\[3pt]
XTE J1855-026 &    &    & 6.07$\pm$4e-3~[158] &    &    & 361.1$\pm$0.4~[158]\\[3pt]
XTE J1858+034 &    &    & 81.0~[159] &    &    & 218.393$\pm$2e-3~[159]\\[3pt]
XTE J1859+083 &    & 12.5~[10]$\dagger$ & 37.97$\pm$0.09~[160] &    & 0.13$\pm$9e-3~[160] & 9.79156$\pm$1e-5~[161]\\[3pt]
4U 1901+03 &    &    & 22.58$\pm$2e-4~[162] &    & 0.04$\pm$3e-4~[162] & 2.761$\pm$1e-3~[163]\\[3pt]
XTE J1906+090 &    &    & 81.4$\pm$0.1~[88] &    &    & 89.17$\pm$0.02~[164]\\[3pt]
4U 1907+097 &    & 27.8~[23]$\dagger$ & 8.38$\pm$3e-4~[165] &    & 0.28$\pm$0.1~[165] & 437.5~[166]\\[3pt]
AX J1910.7+0917 &    &    &    &    &    & 36200.0$\pm$110.0~[167]\\[3pt]
4U 1909+07 &    & 32.0~[23]$\dagger$ & 4.4$\pm$9e-4~[168] & 15.18$\pm$3e-3~[7] & 0.02$\pm$0.039~[168] & 603.6$\pm$0.1~[169]\\[3pt]
IGR J19113+1533 &    &    &    &    &    &   \\[3pt]
SS 433 & 4.2$\pm$0.4~[170] & 11.3$\pm$0.6~[170] & 13.08~[171] & 23.23$\pm$5e-3~[172] & 0.05$\pm$1e-2~[173] &   \\[3pt]
IGR J19140+0951 &    &    & 13.56$\pm$4e-3~[174] &    &    & 5937.0~[175]\\[3pt]
IGR J19149+1036 &    &    & 22.25$\pm$0.05~[176] &    &    &   \\[3pt]
IGR J19294+1816 &    & 12.5~[10]$\dagger$ & 117.2$\pm$0.2~[177] &    &    & 12.44~[178]\\[3pt]
1RXS J194211.9+255552 &    &    & 166.5$\pm$0.5~[179] &    &    &   \\[3pt]
XTE J1946+274 &    & 15.0~[10]$\dagger$ & 172.7$\pm$0.6~[180] &    & 0.25$\pm$9e-3~[180] & 15.78801$\pm$4e-5~[181]\\[3pt]
KS 1947+300 &    & 17.5~[10]$\dagger$ & 40.42$\pm$7e-3~[182] &    & 0.03$\pm$7e-3~[182] & 18.7~[183]\\[3pt]
IGR J19498+2534 &    &    &    &    &    &   \\[3pt]
4U 1954+319 &    & 9.0$\pm$4.0~[184] & 1296.64~[184] &    &    & 18612.0~[185]\\[3pt]
Cyg X-1 & 21.2$\pm$2.2~[186] & 40.6$\pm$7.7~[186] & 5.6$\pm$1e-4~[187] &    & 0.0~[187] &   \\[3pt]
IGR J20006+3210 &    &    &    &    &    & 889.7$\pm$4.7~[188]\\[3pt]
W63 X-1 &    &    &    &    &    & 36.0~[189]\\[3pt]
RX J2030.5+4751 &    & 14.6~[10]$\dagger$ & 46.02~[14] &    & 0.41~[14] &   \\[3pt]
EXO 2030+375 &    & 17.5~[10]$\dagger$ & 46.02$\pm$5e-4~[190] &    & 0.42$\pm$2e-3~[190] & 41.306$\pm$3e-3~[191]\\[3pt]
Cyg X-3 & 7.2~[192] &    & 0.2$\pm$3e-8~[193] &    &    &   \\[3pt]
GRO J2058+42 &    & 18.0~[72] & 55.0~[194] &    &    & 195.25$\pm$0.02~[195]\\[3pt]
SAX J2103.5+4545 &    & 17.5~[10]$\dagger$ & 12.67$\pm$9e-4~[196] &    & 0.41$\pm$3e-3~[196] & 358.61$\pm$0.03~[197]\\[3pt]
IGR J21347+4737 &    & 12.5~[10]$\dagger$ &    &    &    & 322.7$\pm$0.6~[198]\\[3pt]
Cep X-4 &    & 10.8~[10]$\dagger$ & 20.85$\pm$0.05~[199] &    &    & 65.3508$\pm$1e-4~[200]\\[3pt]
1H 2202+501 &    &    &    &    &    &   \\[3pt]
4U 2206+543 &    & 18.0~[72] & 9.56$\pm$0.04~[201] &    & 0.3$\pm$0.02~[201] & 392.0~[202]\\[3pt]
SAX J2239.3+6116 &    & 17.5~[10]$\dagger$ & 262.0$\pm$5.0~[203] &    &    & 1247.2$\pm$0.7~[204]\\[3pt]
MWC 656 &    & 7.8$\pm$2.0~[205] & 60.37$\pm$0.04~[205] &    & 0.1~[206] &   \\[3pt]
2MASS J22535512+6243368 &    &    &    &    &    & 46.753$\pm$3e-3~[207]\\[3pt]
\hline
\end{longtable}
\tablebib{[1]\cite{2014AA...563A...1G}; [2]\cite{2021PASJ...73.1389U}; [3]\cite{2000AA...364L..85H}; [4]\cite{2012AA...537A..59N}; [5]\cite{2017ApJ...844...16H}; [6]\cite{1985ApJ...299..839C}; [7]\cite{2013ApJ...778...45C}; [8]\cite{2006AA...458..513K}; [9]\cite{1992AA...262L..25F}; [10]\cite{1996MNRAS.280L..31P}; [11]\cite{2010MNRAS.406.2663R}; [12]\cite{1978IAUC.3163....1C}; [13]\cite{2010ATel.3079....1C}; [14]\cite{2018MNRAS.481.2779S}; [15]\cite{1998AA...330..189H}; [16]\cite{2008MNRAS.386.2253K}; [17]\cite{2005MNRAS.360.1105C}; [18]\cite{2002ApJ...575..427G}; [19]\cite{2016AA...585A.123M}; [20]\cite{2009ApJ...698..514A}; [21]\cite{2018AA...613A..19D}; [22]\cite{2017ATel10812....1J}; [23]\cite{2005AA...436.1049M}; [24]\cite{1985ApJ...288L..45S}; [25]\cite{1997MNRAS.286..549L}; [26]\cite{2001ApJ...546..455D}; [27]\cite{2006ASPC..355..305B}; [28]\cite{2013AA...553A.103F}; [29]\cite{1999MNRAS.306..100R}; [30]\cite{2001AA...377..161O}; [31]\cite{1996ApJ...459..288F}; [32]\cite{2005ATel..557....1S}; [33]\cite{2018AA...613A..52R}; [34]\cite{2010ApJ...724..306A}; [35]\cite{2021ApJ...923..241A}; [36]\cite{2018PASJ...70...61M}; [37]\cite{2000ApJ...542L..41K}; [38]\cite{2000ApJ...528L..25C}; [39]\cite{2015AARv..23....2W}; [40]\cite{2006AA...451..267M}; [41]\cite{2016ATel.9823....1C}; [42]\cite{2020RAA....20..155R}; [43]\cite{2021AA...647A.165D}; [44]\cite{2019MNRAS.488.4427Z}; [45]\cite{1997ApJ...479..388W}; [46]\cite{1993AA...271..487B}; [47]\cite{2015AA...583L...4G}; [48]\cite{2015AA...577A.130F}; [49]\cite{1997MNRAS.286L..21S}; [50]\cite{1976ApJ...206L..99M}; [51]\cite{2007MNRAS.378.1427C}; [52]\cite{1993IAUC.5836....1S}; [53]\cite{2015ApJ...805...18W} and \cite{2015ApJ...813L..26S}; [54]\cite{2022MNRAS.515.1078V}; [55]\cite{2021ApJ...913...48G}; [56]\cite{1975Natur.254..578I}; [57]\cite{2007AA...473..523V}; [58]\cite{2021JApA...42...58S}; [59]\cite{2010MNRAS.401.1532R}; [60]\cite{2020AA...638A..71S}; [61]\cite{2013arXiv1305.3916L}; [62]\cite{2005ATel..377....1C}; [63]\cite{2004ATel..362....1I}; [64]\cite{1987PASP...99..420H}; [65]\cite{2002ApJ...581.1293R}; [66]\cite{1980ApJ...239..651L}; [67]\cite{1987MNRAS.225..369C}; [68]\cite{2014ApJ...793...77C}; [69]\cite{2006AA...457..595K}; [70]\cite{1976ApJ...209L.119W}; [71]\cite{1985MNRAS.212..219B}; [72]\cite{2005AA...441..235Z}; [73]\cite{2012ApJ...755...64S}; [74]\cite{2001AA...377..148T}; [75]\cite{1977ApJ...216L..15M}; [76]\cite{2005MNRAS.364..455C}; [77]\cite{2018MNRAS.479.4849M}; [78]\cite{1992ApJ...387L..37J}; [79]\cite{2002MNRAS.337.1245L}; [80]\cite{2008AA...483..249M}; [81]\cite{2022MNRAS.510.2286N}; [82]\cite{2007ApJ...659..407S}; [83]\cite{2019ApJ...884...93C}; [84]\cite{2020MNRAS.496.1768D}; [85]\cite{2010ATel.3060....1K}; [86]\cite{1996AAS..120C.209F}; [87]\cite{1981ApJ...243..251K}; [88]\cite{2017ApJ...846..161C}; [89]\cite{2007MNRAS.374..999J}; [90]\cite{2004ARep...48...89A}; [91]\cite{2019ApJ...873...62H}; [92]\cite{2021ApJ...906...13C}; [93]\cite{2001ApJ...553L.165I}; [94]\cite{2015MNRAS.453..976F}; [95]\cite{1983ApJ...274..765K}; [96]\cite{2022ApJ...927..194M}; [97]\cite{2011ATel.3785....1J}; [98]\cite{2010ApJ...711.1306B}; [99]\cite{2013ApJ...775L..25C}; [100]\cite{2017MNRAS.471..355I}; [101]\cite{2005ATel..649....1C}; [102]\cite{2018AA...618A..61G}; [103]\cite{2005AA...433L..41L}; [104]\cite{2013ApJ...762...19F}; [105]\cite{2015MNRAS.446.4148I}; [106]\cite{2006ATel..779....1C}; [107]\cite{2006AA...453..133W}; [108]\cite{2009MNRAS.397L..11J}; [109]\cite{2019ApJ...873...86P}; [110]\cite{2019ApJ...879...34C}; [111]\cite{1999ApJ...523L..65T}; [112]\cite{2010ATel.2564....1M}; [113]\cite{2012MNRAS.422..199M}; [114]\cite{2008AA...486..293B}; [115]\cite{2011AIPC.1379..212J}; [116]\cite{2022MNRAS.509.5747S}; [117]\cite{2016MNRAS.461..816I}; [118]\cite{2006ApJS..163..372W}; [119]\cite{2007ApJ...661..447T}; [120]\cite{2006AA...448..261Z}; [121]\cite{2011MNRAS.417..573S}; [122]\cite{2020MNRAS.493.2694S}; [123]\cite{2017ApJ...843L..10S}; [124]\cite{2006AA...454..265L}; [125]\cite{2008AA...489..657Z}; [126]\cite{1998ApJ...508..854T}; [127]\cite{2009MNRAS.393..419S}; [128]\cite{2017AstL...43..664B}; [129]\cite{2013BCrAO.109...27N}; [130]\cite{2012AA...539A..21D}; [131]\cite{2002ApJ...578L.129S}; [132]\cite{2011AA...532A.124M}; [133]\cite{2005AA...439..255H}; [134]\cite{2019AA...622A.198N}; [135]\cite{2001ApJ...555..489O}; [136]\cite{2005MNRAS.363..882L}; [137]\cite{2013AA...558A..99S}; [138]\cite{1998ApJ...495..435K}; [139]\cite{2020MNRAS.498.2750C}; [140]\cite{2013ApJ...775L..24L}; [141]\cite{2022ApJ...927..139O}; [142]\cite{2008ATel.1679....1M}; [143]\cite{2005MNRAS.364..899C}; [144]\cite{2011ASSP...21..559C}; [145]\cite{2022ATel15614....1C}; [146]\cite{2022ATel15556....1W}; [147]\cite{2008ATel.1392....1G}; [148]\cite{2015arXiv150301087G}; [149]\cite{2001PASJ...53.1179B}; [150]\cite{1999NuPhS..69..220P}; [151]\cite{2013AA...556A..27S}; [152]\cite{2012ApJ...753....3B}; [153]\cite{1999ApJ...517..449F}; [154]\cite{2022AA...657A..58N}; [155]\cite{2011ApJS..196....6L}; [156]\cite{2010MNRAS.401.1564R}; [157]\cite{2007AA...467..249S}; [158]\cite{1999ApJ...517..956C}; [159]\cite{2021ApJ...909..153M}; [160]\cite{2016PhDT.......406B}; [161]\cite{2022MNRAS.509.5955S}; [162]\cite{2005ApJ...635.1217G}; [163]\cite{2019ATel12556....1H}; [164]\cite{1998ApJ...502L.129M}; [165]\cite{1998ApJ...496..386I}; [166]\cite{1984PASJ...36..679M}; [167]\cite{2017MNRAS.469.3056S}; [168]\cite{2004ApJ...617.1284L}; [169]\cite{2020MNRAS.498.4830J}; [170]\cite{2020AA...640A..96P}; [171]\cite{2007AA...474..903B}; [172]\cite{1997ApSS.252..439F}; [173]\cite{2021MNRAS.507L..19C}; [174]\cite{2004ATel..269....1C}; [175]\cite{2016MNRAS.460.3637S}; [176]\cite{2015MNRAS.446.1041C}; [177]\cite{2009ATel.2008....1C}; [178]\cite{2009ATel.2002....1S}; [179]\cite{2015MNRAS.451.2835D}; [180]\cite{2015ApJ...815...44M}; [181]\cite{2001AA...370..529P}; [182]\cite{2004ApJ...613.1164G}; [183]\cite{1995ApJ...446..826C}; [184]\cite{2020ApJ...904..143H}; [185]\cite{2006AA...460L...1M}; [186]\cite{2021Sci...371.1046M}; [187]\cite{1998MNRAS.301..285L}; [188]\cite{2013MNRAS.436..945P}; [189]\cite{2004HEAD....8.1730R}; [190]\cite{2002ApJ...570..287W}; [191]\cite{2021JApA...42...33J}; [192]\cite{2022ApJ...926..123A}; [193]\cite{2017ApJ...849..141B}; [194]\cite{1998ApJ...499..820W}; [195]\cite{2020MNRAS.497.1059K}; [196]\cite{2007MNRAS.374.1108B}; [197]\cite{1998AA...337L..25H}; [198]\cite{2020ATel14291....1P}; [199]\cite{2007AA...470.1065M}; [200]\cite{2021ApJ...920..139M}; [201]\cite{2014AN....335.1060S}; [202]\cite{1992ApJ...401..678S}; [203]\cite{2000AA...361...85I}; [204]\cite{2001AA...380L..26I}; [205]\cite{2010ApJ...723L..93W}; [206]\cite{2014Natur.505..378C}; [207]\cite{2021AA...649A.118L}; }

\end{document}